\documentclass{aa}  
\usepackage{natbib,twoopt}
\usepackage[breaklinks=true]{hyperref} 
\bibpunct{(}{)}{;}{a}{}{,}             
\makeatletter
  \newcommandtwoopt{\citeads}[3][][]{\href{http://adsabs.harvard.edu/abs/#3}%
    {\def\hyper@linkstart##1##2{}%
     \let\hyper@linkend\@empty\citealp[#1][#2]{#3}}}
  \newcommandtwoopt{\citepads}[3][][]{\href{http://adsabs.harvard.edu/abs/#3}%
    {\def\hyper@linkstart##1##2{}%
     \let\hyper@linkend\@empty\citep[#1][#2]{#3}}}
  \newcommandtwoopt{\citetads}[3][][]{\href{http://adsabs.harvard.edu/abs/#3}%
    {\def\hyper@linkstart##1##2{}%
     \let\hyper@linkend\@empty\citet[#1][#2]{#3}}}
  \newcommandtwoopt{\citeyearads}[3][][]%
    {\href{http://adsabs.harvard.edu/abs/#3}
    {\def\hyper@linkstart##1##2{}%
     \let\hyper@linkend\@empty\citeyear[#1][#2]{#3}}}
\makeatother

\usepackage{adjustbox}
\usepackage{xcolor}
\usepackage{graphicx}
\usepackage{txfonts}
\usepackage{tabularx, blindtext}
\usepackage{rotating}
\usepackage{pdflscape}
\usepackage{inputenc}
\usepackage{caption}
\usepackage{longtable}
\usepackage{lscape}
\DeclareCaptionFormat{continued}{#1 (continued)#2#3\par}
\DeclareGraphicsExtensions{.jpg,.png, .pdf}

\begin{document} 

\title{Unveiling the nature of 12 new low-luminosity Galactic Globular Cluster Candidates}
   \author{E. R. Garro \inst{1} 
          \and
          D. Minniti\inst{1,2}
          \and
		B.  Alessi\inst{3}
		\and 
		D. Patchick\inst{3}
		\and
		M. Kronberger\inst{4}
		\and
		 J.  Alonso-García\inst{5,6}    
		\and 
         J.~G.~Fernández-Trincado\inst{7}
         \and
         M. Gómez\inst{1} 
         \and 
		M. Hempel\inst{1}
		\and
		J. B. Pullen\inst{1}
		\and 
		R.K. Saito \inst{8}
		\and
		V. Ripepi \inst{9}
		\and
		R. Zelada Bacigalupo\inst{10}
          }
   \institute{Departamento de Ciencias Físicas, Facultad de Ciencias Exactas, Universidad Andres Bello, Fernández Concha 700, Las Condes, Santiago, Chile
   \and
 Vatican Observatory, Vatican City State, V-00120, Italy
 \and
Deepskyhunters Collaboration, e-mail: deepskyhunters@groups.io
 \and
 EBG MedAustron GmbH, Marie-Curie-Strasse 5, A-2700 Wiener Neustadt, Austria
  \and
 Centro de Astronomía (CITEVA), Universidad de Antofagasta, Av. Angamos 601, Antofagasta, Chile
 \and
 Millennium Institute of Astrophysics , Nuncio Monse\~nor Sotero Sanz 100, Of. 104, Providencia, Santiago, Chile
 \and
 Instituto de Astronom\'ia, Universidad Cat\'olica del Norte, Av. Angamos 0610, Antofagasta, Chile
 \and
  Departamento de Física, Universidade Federal de Santa Catarina, Trindade 88040-900, Florianópolis, SC, Brazil
  \and
INAF-Osservatorio Astronomico di Capodimonte, Via Moiariello 16, 80131 Napoli, Italy
  \and
North Optics Instrumentos Científicos, La Serena, Chile
}
  \date{Received XXX; Accepted YYY}
 
  \abstract
   {
The Galactic globular cluster system is incomplete, especially in the low latitude regions of the Galactic bulge and disk. We report the physical characterization of twelve star clusters in the Milky Way, most of which are explored here for the first time.}
   {
   Our primary aim is determining their main physical parameters, such as reddening and extinction, metallicity, age, total luminosity,  mean cluster proper motions (PMs), distances, in order to unveil the physical nature of these clusters.
   }
   {
We study the clusters using optical and near-infrared (NIR) datasets. In particular, we use the Gaia Early Data Release 3 (EDR3) PMs in order to perform a PM-decontamination procedure and build final catalogues with probable-members.  We match the Gaia EDR3 with the VISTA Variables in the Vía Láctea extended (VVVX) survey and Two Micron All Sky survey (2MASS) in the NIR, in order to construct complete NIR and optical colour-magnitude diagrams (CMDs) and investigate the clusters properties.
   }
   {
The extinctions are evaluated using existing reddening maps. We find ranges spanning $0.09 \lesssim A_{Ks}\lesssim 0.86$ mag and $0.89 \lesssim A_{G}\lesssim 4.72$ mag in the NIR and optical, respectively. Adopting standard intrinsic red clump (RC) magnitudes and extinction values, we obtain first the distance modulus for each cluster and thereafter their heliocentric distances,  that range from about 4 to 20 kpc. Therefore, we are able to place these clusters at $3 \lesssim R_{G}\lesssim 14$ kpc from the Galactic centre.  The best PARSEC isochrone fit yields a metallicity range of $-1.8<$~[Fe/H]~$<+0.3$ and an approximative age range of $2<$~Age~$<14$ Gyr. Finally, we find that all clusters have low-luminosities, with $-6.9<M_{V}<-3.5$ mag.
}
{
Based on our photometric analysis, we find both open clusters (OCs) and globular clusters (GCs) in our sample. In particular, we confirm the OC nature for Kronberger 100, while we classify Patchick 125 as a metal-poor GC,  Ferrero 54 as a metal-rich GC, and ESO 92-18 as a possible old OC or young GC. The classification as GC candidates is also suggested for Kronberger 99,  Patchick 122, Patchick 126, Riddle 15, FSR 190 and Gaia 2. We also conclude that Kronberger 119 and Kronberger 143 might be either old OCs or young GCs.}
   \keywords{Galaxy: bulge – Galaxy: center -- Galaxy: stellar content – Stars Clusters: globular – Infrared: stars – Surveys}
  
   \maketitle
   
\section{Introduction}
\label{Introduction}
One of the most important questions in astrophysics that still needs an explanation is how the Milky Way (MW) bulge was formed and how it has evolved to the present-state \citep{DiMatteo2015,Barbuy2018}.  Existing models (e.g., \citealt{Kormendy2004,Gadotti2009}) suggest that bulges can be classified as \textit{classical bulges}, if they result from a violent event (i.e. galaxy mergers or sinking of giant gas clumps); or \textit{disk-like bulges}, which may be the consequence of internal processes (e.g., disk instabilities) on longer time-scales. 
In the first case, older stellar populations are expected within a spherical structure; in the second, a wide range of ages for stellar populations are awaited in a flattened structure.\\

Very useful tools to reconstruct the MW history as well as to characterise its current evolutionary stage are the globular clusters (GCs). Being the first stellar associations (as well as dwarf spheroidal galaxies) formed in the early Universe with typical lifetimes of $12-14$ Gyr \citep{Salaris2002}, GCs retain fingerprints of past events that have occurred in their host galaxy (e.g., mergers/interactions). Thus, they provide a channel for tracing the MW's properties,  such as its structure,  dynamics,  compositions as well as the dark matter morphology.  However, the number of known MW GCs seems to be small in comparison with the Andromeda galaxy \citep{Harris_2013,Minniti_2017} or the Galactic open clusters (OCs; \citealt{Bonatto2010,Cantat_Gaudin2018,Cantat_Gaudin2020}),  indicating that our Galaxy has been inefficient at forming GCs, or strong processes have destroyed these objects, or many GCs have yet to be discovered.  Indeed, during the last two decades, many new GCs have been detected in the Galactic halo thanks to optical photometric surveys,  like the Sloan Digital Sky Survey (SDSS; \citealt{York_2000}), and also across the Galactic bulge and disk with surveys, such as the near-infrared (NIR) Two Micron All Sky Survey (2MASS, \citealt{Skrutskie2006}) and the VISTA Variables in the Via Lactea (VVV; \citealt{Minniti2010,Saito2012}) and its extension (VVVX; \citealt{Minniti2018}) surveys. Already more than 300 star cluster candidates have been discovered within the VVV/VVVX images, but a dedicated analysis needs to be done in order to confirm their physical nature (e.g., \citealt{Minniti_2017,Minniti2017_FSR}). In fact,  the discovery of new GCs towards the galactic inner regions can be useful to constrain formation and evolution models of the MW, especially when accurate physical parameters are estimated.  Significant effort has been devoted in the last years to studying the real nature of GC candidates and indeed many of them have been confirmed.  A few examples are VVV-CL001 \citep{Minniti2011, FernandezTrincado2021}, VVV-CL002 and VVV-CL003 \citep{MoniBidin2011},  FSR 1716 (a.k.a. Minni 22; \citealt{Minniti2017_FSR}), RLGC~1 and 2 \citep{Ryu2018}, Camargo 1107, 1108, and 1109 \citep{Camargo2019}, FSR 1758 \citep{Barba2019,RomeroColmenares2021},  Garro~01 \citep{Garro2020}, UKS~1 \citep{Fernandez_Trincado2020_UKS1},  Patchick~99 \citep{Garro2021},  FSR 19 and FSR 25 \citep{Obasi2021},  VVV-CL160 \citep{Minniti2021_CL160}, Minni~48 \citep{Minniti2021_M48},  FSR 1776 (a.k.a. Minni 23; \citealt{Dias:submitted-a}), Ton 1 \citep{JFT_TON1:submitted}, and other 9 GCs in \cite{Garro2021b:submitted-a}. \\

The present study provides the photometric analysis of 12 GC low-latitude candidates, shown in Fig. \ref{position},  including measurements of physical parameters, in order to reveal their real nature.\\
In Section \ref{clastercandidates}, we introduce these new GC candidates, listed in Table \ref{DP}. In Section \ref{datasets} we briefly quote the datasets adopted.  In Section \ref{methods}, we describe the methods used to investigate them and how we derive their main physical parameters.  We also add individual notes about each of these clusters in Section \ref{individualnotes}.  Finally, a summary and discussion are presented in Section \ref{end}.

\section{Cluster candidates}
\label{clastercandidates}
\begin{figure*}[!htb]
\centering
\includegraphics[width=14cm, height=8cm]{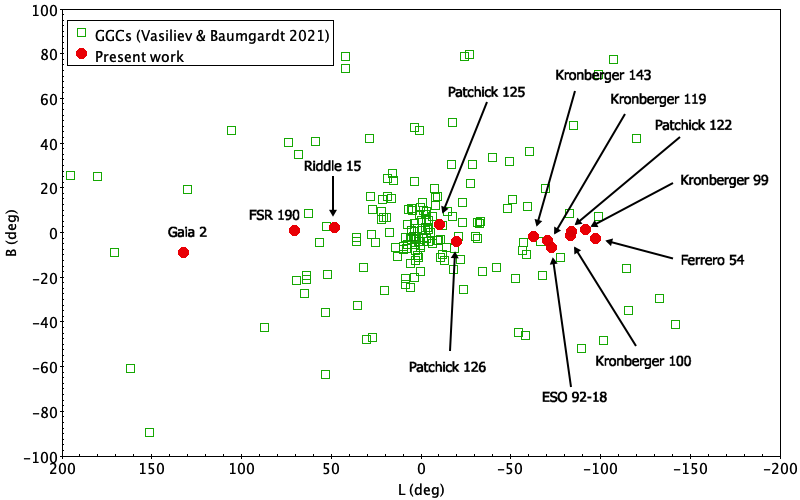}
\caption{Galactic distribution of GCs toward the MW bulge. Star clusters investigated in this work are highlighted with red points, while open green squares represent the Galactic GCs by \cite{Vasiliev2021} in the Gaia EDR3 footprint.}
\label{position}
\end{figure*}

The sample targets, listed in Table \ref{DP} and displayed in Figure \ref{position}, include 12 GC candidates. These were selected mostly from the clusters discovered by members of the Deep Sky Hunters collaboration, a world-wide group of amateur astronomers dedicated to finding new deep sky objects such as open clusters and planetary nebulae \citep{Kronberger2006,Kronberger2016}. The new clusters were originally discovered through a visual inspection of the DECam Plane Survey (DECaPS;  \citealt{Schlafly2018}) as concentrated local enhancements of the stellar density, as shown by Fig. \ref{images}.   Besides the recent DECaPS finds, we are also including some other older discoveries that have yet to be properly analysed: ESO 92-18 was discovered in the early 80's as part of the ESO/Uppsala Survey catalogues \citep{1982euse.book.....L}, but it also was detected by \cite{Froebrich2008} and was also listed as MWSC 3250 by \cite{Kharchenko2013};  Gaia 2 was an independent Digitized Sky Survey (DSS) discovery by \cite{Kronberger2012},  but also detected by \cite{Koposov2017} from Gaia Data Release 2 \citep{GaiaDR2_2018} observations; Riddle 15 was discovered by \cite{Kronberger2006} and also listed as MWSC 3063 by \cite{Kharchenko2013}.\\
These clusters are analysed here for the first time, and we have kept their original names after their discoverers (Kronberger, Patchick, Riddle, and Ferrero).

\section{The optical and NIR datasets}
\label{datasets}
Investigating these 12 clusters appears a tricky task to perform as many of these objects are located in obscured regions, behind dust clouds, or they can be so faint or diffuse that it is difficult to distinguish them against the background.  Also, the presence of patchy differential reddening in front of some of these clusters represents an additional obstacle to their detailed analysis.  Hence, it is necessary to employ distinct datasets at different wavelengths, in order to overcome these complications and obtain trustworthy results. 
We especially rely on NIR datasets, as in these bands the effect of extinction due to the dust is greatly reduced with respect to the optical. Also, since giant stars appear very luminous at these wavelengths, this can help to derive reliable physical parameters, such as distance and metallicity.\\
A brief description of the available datasets is presented below.

\subsection{VVV/VVVX photometry}
We use the NIR datasets of the VVV and VVVX surveys. The data were obtained using the 4.1m wide-field Visible and Infrared Survey Telescope for Astronomy
(VISTA; \citealt{2010Msngr.139....2E}) at ESO Paranal Observatory.  The VISTA Telescope tile field of view is 1.501 deg$^{2}$, and 196 tiles are used to map the bulge area ($-10^{\circ}<l<+10^{\circ}$, $-10^{\circ}<b<+5^{\circ}$) and 152 tiles for the disk ($-65^{\circ}<l<-10^{\circ}$, $-2.25^{\circ}<b<+2.25^{\circ}$).  Also, the extended VVVX campaign is sampling a larger area of the sky with the goal of extending the footprint of the VVV survey by surrounding this previous area in the bulge and the mid-plane areas at longitudes from 295$^{\circ}$ to 10$^{\circ}$.  Both the VVV and VVVX photometric datasets are divided into several bulge+disk tiles: $d001$ to $d152$ in the disk and $b201$ to $b396$ in the bulge for the VVV, whereas $b401$ to $b512$ in the bulge and $e601$ to $e988$ in the disk regions for the VVVX regions.  The VVVX area also includes an extension from 230$^{\circ}$ to 295$^{\circ}$ longitude and -2$^{\circ}$ to +2$^{\circ}$ latitude, those are sampled in the tiles $e1001$ to $e1180$. \\
The VVV data was reduced at the Cambridge Astronomical Survey Unit \citep{Irwin2004} and further processing and archiving was performed with the VISTA Data Flow System \citep{Cross2012}.  For the VVV clusters, we extract the photometry from the VVV catalog published by \cite{AlonsoGarcia2018}. This catalog was obtained using point-spread function techniques, and is calibrated into the VISTA photometric system \citep{Gonzalez_Fernandez2018}. A preliminary catalog from the VVVX data, obtained following a similar approach (Alonso-García et al., in prep), was used to extract the photometry for the clusters in the VVVX footprint. \\

The advantage of using this public survey is that its passbands ($ZYJHK_s$ in VVV and $JHK_s$ in VVVX) allows us to study in detail evolved stars in the colour-magnitude diagram (CMD), in particular red clump (RC) stars, which are good metallicity and distance indicators.  Unfortunately, not all star clusters considered here fall under the VVVX coverage. We detected in the VVV/VVVX images the following star clusters:
\begin{itemize}
\item Ferrero 54 in the VVVX tiles e706,705;
\item Kronberger 100 in the VVVX tile e1032;
\item Kronberger 143 in the VVV tile d003;
\item Patchick 125 in the VVVX tile e932;
\item Patchick 126 in the VVVX tile e676.
\end{itemize}

\subsection{2MASS photometry: brighter stars}
The 2MASS  is an all sky survey in $J$, $H$ and $K_s$ NIR bands. We recovered that photometry basically for two reasons. For all clusters that have a VVV/VVVX counterpart, the 2MASS photometry allows to build a CMD with stars from the MS to the RGB-tip, while VVV/VVVX stars with $K_s< 11$ mag are saturated.  To do this, since the 2MASS and VISTA photometric systems are tied but are different,  we transformed the 2MASS photometry into the VISTA magnitude scale following \cite{Gonzalez_Fernandez2018}. On the other hand, for the other clusters, such as Kronberger 99, Kronberger 119, Patchick 122, Riddle 15, FSR 190, ESO 92-18 and Gaia 2, which are outside the VVVX coverage, we obtain the NIR information for the brighter stars. Although in a few cases the 2MASS star sample appears to be small, we are still able to establish some of the physical parameters of our targets, such as their metallicity.

\subsection{Gaia EDR3 photometry}
The Gaia Early Data Release 3 (EDR3,  \citealt{GaiaEDR3_2021}) photometry is handled to build vector proper motion (VPM) diagrams, used to separate star cluster members from background/foreground stars and confirm the cluster nature of our sample.  Also, its precise astrometry is used to eliminate all nearby stars with distances $D<2$ kpc (Section \ref{decoproc}) from our source catalogues.  Additionally,  this photometry gives the possibility to construct deeper CMDs for those clusters without VVV/VVVX data, as we can appreciate for ESO 92-18 and Gaia 2 CMDs (Fig. \ref{CMD}).  No photometric colour or magnitude cuts were applied to the Gaia EDR3 data.

\section{Methods}
\label{methods}
\subsection{Decontamination procedure}
\label{decoproc}
\begin{figure*}[!htb]
\centering
\includegraphics[width=6.5cm, height=6cm]{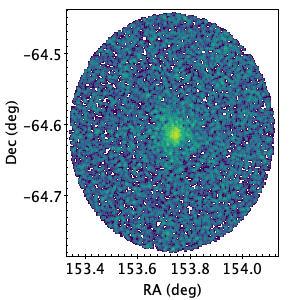} 
\includegraphics[width=6.5cm, height=6cm]{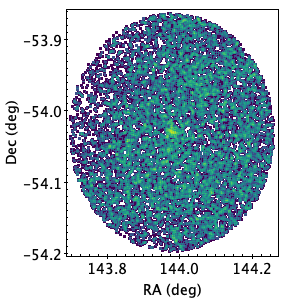} 
\caption{Gaia EDR3 density maps, cleaned from nearby stars, centred on ESO 92-18 (on the left) and Kronberger 100 (on the right) fields. These two clusters are used as representative in order to visually show the different target sizes.  The greener areas are overdensities, while the bluer areas are lower densities.}
\label{radiusDP}
\end{figure*}

\begin{figure*}[!htb]
\centering
\includegraphics[width=18cm, height=4.5cm]{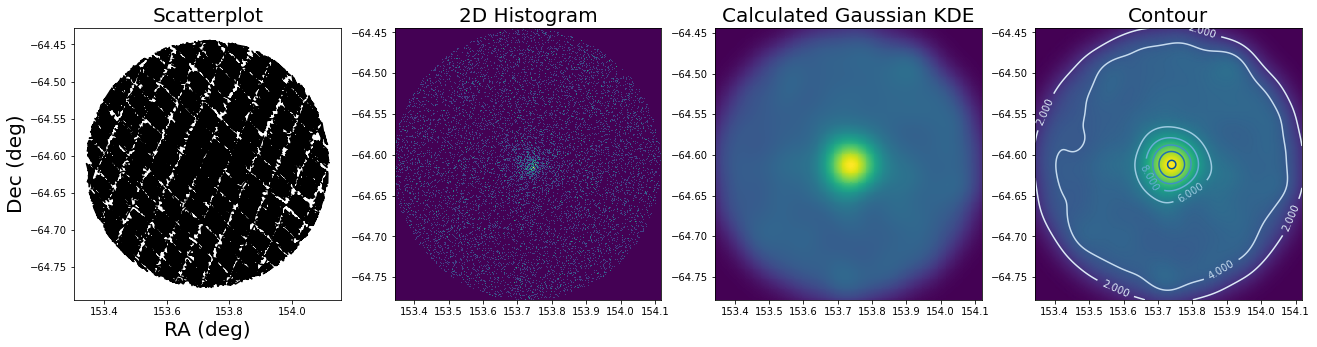} 
\includegraphics[width=18cm, height=4.5cm]{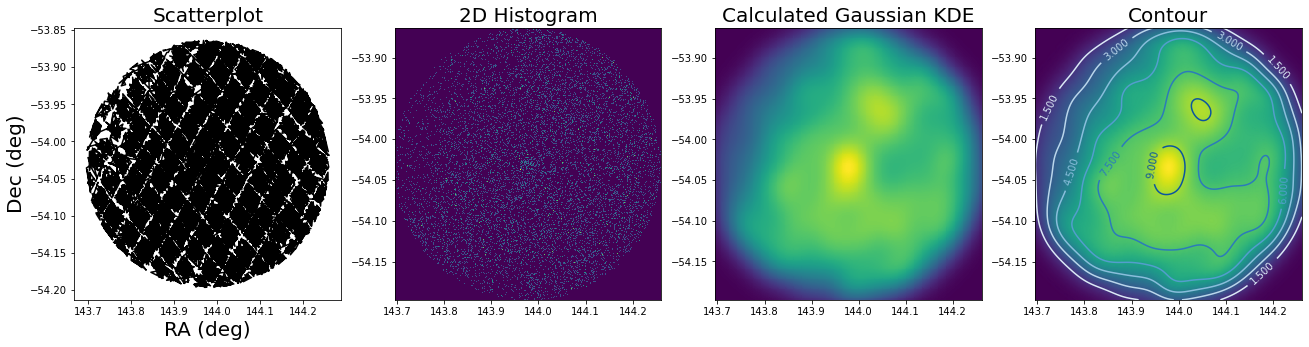} 
\caption{KDE technique applied to select the likely size of a cluster during the decontamination procedure.  From left to right: a scatterplot, 2D Histogram, Gaussian KDE and 2D density with contours (or iso-densities) are shown for ESO 92-18 (top panels) and Kronberger 100 (bottom panels) fields, used as representative.  Green and yellow areas are illustrative of overdensities, while the blue areas show lower densities.}
\label{KDEfig}
\end{figure*}

This analysis employs the combination of the VVV/VVVX and 2MASS photometry in $J$, $H$ and $K_s$ bands, and the optical Gaia EDR3 dataset in order to investigate the nature of these candidates.  \\
First, we join the VVV+Gaia EDR3 and/or the 2MASS+Gaia EDR3 datasets using a $0.5''$ matching radius. We treat separately the two NIR datasets, making a simple magnitude cut at $K_s=11$ mag.  After that, we apply a decontamination procedure in order to build a clean catalogue with likely-star members of every GC candidate, adopting the same procedure applied by \cite{Garro2020,Garro2021}.  As a preliminary step, we exclude all nearby stars with parallax values larger than $0.5$ mas. 
Subsequently, we select stars within a given radius $r$, listed in Table \ref{DP}, from the cluster centre.  We follow two methods in order to derive the cluster size: first, we inspect the density diagrams as a function of the sky position,  as shown in Fig. \ref{radiusDP},  in order to visually individuate as an overdensity the cluster dimensions; second, we apply the Gaussian Kernel Density Estimate (KDE; e.g., \citealt{Rosenblatt1956,Parzen1962}). Briefly, considering the scatterplot on the left hand side of Fig. \ref{KDEfig},  overlapping points make the figure hard to read. Even worse, it is impossible to determine how many data points are in each position.  In this case, a possible solution is to cut the plotting window into several bins (100-300 bins), and represent the number of data points in each bin by a colour. Following the shape of the bin, this makes  2D histogram. Then, it is possible to make a smoother result using Gaussian KDE. Its representation is called a 2D density plot, and we add a isodensity contours to denote each step.  Therefore, we balance both the KDE contours and the visual overdensities in order to select the stars within a given radius (Table \ref{DP}).  We adopt a GCs radius of $\sim 3'$ for Kronberger~99, Kronberger 143, Ferrero 54 and ESO 92-18, Gaia 2; while a larger size ($r\approx 10.2'$) for FSR 190, and smaller dimensions ($r<2.5'$) for Kronberger 100, Kronberger 119, Patchick 122, Patchick 125,  Patchick 126, Riddle 15. At this level, we benefit from exquisite Gaia PMs measurements, constructing the vector PMs diagrams (Fig. \ref{VPM:fig}),  in order to keep only stars with high probability of cluster membership. We are able to estimate the mean cluster PMs, listed in Table \ref{DP}, using the $\sigma$-clipping technique.  In our final datasets, we include only stars within 1 mas yr$^{-1}$ from the mean cluster PM values. Even though the assumption of a fixed PM radius value to select the final star sample is rigorous, it is linked to the fact that the PM overdensities, shown in the VPM diagrams (Fig. \ref{VPM:fig}), which indicate the presence of a cluster, are very small and peaked in most cases.  Also, this allowed us to keep only stars that are likely members of the clusters (with >50\% probability), minimizing the likelihood of including outliers as well as of excluding member stars.  For that reason, we also calculate the PM probability for each star within the cluster size, as shown in Fig. \ref{VPM:fig}.  However,  a very small fraction of contaminants in the final catalogues may be expected since these regions are characterised by high stellar crowding and by disk stars with similar PMs,  as demonstrated by \cite{Garro2021}.  Even if this occurs, it does not have any significant effect on our final photometric analysis, as the over-density peaks are clear in the VPM diagrams (Fig. \ref{VPM:fig}) against the background noise,  ensuring that the final catalogues include likely cluster members. \\
The resulting values are listed in Table \ref{DP}, where we summarise: the cluster identification (ID), the position, the radius cuts, the number of star members ($N_{\star}$) and the mean cluster PMs both in RA ($\mu_{\alpha_{\ast}}=\mu_{\alpha}\times \cos \delta$) and in Dec ($\mu_{\delta}$).

\begin{figure*}[!htb]
\centering 
\includegraphics[width=15cm, height=7.5cm]{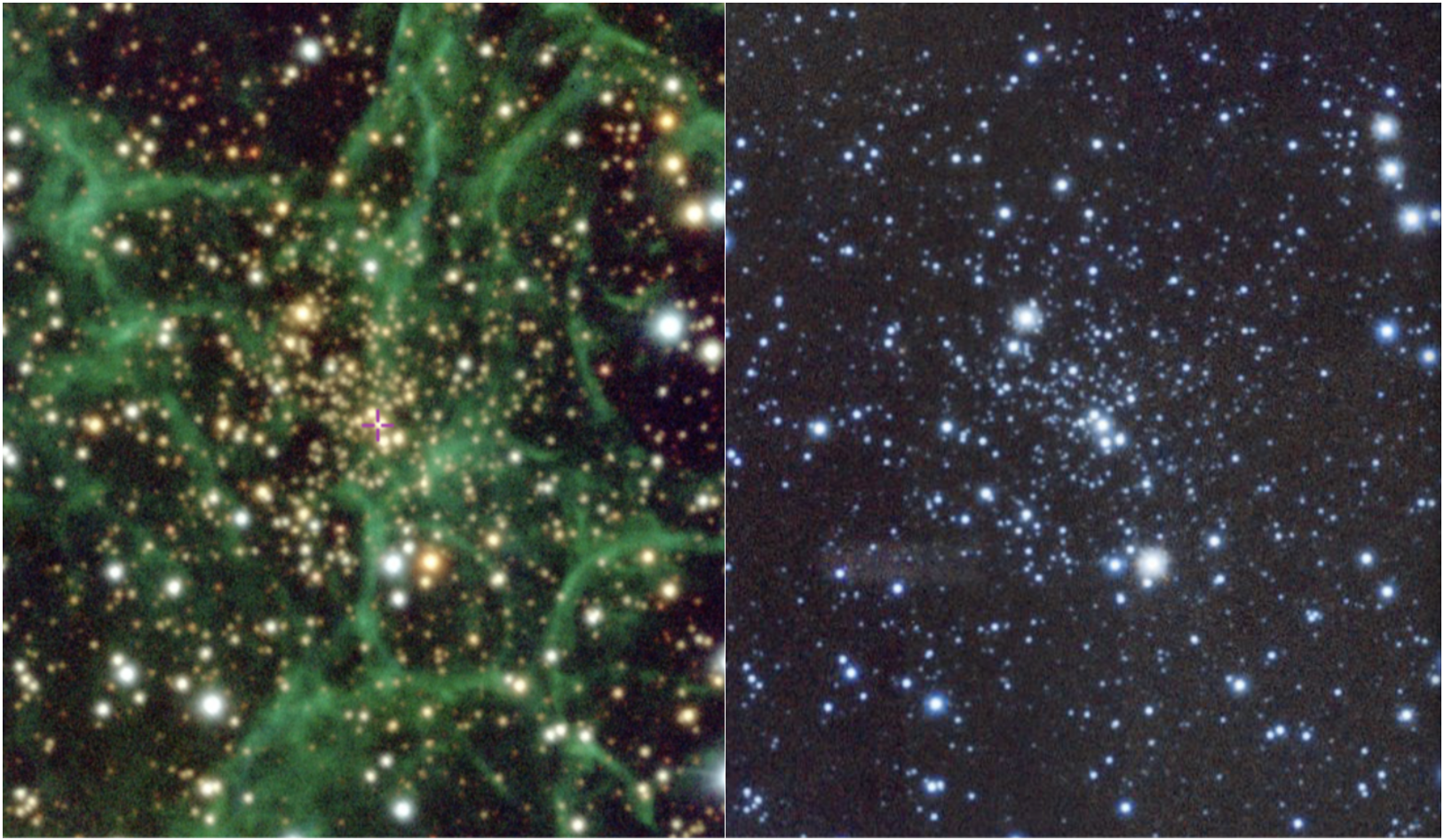} 
\caption{Optical DECaPS (left) and NIR VVVX (right) images of the sky region centred on Ferrero 54.  The field of view is $200'' \times 200''$, oriented along the Galactic coordinates, with Galactic longitude increasing to the left, and Galactic latitude increasing to the top of the images.  
Veils of gas and dust are evident as green filaments in the optical passband. They are part of the Vela supernova remnant complex.}
\label{ferrerofig}
\end{figure*}

\begin{table*}
\centering 
\caption{Positions, selection cuts applied for the decontamination procedure, number of members and mean cluster PMs.}
\begin{adjustbox}{max width=\textwidth}
\begin{tabular}{lcccccc}
\hline\hline
Cluster ID 	             &   RA        &    Dec	       &     r	   &      $N_{\star}$     &   $\mu_{\alpha_{\ast}}$	         &        $\mu_{\delta}$	  \\
			    & [hh:mm:ss] & [dd:mm:ss] & [arcmin] &  &[mas yr$^{-1}$] & [mas yr$^{-1}$] \\
\hline
Kronberger 99	    &  09:11:17.6	   &  $-$46:23:27    &   $3.0'$   &  	 21	   & $ -3.62\pm0.18	  $ & $ +4.46\pm 0.19	   $\\ 
Kronberger 100	    &  09:35:54.7	   &  $-$54:01:49   &   $1.7'$   &  	 187      & $ -3.79\pm0.33	  $ & $ +3.02\pm0.37	   $\\
Kronberger 119	    &  10:47:15.7	   &  $-$63:19:42    &   $2.4'$   &  	182      & $ -3.94\pm0.22	  $ & $ +1.80\pm0.24	   $\\
Kronberger 143	    &  11:57:42.6	   &  $-$64:10:40    &   $3.6'$   &  	 148      & $ -7.76\pm0.21	  $ & $ +0.66\pm 0.22	   $\\
Patchick 122	    &  09:42:30.7	   &  $-$52:25:41    &   $1.6'$   &    28	   & $ -3.72\pm0.12	  $ & $ +3.81\pm0.12	   $\\ 
Patchick 125	    &  17:05:00.7	   &  $-$35:29:41    &   $2.0'$   &  	146      & $ -3.85\pm0.50	  $ & $ +0.64\pm0.39	   $\\
Patchick 126	    &  17:05:38.6	   &  $-$47:20:32    &   $0.9'$   &  	 108      & $ -4.75\pm 0.46	  $ & $ -6.68\pm 0.62   $\\ 
Riddle 15	       &  19:11:08.9	   &  $+$14:49:59    &   $ <1'$   &  	83      & $ -1.03\pm0.32	  $ & $ -1.64\pm 0.27	$\\
FSR 190	          &  20:05:29.3     &  $+$33:34:01    &   $10.2'$  &  	52      & $ -2.41\pm 0.23	  $ & $ -3.50\pm 0.24	$\\
Ferrero 54	 &  08:33:48.3		&  $-$44:26:49 	 &   $3.0'$	  &  	 122	   & $ -1.33\pm0.27	  $ & $ +1.31\pm0.34	   $\\
ESO 92-18	       &  10:14:55.2	   &  $-$64:36:40    &   $3.0'$   &  	 557      & $ -3.54\pm 0.19	  $ & $ +2.72\pm0.18	   $ \\
Gaia 2	  &  01:52:33.0     &  $+$53:02:36   &   $3.0'	 $   &  	 83        & $ -1.31\pm 0.18  $ & $ +1.21\pm0.19   $\\
\hline\hline
\end{tabular}
\end{adjustbox}
\label{DP}
\end{table*}

\subsection{Estimation of physical parameters}
\begin{figure}[htpb]
\centering
\includegraphics[width=9cm, height=9cm]{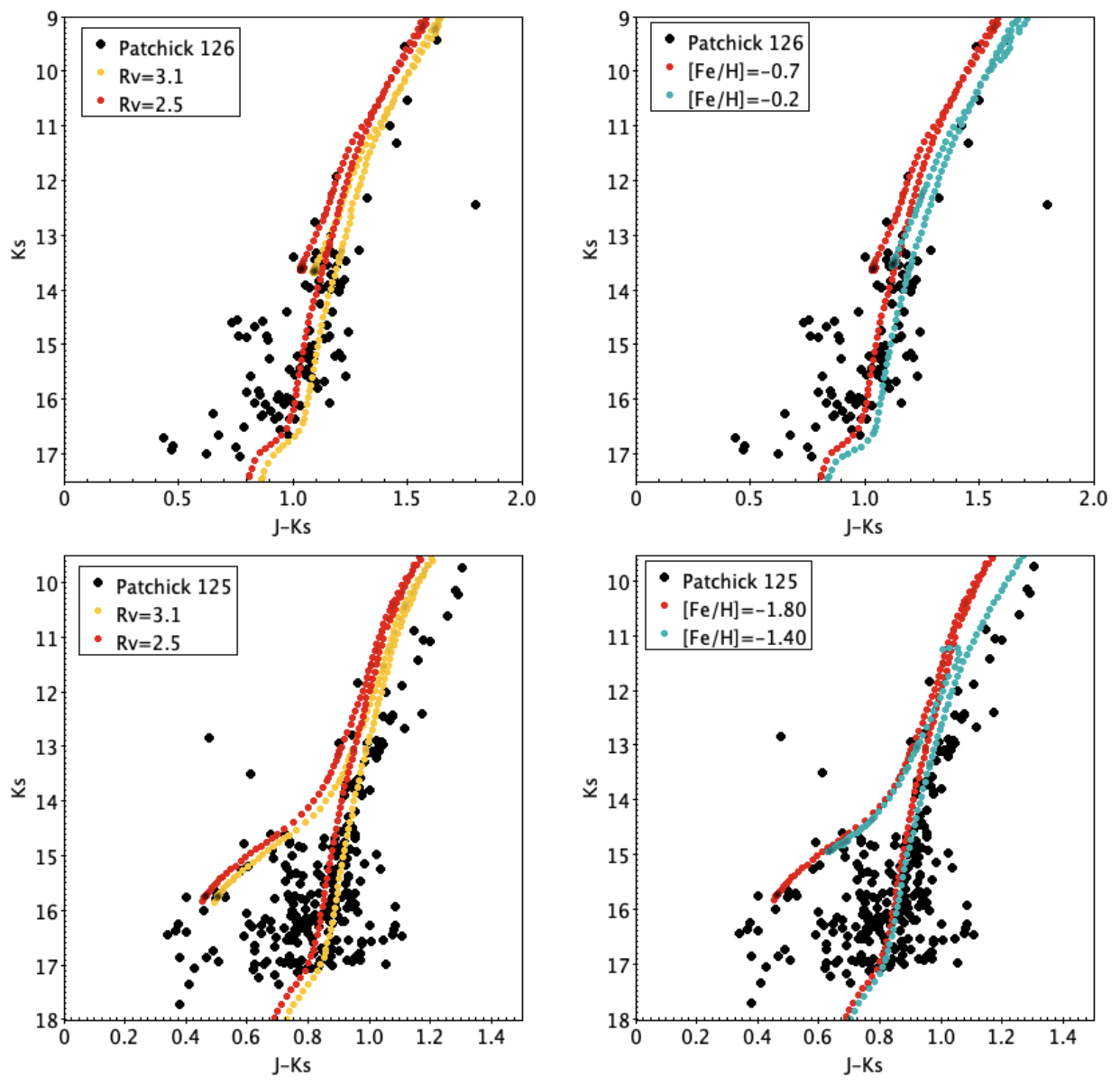} 
\caption{Patchick 125 (bottom panels) and Patchick 126 (top panels) NIR CMDs. On the left panels, we compare two isochrones using two different extinction coefficient $R_V=3.1$ (yellow points) and $R_{V}=2.5$ (red points). On the right panels, we fix the $R_{V}=2.5$ and the ages as listed in Table \ref{parameters}, and we change the metallicity as specified in the legend.}
\label{pat125_126}
\end{figure}

Once the final catalogues were obtained, we derived the main physical parameters for each candidate, such as: the reddenings and extinctions, the distances, the total luminosities,  the ages and the metallicities.  The derivation of these parameters is challenging because in those areas crowded with additional dust and gas (see Fig. \ref{ferrerofig}), the observations and precise measurements are made more complicated.
Moreover,  towards the innermost regions the differential reddening becomes more prominent increasing the uncertainties in the determination of distances.  Then, the extinction and distances are estimated following the same procedures described in our previous works e.g., \citealt{Garro2021} and \citealt{Minniti2021_M48}. We adopt existing reddening maps,  such as those by \cite{Gonzalez2011}, \cite{Soto2019} and \cite{Surot2019} in the NIR and by \cite{Schlafly2011} in the optical passbands.  
Therefore, we use the following relations to derive the color excess and the extinction:  $A_{V}=3.1\times E(B-V)$,  $A_{Ks}=0.11\times A_{V}$,  $A_{G}=0.86\times A_{V}$,  $A_{Ks}=0.75\times E(J-Ks)$,  $A_G=10.115\times A_{Ks}$,  and $A_{G}=2.0 \times E(BP-RP)$.  
We find extinction spans of $0.09\lesssim A_{Ks}\lesssim 0.86$ mag and $0.89\lesssim A_{G}\lesssim 4.72$ mag.  The extinction values are mandatory to compute reliable heliocentric distances, using the position of the red clump (RC).  The advantage of using $M_{Ks}^{RC}$ arises from its simple identification and low sensitivity to extinction by interstellar dust (e.g., \citealt{SalarisGirardi2002}). Therefore, we adopt the intrinsic RC magnitude $M_{Ks}^{RC} = -1.606 \pm 0.009$ mag and  $M_{G}^{RC}  = 0.383 \pm 0.009$ mag from \cite{RuizDern2018}, the observed NIR and optical RC magnitude listed in Table \ref{parameters} (column 2 and 3),  and the extinction values (column 3 and 4), in order to derive the distance moduli (column 6). These parameters can be translated into heliocentric distances (column 7 and 8), finding $4.4 \lesssim D_{NIR} \lesssim 18.1$ kpc from the NIR photometry.  Also,  both NIR and optical distance values agree well within the errors,  since we find $3.7 \lesssim D_{GEDR3}\lesssim 19.5$ kpc from the Gaia photometry.  However, clusters with blue horizontal branches usually do not show a defined RC,
especially in poorly-populated CMDs.  Therefore, we derive the color excess for the optical passband comparing
the cluster HB and RGB colours with the absolute colours by \cite{Babusiaux2018}. \\
The extinction coefficient that we use in our analysis is $R_{V}=3.1$, which is the standard value for diffuse interstellar medium (e.g., \citealt{Sneden1978}).  Recent works \citep{Nataf2013,Pallanca2021a, Pallanca2021b,Souza2021} have demonstrated that a low $R_{V}\approx 2.5 - 2.7$ is more appropriate to reproduce the bulge populations, especially in the regions where Patchick 125 and 126 are located.  Clearly,  a different extinction value has an effect on the colour excess and distance estimates.  We find that the distances increase of $\sim 0.1$ kpc for Patchick 125 and $\sim 0.16$ kpc for Patchick 126 when a low $R_{V}$ is adopted.  Nonetheless, both distance values, obtained adopting $R_V=3.1$ and 2.5, are in good agreement within the errors. Additionally,  we note that when we use $R_V=2.5$, and we keep the metallicities and ages, as listed in Table \ref{parameters},  the isochrones do not fit properly the points (especially along the RGB sequences) in the CMD,  unless a higher metallicity is used, as shown in Fig. \ref{pat125_126}.  However, we argue that $R_{V}=3.1$ is a good compromise,  if we base on the Patchick 125 metallicity of [Fe/H] $=-1.70$, spectroscopically found by \cite{JFT_Pat125:submitted} (read Sect. \ref{Patchick125}). \\

\begin{figure}[!htb]
\centering
\includegraphics[width=8cm, height=8cm]{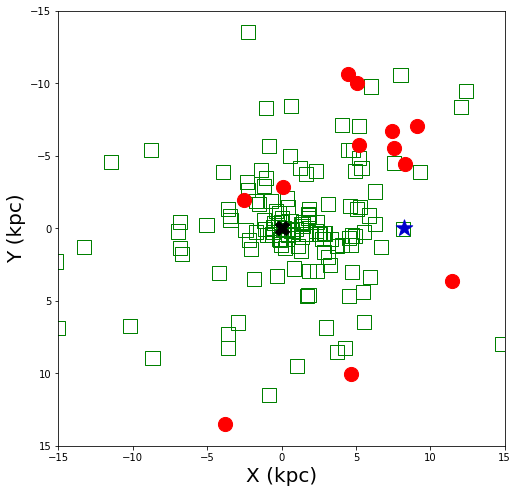}
\caption{Galactocentric distribution of star clusters analysed in the present work (red points) and Galactic GCs (green open squares) by \cite{Baumgardt2021}. The black cross represents the position of the Galactic centre, while the blue star is the position of the Sun.}
\label{galdistances}
\end{figure}

Moreover, we derive their Galactocentric X,Y, Z coordinates in kpc using the NIR distances.  As shown by Fig. \ref{galdistances},  we place our candidates at their respective distances $R_{G}$ from the Galactic centre, assuming $R_{\odot}=8.2$ kpc \citep{Gravity2019}. We find that they are located at Galactocentric distances ranging between $2.94$ kpc (for Patchick 126) and $14.06$ kpc (for Riddle 15), as listed in Table \ref{parameters}.  Additionally, we calculate the distance above or below the Galactic plane, using the relation $Z=D \times \sin(b)$ and assuming $Z_{\odot}=0$ kpc,  spanning $-1.23 \lesssim Z \lesssim 0.78$ kpc. 

\begin{figure}[!htb]
\centering
\includegraphics[width=6cm, height=6cm]{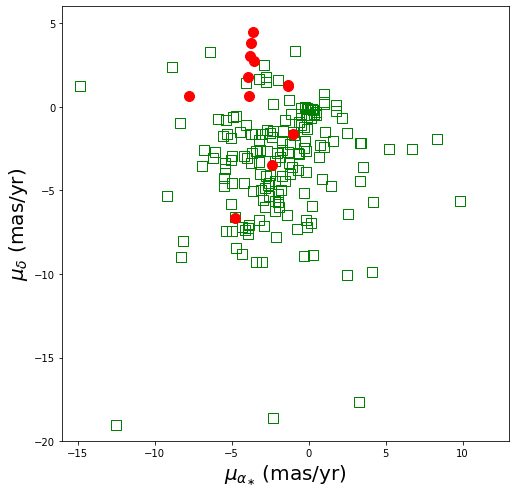} 
\includegraphics[width=6cm, height=6cm]{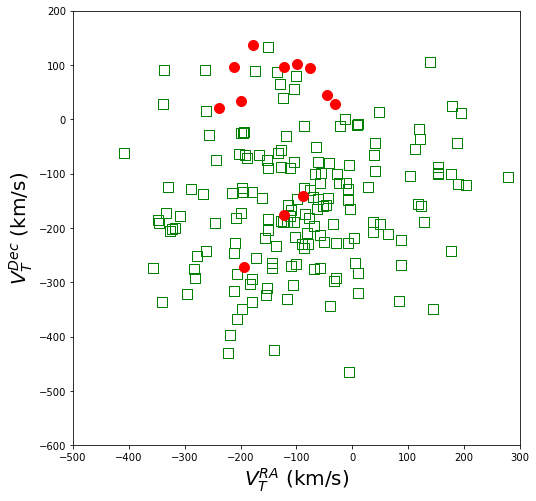} 
\caption{Gaia EDR3 PM-distribution of star clusters in the present work (red points) and Galactic GCs (open green squares) by \cite{Vasiliev2021}. }
\label{pmdistr}
\end{figure}

Once distances and mean PMs are obtained, we can compute their tangential velocities ($V_{T}^{RA}$ and $V_{T}^{Dec}$ ). Moreover, we compare the PMs and $V_{T}$ distributions with those of known and well-characterised MW GCs by \cite{Vasiliev2021} as shown in Fig. \ref{pmdistr}. We obtain negative $\mu_{\delta}$ for  Riddle 15, FSR 190 and Patchick 126, which show $-8.0<\mu_{\delta}<0$ mas yr$^{-1}$ and $-10.0<\mu_{\alpha_{\ast}}<0$ mas yr$^{-1}$; while all the others display $\mu_{\delta}>0$ mas yr$^{-1}$ and the same $\mu_{\alpha_{\ast}}$ range. Also, regarding $V_{T}$, we clearly obtain a trend similar to that of the PMs,  since we find negative $V_{T}^{Dec}$ for Riddle 15, FSR 190 and Patchick 126, which are $-300<V_{T}^{Dec}<-100$ km s$^{-1}$, while for the other clusters $V_{T}^{Dec}$ is always positive between 0 and 200 km s$^{-1}$, while the $-300<V_{T}^{RA}<0$ km s$^{-1}$ for all clusters. Obviously the clusters exhibit the longitude dependence due to the Solar motion around the Galactic center, but as a whole the new GCs identified here are consistent with the bulk motions observed for the rest of the MW globular cluster system.\\

\begin{figure*}[!htb]
\centering
\includegraphics[width=4.5cm, height=4.9cm]{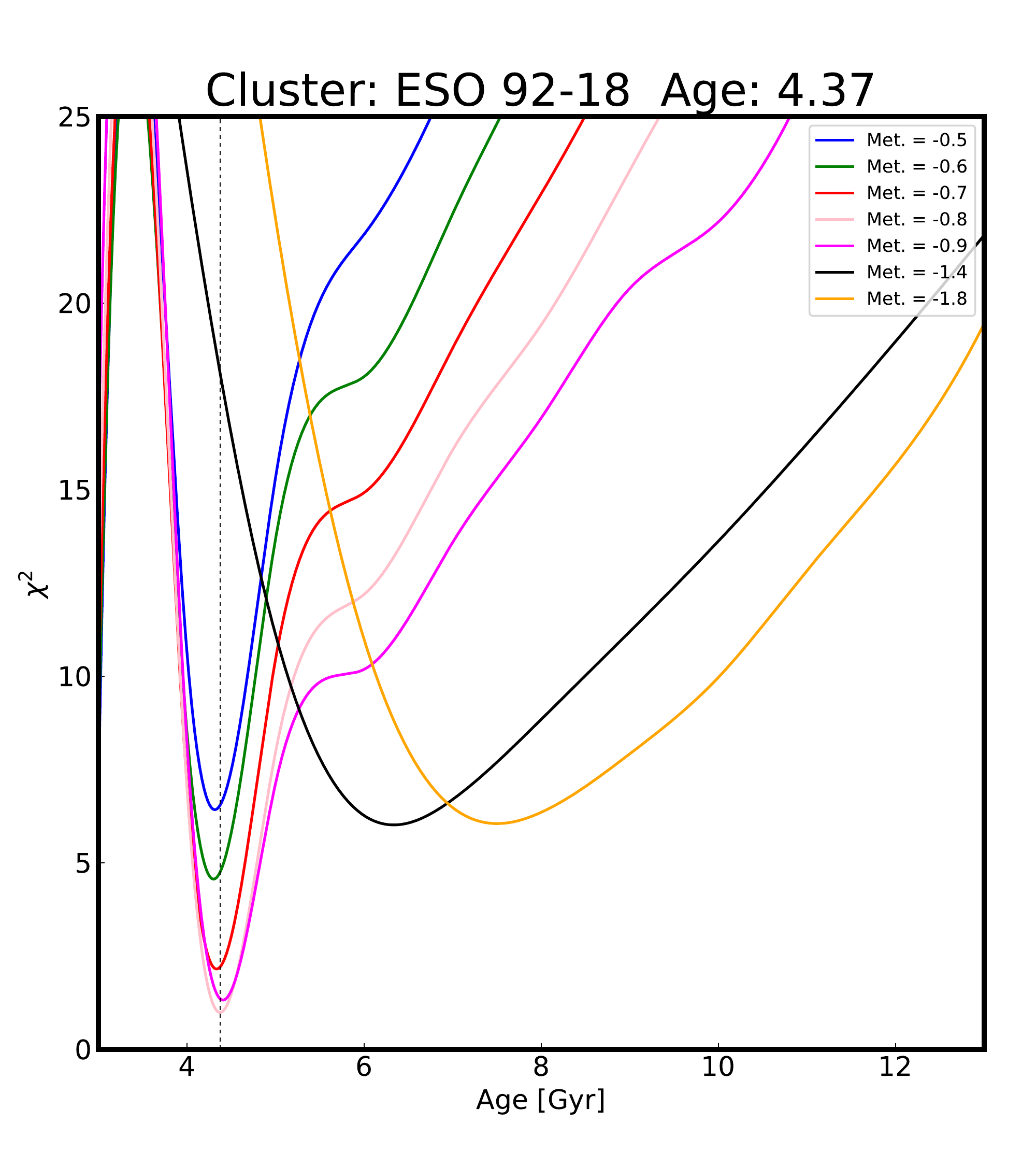} 
\includegraphics[width=4.5cm, height=4.9cm]{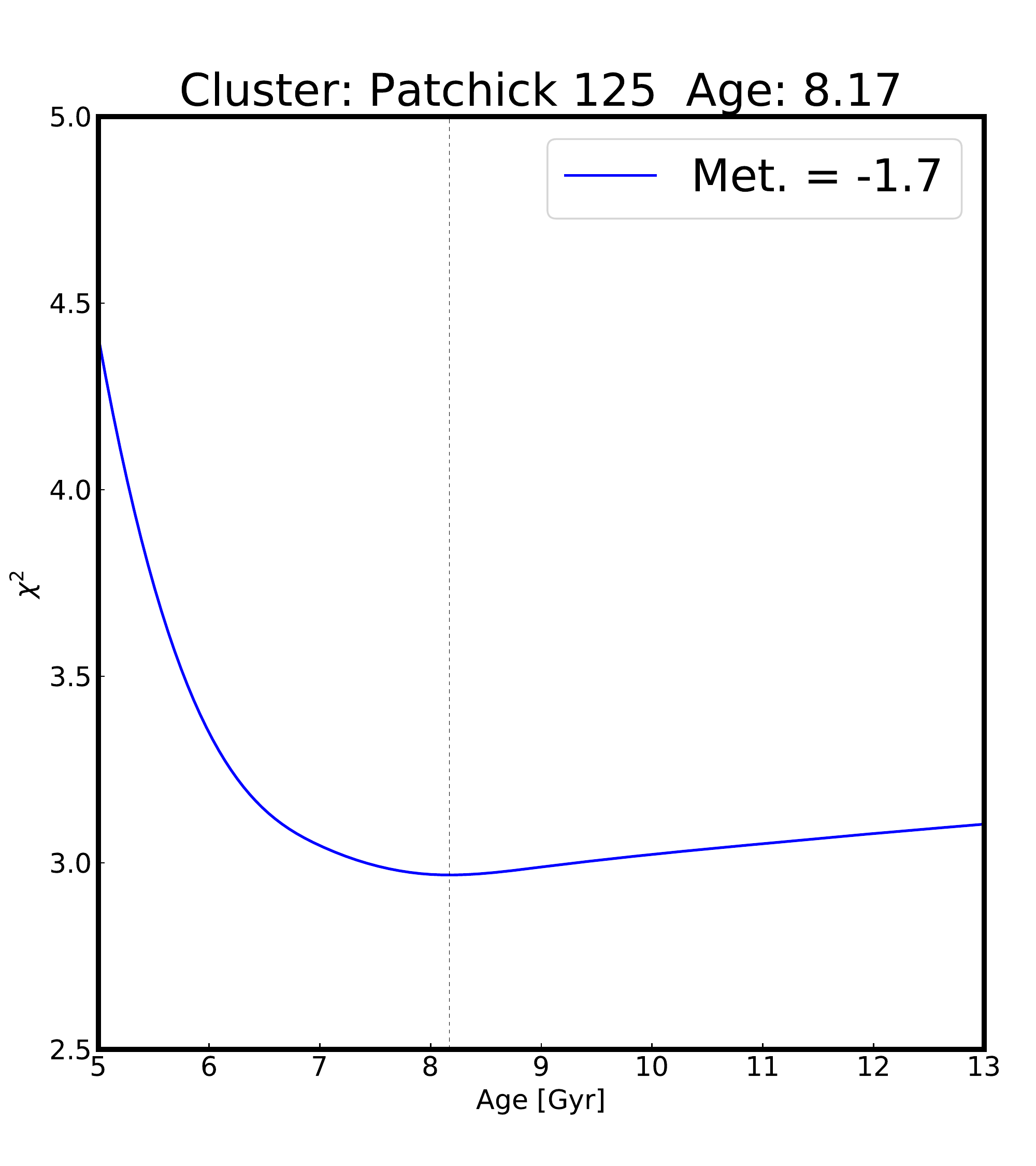} 
\includegraphics[width=4.5cm, height=4.9cm]{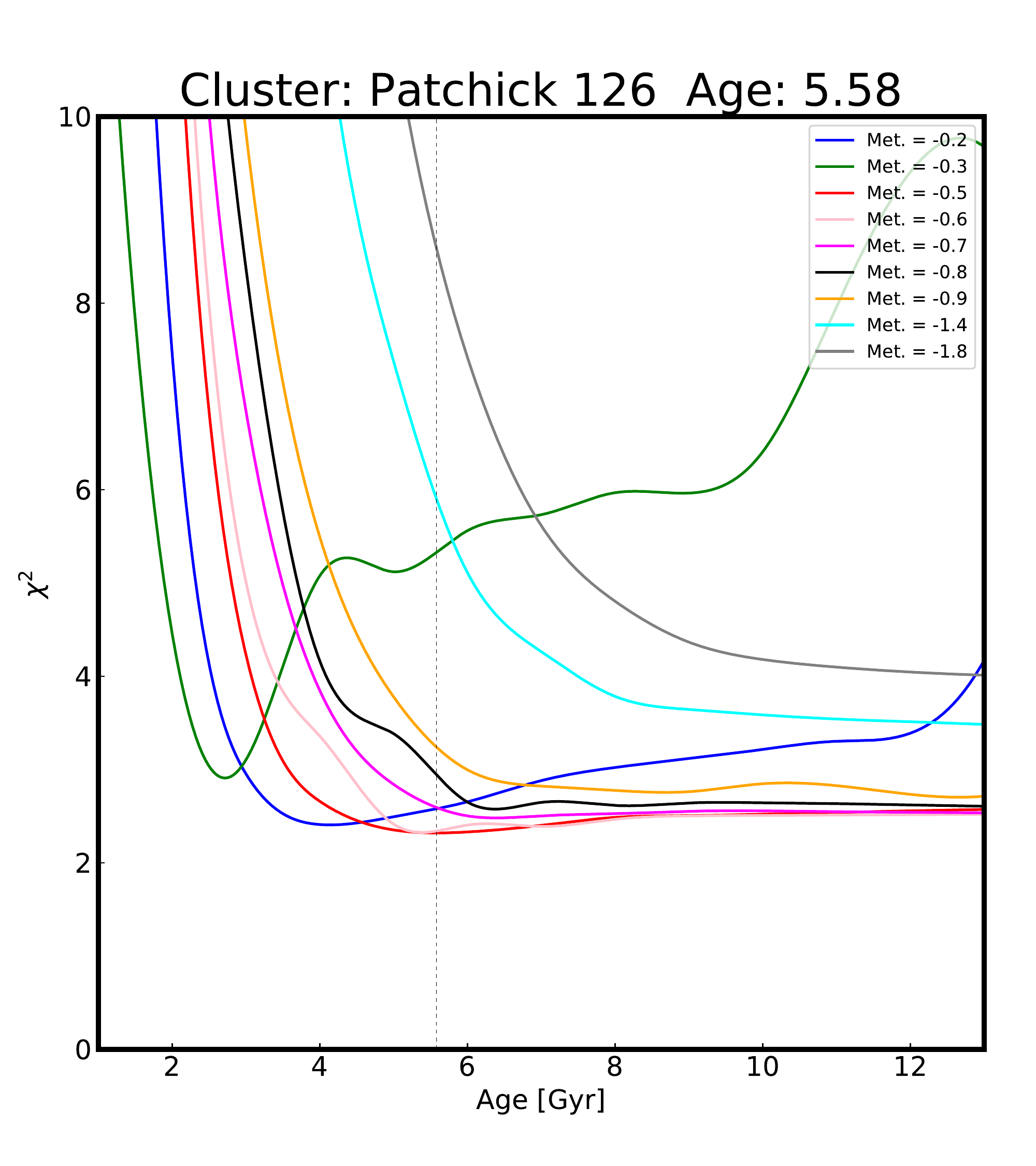} 
\includegraphics[width=4.5cm, height=4.5cm]{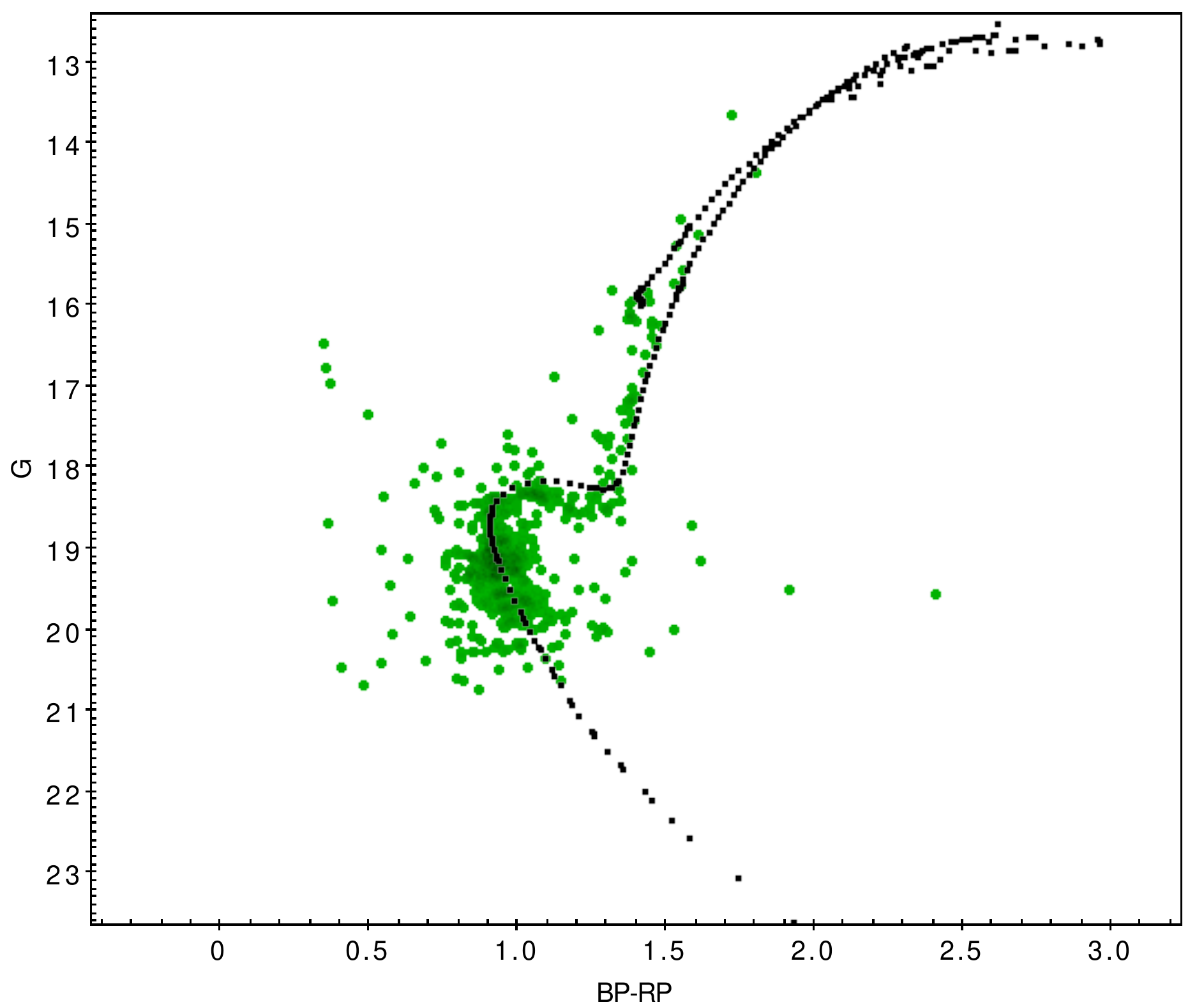} 
\caption{$\chi^{2}$ estimate as a function of the age  for three representative clusters: ESO 92-18, Patchick 125, Patchick 126. The colour of the lines depends on the metallicity, specified in the legend. The black dashed lines indicate the minimum $\chi^{2}$ value at the relative age. The age and metallicity that correspond to the $\chi^{2}_{min}$ are listed in Table \ref{chi2}.  On the right panel, we show the Gaia EDR3 CMD for ESO 92-18. We fit the PARSEC isochrone (black dotted line) using the age and metallicity obtained by the $\chi^{2}$ analysis and listed in Table \ref{chi2}. We clearly used the reddening, distance modulus and extinction as listed in Table \ref{parameters}. }
\label{chi2:fig}
\end{figure*}

Reddenings and distances are crucial to derive the cluster metallicities and ages.  One of the methods that allows to determine these parameters with reasonable reliability is the isochrone-fitting method.  Our CMDs in Fig. \ref{CMD} are populated by evolved stars, such as RC,  blue horizontal branch (BHB) and red giant branch (RGB). These stars are excellent indicators of the global metallicity content of a stellar population, as they are more sensitive to the metallicity variations \citep{McQuinn2019}.  However, for Kronberger 100, ESO 92-18 and Gaia 2,  fainter main sequence stars are detected by Gaia EDR3 photometry. We use the PARSEC models \citep{Bressan2012,Marigo2017} in  order to obtain the metallicity for each cluster, finding a range of $-1.8\lesssim$ [Fe/H] $\lesssim +0.3$ with a mean error of $\pm 0.2$ dex. On the other hand,  it is well-known that the main sequence turn-off (MSTO) is the best point to derive the absolute age of a cluster.  Yet,  some of our Gaia and VVVX/2MASS CMDs are not deep enough to reach (or go down $\sim 2$ mag below) the MSTO point, as we can see in Fig. \ref{CMD}, for example,  for Riddle 15, Patchick 126 and Kronberger 119.  Therefore,  our determination of the ages of these clusters is approximate, and deeper observations are necessary to confirm our results. However, we find a range of $2<$ Age $<14$ Gyr, well in agreement with old and intermediate-age Galactic GCs.  More precise ages are evaluated for Kronberger 100, Kronberger 119, Kronberger 143, Patchick 125, ESO 92-18 and Gaia 2, as we detail in Section \ref{individualnotes}.  Additionally,  we determine the errors in age and in metallicity by testing isochrones at a fixed metallicity and a fixed age, respectively, until they no longer fit the cluster sequence simultaneously in the NIR and optical CMDs (see Fig. \ref{CMD}).  \\
On the other hand, we apply the $\chi^{2} $ analysis in order to derive the age and the metallicity as an independent method.  We perform this analysis for all clusters on the Gaia CMDs, since in many cases optical CMDs are deeper than NIR CMDs, using a family of PARSEC isochrones. We adopt ages between 1 and 13 Gyr, and we change the metallicity values, depending on the [Fe/H] estimates, listed in Table \ref{parameters}.
Figure \ref{chi2:fig} shows the $\chi^{2}$ values as a function of the ages, whereas we list the likely age and metallicity corresponding to the minimum $\chi^{2}_{min}$ value in Table \ref{chi2}.  
We obtain a good agreement between the metallicity estimates obtained with the isochrone-fitting and $\chi^{2}$ methods, except for Kronberger 99, for which the $\chi^{2}$ method yields a lower metallicity ([Fe/H] $=-1.4$).  Nonetheless,  the age values represent a lower-limit again, especially for those clusters with few points,  as i.e.  Patchick 122.  This was expected,  since neither the MS nor the MS-TO are reached in some CMDs. Indeed,  for the ESO 92-18 (see Fig. \ref{chi2:fig}, right panel) and Gaia 2 cases,  we obtain an excellent agreement (within $1\sigma$) between the two methods,  due to the fact that the MS-TO and the higher part of the MS are above the detection limit.  Conversely,  the FSR 190 CMD is very noisy, therefore our procedure does not show a clear minimum for the age.  For that reason, we suggest that the estimates for that cluster found by the previous method and by the comparison with the literature (see Sect.  \ref{SectFSR190}) are more reliable. 

\begin{figure*}[!htb]
\centering
\includegraphics[width=6cm, height=6cm]{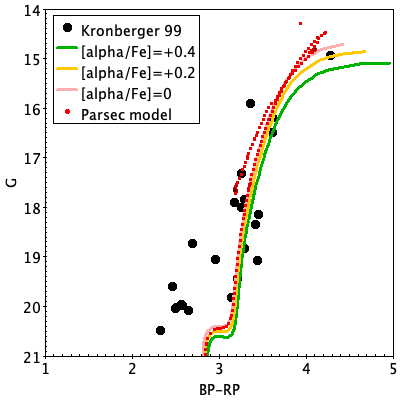} 
\includegraphics[width=6cm, height=6cm]{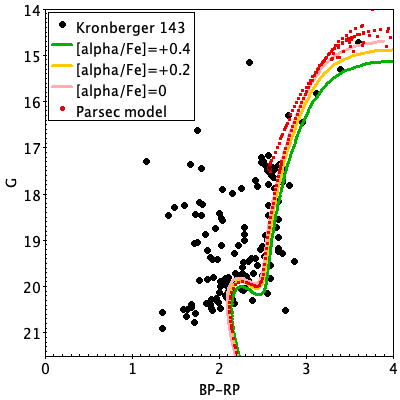} 
\includegraphics[width=6cm, height=6cm]{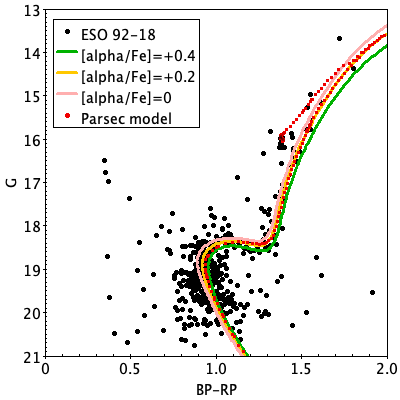} 
\caption{Gaia EDR3 optical CMDs for the clusters: Kronberger 99 (left panel), Kronberger 143 (middle panel), and ESO 92-18 (right panel). Green, yellow and pink lines are the Dartmouth isochrones with $[\alpha /Fe]=+0.4$, $+0.2$, $0$ respectively. The dotted red lines are the PARSEC isochrones. We keep the age and metallicity fixed as listed in Table \ref{parameters}. }
\label{isochrones}
\end{figure*}

Furthermore, it is important to note that the PARSEC isochrones are available for $[\alpha /Fe]=0$,  a typical value associated to OCs and metal-rich GCs. In the case of the metal-poor clusters, such as Kronberger 99, Kronberger 143 and ESO 92-18, an enhanced $[\alpha / Fe]$ value should be adopted. Therefore, for these clusters we also used the Dartmouth Stellar Evolution Database \citep{Dotter2008},  adopting enhanced $[\alpha /Fe]$ values, and keeping age and metallicity fixed, as listed in Table \ref{parameters}. In Fig. \ref{isochrones}, we compare the Dartmouth and PARSEC isochrones, finding a good agreement between these models. However, we suggest that these clusters should have $[\alpha /Fe] =+0.2$. \\
We categorise all clusters with Age $\gtrsim 8$ Gyr and metallicity [Fe/H] $\lesssim -0.1$ dex as GCs,  and the rest as OCs.\\

Finally, the total luminosity in $K_{s}-$band for each cluster is derived following the same approach of \cite{Garro2021_SGR}. We first derive the absolute magnitude ($M_{Ks}$) measuring the cluster total flux.  Although \cite{Kharchenko2016} have demonstrated that the larger contribution in luminosity arises from the brightest stars, since the first 10-12 members accumulate more than half of the integrated luminosity of the cluster,  we are aware that the resulting $M_{Ks}$ are a lower limit, because our photometry does not include fainter and lower mass stars.  Therefore, in order to quantify the fraction of luminosity missed we adopt the same procedure of \cite{Garro2021_SGR}, comparing the luminosity of our sample with those of known Galactic GCs. Once this luminosity fraction is obtained, we add this to the absolute magnitude and convert it into the absolute magnitude in $V-$band. In this way, we obtain an empirical correction to our clusters' luminosities under the assumption of similarity among Galactic GCs. We find that all clusters have low-luminosities with $-3.5<M_{V}<-6.9$ mag,  much fainter than the peak of the MW GC luminosity function (GCLF) is $M_{V}^{MW}=-7.4\pm 0.2$ mag from \cite{Harris_1991,ashman_1998},  also in agreement with the value found by \cite{Garro2021_SGR} $M_{V}^{MW}=-7.46\pm 1.04$ mag. \\


\begin{table*}[!htb]
\centering 
\caption{Main physical parameters for each star cluster candidate, using the VVV/VVVX, 2MASS and Gaia EDR3 photometry. }
\begin{adjustbox}{max width=\textwidth}
\begin{tabular}{lcccccccccccccc}
\hline\hline
Cluster ID             &$K/K_{s}$\tablefootmark{a}		&$  G$\tablefootmark{a}		    &$ A_{Ks}	$&$  A_{G}	 $&$ (m-M)_{NIR}	$&$D_{NIR}$\tablefootmark{b}&$ D_{GEDR3}$\tablefootmark{b}&$R_{G}$& Z &$ M_{Ks}$	 &  $M_{V}$\tablefootmark{c} &  Age\tablefootmark{d}	    & [Fe/H]    &Type\tablefootmark{e}	\\
   &[mag] &[mag]&[mag]&[mag] &[mag]&[kpc] &[kpc] &[kpc]&[kpc]&[mag]& [mag]& [Gyr] & [dex] & \\
   \hline
Kronberger 99  &$12.60\pm 0.03	 $&$ 17.9\pm  0.002	 $&$  0.62\pm 0.01	$&$  4.48\pm 0.09	 $&$ 13.23\pm 0.03  $&$ 4.42\pm 0.4        $&$ 3.70\pm 0.4	     $& $9.42\pm 0.4$ & 0.09 &$ -6.02\pm 1.9   $& $-3.6$&$ >6-8\ [8]	    $&$ -0.8\pm 0.2	$ &GC? \\  
Kronberger 100	&$12.92\pm 0.03   $&$ 17.37\pm 0.001   $&$  0.37\pm 0.01	$&$  2.76\pm 0.10	 $&$ 14.14\pm 0.03	$&$6.76\pm 0.4       $&$ 6.65\pm 0.5     $&$10.0\pm 0.5$& $-0.17$&$ -4.88\pm 1.8  $& -- &$	   2-5\ [3] $&$	+0.3\pm0.3   $ &OC\\  
Kronberger 119	&$13.90\pm 0.06	 $&$ 17.4\pm 0.003	 $&$  0.24\pm 0.01	$&$  1.85\pm 0.03	 $&$ 15.27\pm 0.06	$&$11.30\pm 0.5        $&$ 9.8\pm 0.5	     $&$11.54\pm 0.5$&$-0.74$&$ -6.27\pm 1.2   $& $-4.3$ &$ 6\pm 1	 $&$-0.5\pm0.2	$ &OOC\\  
Kronberger 143	&$12.90\pm 0.01	 $&$ 17.7\pm 0.001	 $&$  0.45\pm 0.01	$&$  3.12\pm 0.10	 $&$ 14.06\pm 0.02	$&$6.50\pm 0.4	       $&$ 6.30\pm 0.4      $&$7.81\pm 0.4$& $-0.22$&$ -6.86\pm 1.6   $& $-4.9$ &$  >5-8\ [5]   $&$ -0.6\pm0.2    $&OOC\\  
Patchick 122	&$12.75\pm 0.05   $&$ 18.2\pm 0.002	 $&$  0.63\pm 0.01	$&$  4.40\pm 0.10	 $&$ 13.72\pm 0.05	$&$5.60\pm 0.4	       $&$ 4.60 \pm 0.4     $&$9.41\pm 0.4$ &$0.04$&$ -6.12\pm 1.0   $& $-4.2$&$	  >6\ [8]	 $&$ -0.5\pm0.2	$ & GC?\\  
Patchick 125	&$--	 $&$ --	 $&$  0.33\pm 0.01	$&$  2.56\pm 0.05	 $&$ 15.18	\pm 0.05$&$10.90\pm 0.5       $&$ 11.2\pm 0.5     $&$3.23\pm 0.5$& $0.65$&$ -6.1\pm 0.8   $& $-3.8$&$  14\pm 2	 $&$ -1.8\pm0.2	$ & GC\\  
Patchick 126	&$13.50\pm 0.01$   &$	18.0\pm 0.002     $&$  0.44\pm 0.01 $&$  2.66\pm 0.07	 $&$ 14.66\pm 0.02	$&$8.60\pm 0.4	   $&$ 7.40\pm 0.5	     $&$2.94\pm 0.4$& $-0.57$&$ -5.56\pm 0.8   $& $-3.5$&$  > 8	  $&$ -0.7\pm0.3	$ & GC\\  
Riddle 15	   &$	--	    $&$ --      $&$  0.52\pm 0.01$&$  3.09\pm 0.06	 $&$ 16.29\pm 0.02	$&$18.10\pm 0.5        $&$ 19.5    \pm 0.5$&$14.06\pm 0.5$& $0.78$&$ -7.6\pm 0.8  $& $-6.2$ &$	 >10\ [13]	    $&$ -1.4\pm0.2	$& GC\\  
FSR 190\tablefootmark{f}	      &$14.4	 $&$ 20.1?	 $&$  0.86	$&$   4.72?	    $&$ 15.14	$&$10.67        $&$   9.5?     $&11.11& $0.18$&$ -8.36  $& $-6.9?$&$	  >7\ [9]	 $&$ -0.9\pm0.2	$ &GC?\\  
Ferrero 54	&$13.25\pm 0.03   $&$	17.8\pm 0.002	 $&$  0.60\pm 0.01	$&$  3.00	\pm 0.09 $&$ 14.27\pm 0.03 	$&$7.10\pm 0.4         $&$ 7.26\pm 0.5     $&$11.5\pm 0.4$&$-0.32$&$ -5.87\pm 1.7   $& $-4.1$&$	  >10\ [11] $&$ -0.2\pm0.2	$ & GC\\  
ESO 92-18&$13.51\pm 0.05   $&$16.3\pm 0.001	 $&$  0.09\pm 0.002	$&$  0.89	\pm 0.01 $&$ 15.12\pm 0.05	$&$10.60 \pm 0.5       $&$ 9.60\pm 0.5     $&$11.35\pm 0.5$&$-1.23$&$ -7.3\pm 1.5   $& $-5.6$&$	  5\pm 1  	 $&$ -0.9\pm 0.2   $& OOC\\  
Gaia 2	      &$   --   $&$   --    $&$  0.10\pm 0.002	$&$  0.97\pm 0.02	 $&$ 13.43\pm 0.01	$&$4.91	\pm 0.5    $&$ 4.43\pm 0.6     $&$12.03\pm 0.5$&$-0.75$&$ -5.4\pm 2.0 $&$-3.9$&$	 10\pm 1	 $&$ -0.9\pm0.2	$&GC\\  
\hline
\hline\hline
\end{tabular}
\end{adjustbox}
\tablefoot{
	\tablefoottext{a}{We refer to $K$ or $K_s$ magnitudes for all clusters analysed adopting the 2MASS or VVV/VVVX datasets, respectively.  Also, both $K/K_s$ and $G$ magnitudes listed in the table are the average magnitudes of the RC. }
	\tablefoottext{b}{We adopt subscripts \textit{NIR} and \textit{GEDR3} in order to indicate if the distance is derived from the 2MASS and/or VVV/VVVX in NIR or Gaia EDR3  (GEDR3) photometries.}
		\tablefoottext{c}{The absolute magnitude in $V-$band is calculated from the comparison with the well-known Galactic GCs following \cite{Garro2021_SGR}.}
   \tablefoottext{d}{We highlight the age used in the fit of the isochrones (Fig. \ref{CMD}) in square brackets.}
    \tablefoottext{e}{The abbreviations indicate:  \textbf{GC?}=globular cluster candidate;  \textbf{OOC}=old open cluster or young globular cluster; \textbf{GC}=globular cluster; \textbf{OC}=open cluster.}
    \tablefoottext{f}{We mark FSR 190, due to the high extinction which does not allow us to accurately calculate all the physical parameters especially in the optical passband. }    
    }
\label{parameters}
\end{table*}

\begin{table}[!htb]
\centering 
\caption{$\chi_{2}$ analysis performed for the clusters sample in order to statistically estimate age and metallicity values. }
\begin{adjustbox}{max width=\textwidth}
\begin{tabular}{lcc}
\hline\hline
Cluster ID 	           &     [Fe/H] (dex)&Age (Gyr)  \\
\hline
Kronberger 99 & $-1.4$ & 6.0\\
Kronberger 100 & $+0.3$ & 3.6\\
Kronberger 119 & $-0.9$ &4.4\\
Kronberger 143 & $-0.9$ &4.4\\
Patchick 122 &$-0.5$ &4.3\\
Patchick 125 & $-1.7$\tablefootmark{a}& 8.2\\
Patchick 126  &$-0.5\ / -0.6$ & 5.6\\
Riddle 15 &$-0.9\ /-1.4$ &10.4\\
FSR 190  & $-0.9$& --\\
Ferrero 54  & $-0.3$ & 6.5\\
ESO 92-18& $-0.8$ &4.3\\
Gaia 2 & $-0.9$ & 9.8\\
\hline\hline
\end{tabular}
\end{adjustbox}
\tablefoot{
	\tablefoottext{a}{We adopt the metallicity found by \cite{JFT_Pat125:submitted}.}}
\label{chi2}
\end{table}

\section{Fundamental parameters for each cluster}
\label{individualnotes}

\subsection{Kronberger 99}
We analyse Kronberger 99 by matching Gaia EDR3 and 2MASS datasets, because this cluster is outside of the VVVX field.  Kron-berger~99 appears as an overdensity both in the Gaia EDR3 and 2MASS images, especially when foreground stars are excluded.  Inspecting the VPM diagram, we can easily extract the mean cluster PM, as shown in Fig. \ref{CMD}, however after the PM-decontamination procedure, we find a very poorly-populated cluster with 21 giant members.  We select it as a GC candidate, since a scarcely-populated RC can be observed at $K = 12.6$ mag and $G=17.9$ mag.  The analysis yields an extinction and distance modulus for the cluster of $A_{Ks}=0.62$ mag and $(m-M)_{0}~=~13.23$ mag, equivalent to $D_{NIR}=4.42$ kpc, in good agreement with $D_{GDR3}=3.7$ kpc, within 1$\sigma$.  This also places this cluster at $R_{G}=9.42$ kpc from the Galactic centre. The best PARSEC isochrone fit implies a metallicity of [Fe/H]~$= -0.8\pm 0.2$. On the other hand, due to the sparse population and lack of faint stars, it is very difficult to establish the age of this cluster. However, at [Fe/H]$=-0.8$,  we superimpose several isochrones with different age values, finding that all models with Age~$<6$ Gyr deviate considerably from the position of the giant stars in the CMD, in particular the RGB sequence, which is not the case for older ages.  Also, computing the difference in magnitude between the HB and MSTO (e.g., \citealt{SalarisCassisi2005,CassisiSalaris2013}) we find that $ \Delta K $ (HB-MSTO) is greater than 1.5 mag, which can be translated into a lower age limit of $ \sim 6 $ Gyr,  which is also in agreement with that found by the $ \chi^{2} $ analysis.

\subsection{Kronberger 100}
Kronberger 100 is investigated using the matches between Gaia~EDR3+2MASS and Gaia EDR3+VVVX datasets.  Approximately twenty bright stars are evolved off the MS, whereas the bulk of the probable cluster members form a populated MS, as we can appreciate from the CMDs (Fig. \ref{CMD}).  We estimate an extinction of $A_{Ks}=0.37$ mag in the NIR and $A_{G}=2.76$ mag in the optical.  This places Kronberger~100 at a distance modulus of $(m-M)_{0}=14.14$ mag, equivalent to $D_{NIR}=6.76$ kpc, which is in very good agreement with the heliocentric distance obtained from the Gaia EDR3 photometry ($D_{GDR3}=6.65$ kpc.)  Additionally, we obtain a distance from the Galactic centre of $\sim 10.0$ kpc. The MSTO point lies near $K_s=15.5\pm 0.1$ mag. The best PARSEC isochrone fit and also the $\chi^{2}$ analysis yield a mean cluster age of $\sim 3$ Gyr, however ages between 2 and 5 Gyr may be adequate to reproduce the sequences of this cluster,  while the mean metallicity is [Fe/H] $=+0.3\pm 0.3$ dex.  Hence, we classify it as an open cluster (OC), containing 187 stars. 

\subsection{Kronberger 119}
Matching Gaia EDR3 and 2MASS catalogues, we analyse Kron-berger 119, finding a cluster with 182 members.  Both in the 2MASS and Gaia CMDs a clear overdensity depicting the RC is seen at $K=13.9$ mag and $G=17.4$ mag, respectively.  This helps us to derive the extinction of $A_{Ks}= 0.24$ mag and $A_{G}=1.85$ mag, and also the distance modulus of $(m-M)_0=15.27$ mag, corresponding to $D_{NIR}=11.30$ kpc.  However, we found a smaller distance of $D_{GEDR3}= 9.8$ kpc from the Gaia photometry.  Adopting the NIR distance, we place this cluster at $\sim 11.54$ kpc from the Galactic centre.  As we can appreciate from Fig. \ref{CMD},  the MSTO is not reached in the 2MASS CMD, however we have a deeper optical CMD, which allows us to derive a good age estimate, since we evaluated as MSTO point the overdensity at $G=19.9 \pm 0.1$ mag, evident also when cluster luminosity function is used.  Therefore, fitting the PARSEC isochores, we categorise Kronberger 119 as a young GC or an old OC with an Age~$=6\pm 1$ Gyr and mean metallicity of [Fe/H]~$=-0.5\pm 0.2$ dex.  However, we found slightly younger age of 4.4 Gyr and lower metallicity of $-0.9$ with the $\chi^{2}$ analysis.

\subsection{Kronberger 143}
Kronberger 143 falls in the VVVX coverage area, so we examine this cluster using a match of Gaia EDR3+2MASS and Gaia EDR3+VVVX catalogues.  From our decontamination procedure, we find that this cluster contains 148 members. Also, a well-defined RGB and RC, and the TO region are demarcated both in the NIR and optical CMDs (Fig. \ref{CMD}).  We estimate its main cluster parameters: extinction of $A_{Ks}=0.45$ mag and $A_{G}=3.12$ mag, $(m-M)_{0}=14.06$, equivalent to $D_{NIR}=6.5$ kpc, in excellent agreement with $D_{GEDR3}=6.3$ kpc.  This places it at $7.81$ kpc from the Galactic centre. We classify Kronberger~143 as a young GC with an Age~$>5$ Gyr, which represents a solid lower limit due to the fact that when younger isochrones are fitted, they do not reproduce any sequences (especially the MSTO region), while models with ages younger than 8 Gyr are more suitable for this cluster.  Similar results are achieved through the $\chi^{2}$ analysis.  

\subsection{Patchick 122}
Patchick 122 is investigated using a combination of Gaia EDR3 and 2MASS photometry.  This is a small ($r\approx 1.6'$) and poorly populated cluster with 28 star members.  We select it as GC candidate since a clear overdensity is obvious in the VPM diagram (Fig. \ref{CMD}), but also a minute RC can be identified at $K_s=12.75$ mag and $G=18.2$ mag.  We can speculate about it, suggesting that this may be a dissolved cluster, a survivor of strong dynamic processes,  or that we suffer of incompleteness.  Nonetheless, we derive its main physical parameters, such as an extinction of $A_{Ks}= 0.63$ mag and $A_{G}=4.4$ mag,  and a distance modulus of $(m-M)_0=13.72$, corresponding to $D_{NIR}=5.6$ kpc, in agreement within $2\sigma$ with the optical estimate of  $D_{GDR3}=4.6$ kpc.  Using the NIR heliocentric distance, we place this cluster at $R_{G}=9.41$ kpc from the Galactic centre.  Patchick 122 may be a good GC candidate, since when PARSEC isochrones are fitted we obtained an Age~$>6$ Gyr and metallicity of [Fe/H]~$=-0.5\pm 0.2$ dex. Clearly, it is very difficult to reliably pinpoint the age of this cluster given the scarcity of stars detected.  Hence, in order to overcome the age-metallicity degeneration, we estimate the age based on the vertical method.  We find $\Delta G$ (HB-MSTO) $\gtrsim 2.0$ mag and $\Delta K$  (HB-MSTO) $\gtrsim 2.5$ mag, which can be translated into a lower limit of $\sim$ 6 Gyr.  On the other hand, the $\chi^{2}$ analysis suggests a younger age of 4.3 Gyr.  For that reason,  we fit models with Age ~$<6 $ Gyr but keeping the metallicity fixed at [Fe/H]$=-0.5$ (obtained also by the $\chi^{2}$ procedure),  finding that the isochrones shift to lower magnitudes and towards the reddest part of the CMD, deviating significantly from the brightest stars.  Therefore, we can exclude that this cluster is younger than 6 Gyr.

\subsection{Patchick 125}
\label{Patchick125}
We analyse Patchick 125 matching Gaia EDR3+2MASS and Gaia EDR3+VVVX datasets.  After the decontamination procedure, we find a small cluster with 146 star members.  We derive its main physical parameters, obtaining an extinction of $A_{Ks}=0.33$ mag in NIR and $A_{G}=2.56$ mag in optical, a distance modulus of $(m-M)_0=15.18$ mag, equivalent to a heliocentric distance of $D_{NIR}=10.90$ kpc, in good agreement with the optical passbands ($D_{GEDR3}=11.2$ kpc).  The NIR distance places this cluster at $3.23$ kpc from the Galactic centre. Analysing the PM-decontaminated CMDs, we can clearly observe an extended BHB, suggesting a metal-poor content. Indeed, the best PARSEC isochrones fit yields an Age~$=14 \pm 1$ Gyr and [Fe/H]~$=-1.80\pm 0.2$ dex, confirming that Patchick 125 is an old GC. We evaluated its age using two observables: \textit{(i)} calculating the difference in magnitude between the HB and MSTO, finding $\Delta K_s$(HB-MSTO) $>4$ mag and $\Delta G$(HB-MSTO) $>3$ mag, translating it into a solid limit of $\sim 13$ Gyr; \textit{(ii)} we considered the RGB-tip, which is less sensitive to age changes,  but fitting models with Age~$<13$ Gyr the isochrone-tip moves to brighter magnitudes. Therefore, we suggest that its age may be very old, in contrast with the $\chi^{2}$ method. \\
Additionally, we find that Patchick 125 is located within $\sim 10.2''$ from the centre of the recently discovered GC Gran 3 \citep{Gran2021}. Therefore, we believe that both objects are the same cluster. However, we keep the original name,  referring us to the astronomer (Dana
Patchick in 2018, internal communication) who discovered it first.  Comparing the present work with \cite{Gran2021},  we find similar results. In particular, the mean cluster PMs $\mu_{\alpha_{\ast}}=-3.78$ mas yr$^{-1}$ and $\mu_{\delta}=+0.66$ mas yr$^{-1}$,  the Gaia EDR3 extinction $A_{G}=2.60$ mag and the absolute magnitude in $V-$band $M_{V}=-6.02$ mag are in excellent agreement with our results. However, they found a more metal-poor cluster with [Fe/H] $=-2.37\pm 0.18$, relying on the relation between CaT EW and the magnitude of the HB in the Jonhson V filter.  On the other hand, our metallicity is in good agreement with that derived by \cite{JFT_Pat125:submitted}, since they obtained a mean [Fe/H] $=-1.70$, measuring the Fe~I lines from the APOGEE-2S data.  \\

\begin{figure}[htpb]
\centering
\includegraphics[width=9cm, height=9cm]{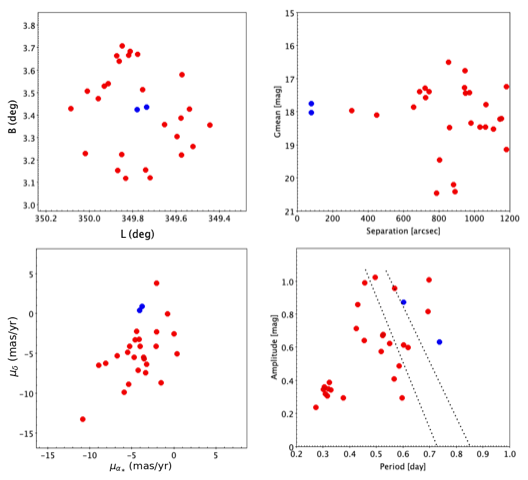} 
\caption{Position (top left panel),  magnitude-separation (top right panel), Gaia EDR3 PM-distribution (bottom left panel) and period-amplitude (bottom right panel) diagrams for the RR Lyrae stars located within $20’$ from the centre of the cluster Patchick 125.  The mean ridge lines for the Oosterhoff type I and II s populations (left and right dotted lines, respectively) are taken from \cite{Clement1999}. We mark as blue points the two RR Lyrae members.  }
\label{rrl_p125}
\end{figure}

Moreover, we searched for RR Lyrae stars in the cluster field. Indeed, according to \cite{Clementini2019},  there are 28 RR Lyrae located within $20’$ from the centre of the cluster Patchick 125. 
Figure \ref{rrl_p125} shows their positions, magnitudes, and period versus amplitude diagrams.
Based on their positions, magnitudes and PMs, we consider that two of them are bona-fide cluster members:
Gaia DR2 5977224553266268928 and DR2 5977223144516980608, located at $79''$ and $81''$ from the cluster centre, respectively.  
These stars are highlighted with blue points in Figure  \ref{rrl_p125}, and 
their mean magnitudes are $G= 18.023 \pm 0.016$ and $17.753 \pm 0.013$ mag, 
their mean PMs are $\mu_{\alpha_{\ast}} = -4.067 \pm 0.330$, $\mu_{\delta}= 0.362 \pm 0.224$ mas yr$^{-1}$, and
$\mu_{\alpha_{\ast}}= -3.793 \pm 0.270$, $\mu_{\delta}= 0.853 \pm 0.185$ mas yr$^{-1}$, respectively.
Their mean periods are $P=0.601940$ and $P=0.738296$ day,
and their amplitudes are $A=0.872$, and $A=0.633$ mag in the optical passband \citep{Clementini2019},  respectively.
The other RR Lyrae in the field, located beyond $2’$ from the cluster centre may not be securely considered cluster members because of their different magnitudes and PMs.  In particular, we exclude the other two variables located within 500 arcsec, because their PMs do not match the mean cluster PMs. Additionally, we derive their distances using the  period-luminosity-metallicity (PL$_K$Z) relation by \cite{AlonsoGarcia2021_PLKZrel}.  Firstly, we transformed the Patchick 125 metallicity (listed in Table \ref{parameters}) into $\log$~Z using the relation shown in \cite{Navarrete2017}, assuming $f=10^{[\alpha/Fe]} = 3$ and $Z_{\odot}=0.017$.  We find $D=11.06 \pm 0.30$ kpc for Gaia DR2 5977224553266268928,  and $D=12.17\pm 0.50$ kpc  Gaia DR2 5977223144516980608, both values in excellent agreement with the cluster distances obtained from the VVVX and Gaia EDR3 photometries.  We also changed the $\alpha$-element enhancement factor $f=1.58$ (adopting the $[\alpha /Fe]=+0.2$),  and we find for Gaia DR2 5977224553266268928 a distance of $10.4$ kpc and for Gaia DR2 5977223144516980608 a distance of $11.5$ kpc, in excellent agreement with the NIR cluster distance.  However, we test our estimates using different PL$_K$Z relations by \cite{Muraveva2015}, \cite{2017A&A...605A..79G} and \cite{Navarrete2017}, yielding slightly larger distance values. However, these relations are still within the mean errors of $0.3$ and $0.5$ kpc obtained by the comparison between these four PL$_K$Z relations.\\
The presence of two RRab members of this cluster also support the conclusion that Patchick 125 is an old and metal-poor GC. 
In particular, the position of these two member RR Lyrae in the period-amplitude diagram (shown in the bottom right panel of Figure  \ref{rrl_p125}) suggest that Patchick 125 is an Oosterhoff type II GC.  The mean ridge lines, used in Fig.  \ref{rrl_p125},  for the Oosterhoff type I and II populations are taken from \cite{Clement1999}, assuming $A_V/A_I \approx 1.4$ mag. 

\subsection{Patchick 126}
We use a combination of Gaia EDR3 and the NIR 2MASS and VVVX datasets in order to investigate Patchick 126.  For this cluster, the CMDs convey a well-defined RGB and a poorly populated RC at $K=13.5$ mag and $G=18.0$ mag.  We estimate an extinction of $0.441$ mag in the NIR and $2.66$ mag in the optical.  From these parameters, we obtain a distance modulus of $(m-M)_{0}=14.66$ mag, which places it at a heliocentric distance of $8.6$ kpc, in agreement with $D_{GEDR3}=7.4$ kpc found from the Gaia photometry,  within $2\sigma$. Also,  we derive its distance from the Galactic centre,  $R_{G}=2.94$ kpc. Our photometric analysis suggests that this may be a GC.   We found the mean metallicity of the cluster is [Fe/H] $=-0.7\pm 0.3$ dex and an age older than 8 Gyr, which were evaluated using the isochones fitting method.  Although the metallicities derived from the two methods are in agreement, we find a younger age of 5.6 Gyr with the $\chi^{2}$ procedure. Therefore, as done for the other clusters, we calculated the difference in magnitude between the HB and MSTO also for Patchick 126, obtaining $\Delta K_s$(HB-MSTO) $>2.5$ mag and $\Delta G$(HB-MSTO) $>2.5$ mag, this can be translated into a lower age limit of 8 Gyr.  Additionally,  taking its relatively low metallicity into account, Patchick 126 could be a younger GC,  based on the 108 stars selected as members.

\subsection{Riddle 15}
Riddle 15 is studied using the match between Gaia EDR3 and 2MASS photometry, because it is outside of the VVVX footprint. It appears confined within a radius smaller than $1'$. That makes this candidate one of the smallest in our sample, suggesting that it may be located at a large distance. Indeed,  we find $D_{NIR}=18.1$ kpc in NIR and $D_{GDR3}=19.5$ kpc in the optical, also its Galactocentric distance is $R_{G}=14.06$ kpc.  The main distance error source is related to the dust effect, since we find higher extinction values of $A_{Ks}=0.52$ mag and $A_{G}=3.09$ in NIR and optical passbands, respectively. The only information that CMDs display is relative to its 83 brighter stars. Fortunately, we are able to reach below the HB, revealing an extended HB.  Indeed, we derive its metallicity using the PARSEC isochrones, yielding [Fe/H]~$=-1.4\pm 0.2$ dex.  Taking into account both the low metal content and the magnitude level of the HB at $G\approx 20.0$ mag,  we believe that this cluster may be older than 10 Gyr, with the best isochrone fit suggesting an age of $\sim 13$ Gyr,  in good agreement with the $\chi^{2}$ analysis.  However,  \cite{Bica2019} classify Riddle 15 as an OC candidate or a compact OC remnant candidate.  The latter based their analysis on 2MASS data whose depth and angular resolution are clearly not appropriate for getting reasonable results for this faint, small, and distant cluster. \\ 

For that reason, we searched for RR Lyrae stars in the cluster field, since they are good tracers of old and metal-poor populations. We found one RR Lyra type-ab variable star located at $43''$ from the Riddle 15 cluster centre.
This star is Gaia DR2 4320480608680062080, with mean magnitude $G=20.254 \pm 0.012$, period $P=0.608333$ day, Epoch (JD) = 2456956.0055, Amplitude $A=0.57$ mag in the optical passband, proper motions $\mu_{\alpha_{\ast}}=-1.90 \pm 1.14$ mas yr$^{-1}$ and $\mu_{\delta}= -1.79 \pm 1.05$ mas/ yr$^{-1}$ \citep{Clementini2019}.  Also, we derive its absolute magnitude in $G-$band adopting the $M_{G}-$[Fe/H] relation by \cite{Muraveva2018}, assuming the [Fe/H] $=-1.4$ as Riddle~15. Hence, this allows us to derive its distance $D=20.0 \pm 0.5$ kpc, in agreement with the cluster distance found from Gaia EDR3 photometry.\\
The position, magnitude and PMs for this RR Lyra are consistent with cluster membership, also suggesting that Riddle 15 is a relatively old and metal-poor GC,  excluding the OC nature. \\
There is another RR Lyra type-c candidate, Gaia DR2 4320478237881151488, located at $10’$ from the Riddle 15 centre. However, this is likely a field star because it is too bright ($G= 15.947 \pm 0.006$ mag) and its PMs are too large ($\mu_{\alpha_{\ast}}=3.54 \pm 0.07$ mas yr$^{-1}$ and $\mu_{\delta}= -13.79 \pm 0.07$ mas yr$^{-1}$) to be considered a cluster member.

\subsection{FSR 190}
\label{SectFSR190}
It is very hard to analyse the field of FSR 190 cluster because of heavy crowding and differential reddening. In fact, this cluster appears to be highly reddened as we can appreciate from its CMDs (Fig. \ref{CMD}).  We derived an extinction of $A_{Ks}=0.86$ mag in NIR and $A_{G}=4.72$ mag in optical.  Additionally, the main information from its CMDs,  is related to 52 brighter star members, using a combination of Gaia EDR3 and 2MASS catalogues.  From our analysis, we find that FSR 190 may be a GC, located at a heliocentric distance of $D_{NIR}=10.67$ kpc and at a Galactocentric distance of $R_{G}=11.11$ kpc. In the optical passband, we find a slightly smaller distance of $D_{GEDR3}=9.5$ kpc, but this value is more uncertain because of the high extinction. The best PARSEC isochrone fit suggests a metallicity of [Fe/H]~$=-0.9 \pm 0.2$ dex. With the metallicity fixed, we searched for the best age, finding Age~$>9$ Gyr.  However, placing a small RC at $K=14.4$ mag and $G=20.1$ mag,  and supposing that all the stars below $K = 16.0$ mag are MSTO stars, the cluster could have an age younger than 7 Gyr, but still in excess of 2 Gyr.  if we use a family of isochrones with Age~$<7$ Gyr, these differ considerably from the observational points of both CMDs.  Therefore, we constrained the age of the cluster to be larger than 7 Gyr.\\ Although the analysis of this cluster turns out to be tricky and with many question marks, we are in full agreement with the results of  \cite{Froebrich2008}.  They derived for FSR 190 similar values, such as metallicity ([Fe/H]~$-0.9\pm 0.4$), age (Age~$>7$ Gyr) and extinction ($A_{K}=0.8\pm 0.1$ mag), as well as similar values of distances ($D=10.0 \pm 1.0$ kpc and $R_{G}=10.5\pm 0.8$ kpc, adopting a $R_{\odot}=7.2$ kpc) within the errors.  Our results also agree with the recent results of \cite{CantatGaudin2018}.  Also,  \cite{Bica2019} suggested that FSR 190 may be an OC or GC candidate,  commenting that it could be a Palomar-like GC.

\subsection{Ferrero 54}
The main obstacle for the study of Ferrero 54 is the presence of a nebulosity of dust and gas in front of this cluster, as shown from the optical DECaPS image (Fig. \ref{ferrerofig}), which disappears in the NIR image.  These wisps are part of the Vela Supernova remnant complex.The decontamination procedure reveals a cluster including 122 giant stars, with a sparsely populated RC at $K~=~13.25$ mag and $G=17.8$ mag. Its extinction is $A_{Ks}=0.60$ mag and $A_{G}=3.0$ mag. Using these values and the position of RC stars, we recover the distance modulus of $(m-M)_{0}=14.27$ mag, equivalent to $D_{NIR}=7.1$ kpc,  which is in excellent agreement with that found from the Gaia photometry of $D_{GEDR3}=7.26$ kpc.  The NIR distance places Ferrero 54 at $R_{G}=11.5$ kpc from the Galactic centre.  The PARSEC isochrones fit yields a metallicity of [Fe/H]~$=-0.2\pm 0.2$ dex and an Age~$\approx 11$ Gyr.  We find that the metallicity estimates derived from the isochrone-fitting and $\chi^{2}$ methods are in good agreement, on the other hand the $\chi^{2}$ procedure yields a younger age of 6.5 Gyr. Nevertheless, we give a solid lower age limit using the vertical method as done for other clusters, finding $\Delta K_s$(HB-MSTO)~$>3.0$ mag and $\Delta G$(HB-MSTO)~$>3.0$ mag, which can be translated into a lower limit of $\sim 10$ Gyr.  Additionally,  GCs with almost solar metallicities are very scarce, and also in general they are located in the Galactic bulge.
So if the high metallicity of Ferrero 54 is confirmed, it would make this an exceptional object among the group of Galactic GCs. Finally, based on our analysis, we confirm the GC nature for Ferrero 54.  

\subsection{ESO 92-18}
As expected from the density diagrams (Figs. \ref{radiusDP} and \ref{KDEfig}), ESO 92-18 is the most populated cluster here examined, containing 557 star members after the Gaia EDR3 PM-decontamination procedure. We use the matched Gaia EDR3+2MASS datasets in order to have the NIR counterpart, even if we lose many fainter stars in the 2MASS CMD. However,  reliable results can be achieved especially when the Gaia EDR3 CMD is analysed, since main evolutionary sequences are visible: the brightest MS stars and the MSTO point (at $G=18.8$ mag),  a compact RC (at $K=13.51$ mag and at $G=16.3$ mag) and a well-defined RGB.  The main parameters that we estimated for this cluster are extinctions of $A_{Ks}=0.09$ mag and $A_{G}=0.89$, a distance modulus of $(m-M)_{0} =15.12$ mag, corresponding to $D=10.6$ kpc, in good agreement with $D_{GEDR3}=9.6$ kpc from the Gaia passband, its Galactocentric distance is $R_{G}=11.35$ kpc. Fitting the PARSEC isochrone models we found that ESO 92-18 may be an old OC or young GC with an Age~$=5\pm 1$ Gyr and a metallicity of [Fe/H]~$=-0.9\pm 0.2$ dex,  which are in excellent agreement with the parameters derived from the $\chi^2$ analysis.  Our results agree with the parameters obtained by \cite{Salaris2004},  and \cite{Shao2019}, especially when ages are compared.  Also \cite{Bica2019} classified ESO 92-18 as an OC.

\subsection{Gaia 2}
Gaia 2 is inspected matching the Gaia EDR3+2MASS datasets.  Although the 2MASS CMD is not so clear to interpret, the Gaia EDR3 CMD shows the MSTO point at $G=17.7$ mag as a clear overdensity,  the highest part of the MS stage, but also a poorly populated (11 stars) RGB. We can use these sequences in order to determine its main physical parameters. Indeed, we estimate an extinction of $A_{Ks}=0.10$ mag and $A_{G}=0.97$ mag, a distance modulus of $(m-M)_{0}=13.43$ mag, which can be translated into a heliocentric distance of $D_{NIR}=4.91$ kpc, in very good agreement with $D_{GEDR3}=4.43$ kpc,  and a Galactocentric distance of $R_{G}=12.03$ kpc.  The best isochrone fit yields an Age~$=10\pm 1$ Gyr and a metallicity of [Fe/H]~$=-0.9\pm 0.2$ dex, which are in good agreement with the values obtained from the $\chi^2$ analysis. This time the uncertainties about the metallicity are higher than age due to the lack of evolved stars (especially of the RC or BHB), but also because the MSTO point is less sensitive to metallicity changes. Additionally, since this cluster is near to the Sun, we calculate its distance using the Gaia EDR3 parallax values.  We use the bayesian distances provided by \cite{Bailer_Jones2021},  finding a mean value of $D=3.57\pm 1.08$ kpc, in good agreement with the previous estimate,  within $1\sigma$. \\ 
Moreover, we compare our results with \cite{Bica2019}, which catalogued this cluster as a ultra-faint ($M_V=-2$ mag) GC, located at heliocentric distance $5.4$ kpc, in agreement within $1\sigma$ with our estimates.  Finally, we classify Gaia 2 as a good GC candidate.


\section{Summary and discussion}
\label{end}
We present the first physical characterization of twelve newly discovered stellar clusters. A dedicated decontamination procedure is carried out in order to select the PM-star members. There are several difficulties  usually found in this kind of study: from the differential reddening that distorts the evolutionary sequences in the CMDs (as shown e.g. by the FSR 190 CMDs in Fig. \ref{CMD}) and introduce uncertainties in the observed parameters (e.g., distance estimates),  to the high stellar crowding, which may contaminate the final catalogue or alter the total luminosity of the cluster, or to the presence of nebulosity as in the case of Ferrero 54 (Fig.\ref{ferrerofig}).  For these reasons, our photometric analysis is based on the combination of the optical Gaia EDR3 and NIR VVVX/2MASS datasets in order to obtain reliable parameters and understand the real nature of these low-luminosity objects. \\

Existing reddening maps in the NIR and in the optical are adopted in order to derive the extinction values. We find a wide range of $0.09\lesssim A_{Ks}\lesssim 0.86$ mag in NIR and between $0.89 \lesssim A_{G}\lesssim 4.72$ mag in the optical.  Using these parameters and intrinsic RC colour and magnitude values, we are able to measure the distance modulus for each cluster, translating them into heliocentric distances. We find a large distance range between 4.4 and 18.1 kpc that place them at 2.94 and 14.06 kpc from the Galactic centre.\\
Ages and metallicities are evaluated adopting the isochrone-fitting method, using the PARSEC model. We find that 8 clusters show rich/intermediate metal content with [Fe/H] between $-0.5$ and $-0.9$,  one young cluster with [Fe/H]~$=+0.3$, and there are two are metal-poor clusters with [Fe/H]~$=-1.4$ and $-1.8$.  Regarding the age, for most of our clusters we are only able to provide a lower limit, using the vertical method since the MSTO point is below the detection limit. 
\begin{figure}[!htbp]
\centering
\includegraphics[width=8cm, height=6cm]{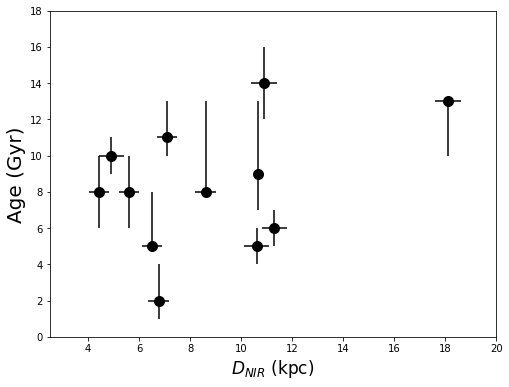} 
\includegraphics[width=8cm, height=6cm]{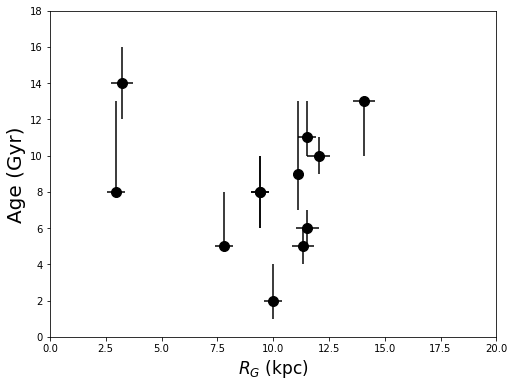} 
\caption{Age as function of NIR heliocentric distance (top panel) and of Galactocentric distance $R_{G}$ (bottom panel) for our sample. We also show the age error bars, even if not all clusters have errorbars in their ages in Table \ref{parameters}.  However, we consider as lower-error the lower limit written in Table \ref{parameters}, whereas as higher-error we adopt the higher limit of 13 Gyr,  the typical age of old Galactic GCs,  therefore we constrain the age error as the difference between our age estimates and 13 Gyr.}
\label{agedist}
\end{figure}

However, we can better constrain the ages of the following clusters: Kronberger 100, Kronberger 119, Kronberger 143, Patchick 125, ESO 92-18 and Gaia~2.
As shown in Fig. \ref{agedist}, we find clusters with a wide range of ages. The majority of clusters populate the Galactic thin and thick disk and exhibit ages from 2 Gyr to 13 Gyr. However, we also identify two old clusters (Patchick 125 and Patchick 126) that are located well inside the Galactic bulge. Their ages of 14 Gyr and $>8$ Gyr, respectively, are inline with the results from previous works implying that the lower age limit for the MW bulge is $\sim 5$ Gyr (e.g., \citealt{Clarkson_2011}) and the mean age is $\sim10$ Gyr (e.g., \citealt{Freeman2008}), suggesting that the clusters are either real tenants of the bulge or, alternatively, intruders originating from the halo (e.g., \citealt{Forbes2010,Massari2019,Minniti2021_CL160,RomeroColmenares2021}).\\
In conclusion, we confirm the OC nature for Kronberger 100, while we classify Kronberger 119 and Kronberger 143 as old OCs based on their comparably young ages and their relative co-movement with the surrounding stellar field, although a classification as young GCs cannot be ruled out. For the same reasons, ESO 92-18, the most populated cluster analysed, is also categorised as old OC or young GC. Five objects are classified as GCs based on our findings: Patchick 125 and Riddle 15 are classified as a metal-poor GCs, Patchick 126 and Gaia 2 as GCs of intermediate metallicity, and Ferrero 54 as metal-rich GC. Finally, we cautiously categorise three objects with poorly constrained ages (Kronberger 99, Patchick 122 and FSR 190) as GC candidates. \\
We note that in all cases, deeper Hubble space telescope (HST) or James Webb space telescope (JWST) observations as well as spectroscopic observations are desired in order to build complete catalogues of likely cluster members, improving the structural parameters and luminosities.

\begin{acknowledgements}
We gratefully acknowledge the use of data from the ESO Public Survey program IDs 179.B-2002 and 198.B-2004 taken with the VISTA telescope and data products from the Cambridge Astronomical Survey Unit. ERG acknowledges support from ANID PhD scholarship No. 21210330.  D.M. gratefully acknowledges support by the ANID BASAL project FB210003. This work has made use of data from the European Space Agency (ESA) mission Gaia (http:/www.cosmos.esa.int/gaia), processed by the Gaia Data Processing and Analysis Consortium (DPAC, http://www.cosmos.esa.int/web/gaia/dpac/consortium). R.K.S. acknowledges support from CNPq/Brazil through project 305902/2019-9.  J.A.-G. acknowledges support from Fondecyt Regular 1201490 and from ANID – Millennium Science Initiative Program – ICN12\_009 awarded to the Millennium Institute of Astrophysics MAS.
\end{acknowledgements}

\bibliographystyle{aa.bst}
\bibliography{bibliopaper5}

\newcommand{\noop}[1]{}
\begin{thebibliography}{92}
\expandafter\ifx\csname natexlab\endcsname\relax\def\natexlab#1{#1}\fi

\bibitem[{{Alonso-Garc{\'\i}a} {et~al.}(2018){Alonso-Garc{\'\i}a}, {Saito},
  {Hempel}, {Minniti}, {Pullen}, {Catelan}, {Ramos}, {Cross}, {Gonzalez},
  {Lucas}, {Palma}, {Valenti}, \& {Zoccali}}]{AlonsoGarcia2018}
{Alonso-Garc{\'\i}a}, J., {Saito}, R.~K., {Hempel}, M., {et~al.} 2018, \aap,
  619, A4

\bibitem[{{Alonso-Garc{\'\i}a} {et~al.}(2021){Alonso-Garc{\'\i}a}, {Smith},
  {Catelan}, {Minniti}, {Navarrete}, {Borissova}, {Carballo-Bello}, {Contreras
  Ramos}, {Fern{\'a}ndez-Trincado}, {Ferreira Lopes}, {Gran}, {Garro},
  {Geisler}, {Guo}, {Hempel}, {Kerins}, {Lucas}, {Palma}, {Pe{\~n}a
  Ram{\'\i}rez}, {Ram{\'\i}rez Alegr{\'\i}a}, \&
  {Saito}}]{AlonsoGarcia2021_PLKZrel}
{Alonso-Garc{\'\i}a}, J., {Smith}, L.~C., {Catelan}, M., {et~al.} 2021, \aap,
  651, A47

\bibitem[{Ashman \& Zepf(1998)}]{ashman_1998}
Ashman, K.~M. \& Zepf, S.~E. 1998, Cambridge Univ. Press, Cambridge (Cambridge
  Astrophysics Series; 30)

\bibitem[{{Bailer-Jones} {et~al.}(2021){Bailer-Jones}, {Rybizki}, {Fouesneau},
  {Demleitner}, \& {Andrae}}]{Bailer_Jones2021}
{Bailer-Jones}, C.~A.~L., {Rybizki}, J., {Fouesneau}, M., {Demleitner}, M., \&
  {Andrae}, R. 2021, VizieR Online Data Catalog, I/352

\bibitem[{{Barb{\'a}} {et~al.}(2019){Barb{\'a}}, {Minniti}, {Geisler},
  {Alonso-Garc{\'\i}a}, {Hempel}, {Monachesi}, {Arias}, \&
  {G{\'o}mez}}]{Barba2019}
{Barb{\'a}}, R.~H., {Minniti}, D., {Geisler}, D., {et~al.} 2019, \apjl, 870,
  L24

\bibitem[{{Barbuy} {et~al.}(2018){Barbuy}, {Chiappini}, \&
  {Gerhard}}]{Barbuy2018}
{Barbuy}, B., {Chiappini}, C., \& {Gerhard}, O. 2018, \araa, 56, 223

\bibitem[{Baumgardt \& Vasiliev(2021)}]{Baumgardt2021}
Baumgardt, H. \& Vasiliev, E. 2021, MNRAS, 505, 5957

\bibitem[{{Bica} {et~al.}(2019){Bica}, {Pavani}, {Bonatto}, \&
  {Lima}}]{Bica2019}
{Bica}, E., {Pavani}, D.~B., {Bonatto}, C.~J., \& {Lima}, E.~F. 2019, \aj, 157,
  12

\bibitem[{{Bonatto} \& {Bica}(2010)}]{Bonatto2010}
{Bonatto}, C. \& {Bica}, E. 2010, \aap, 521, A74

\bibitem[{{Bressan} {et~al.}(2012){Bressan}, {Marigo}, {Girardi}, {Salasnich},
  {Dal Cero}, {Rubele}, \& {Nanni}}]{Bressan2012}
{Bressan}, A., {Marigo}, P., {Girardi}, L., {et~al.} 2012, \mnras, 427, 127

\bibitem[{{Camargo} \& {Minniti}(2019)}]{Camargo2019}
{Camargo}, D. \& {Minniti}, D. 2019, \mnras, 484, L90

\bibitem[{{Cantat-Gaudin} {et~al.}(2020){Cantat-Gaudin}, {Anders},
  {Castro-Ginard}, {Jordi}, {Romero-G{\'o}mez}, {Soubiran}, {Casamiquela},
  {Tarricq}, {Moitinho}, {Vallenari}, {Bragaglia}, {Krone-Martins}, \&
  {Kounkel}}]{Cantat_Gaudin2020}
{Cantat-Gaudin}, T., {Anders}, F., {Castro-Ginard}, A., {et~al.} 2020, \aap,
  640, A1

\bibitem[{{Cantat-Gaudin} {et~al.}(2018{\natexlab{a}}){Cantat-Gaudin}, {Jordi},
  {Vallenari}, {Bragaglia}, {Balaguer-N{\'u}{\~n}ez}, {Soubiran}, {Bossini},
  {Moitinho}, {Castro-Ginard}, {Krone-Martins}, {Casamiquela}, {Sordo}, \&
  {Carrera}}]{Cantat_Gaudin2018}
{Cantat-Gaudin}, T., {Jordi}, C., {Vallenari}, A., {et~al.} 2018{\natexlab{a}},
  \aap, 618, A93

\bibitem[{{Cantat-Gaudin} {et~al.}(2018{\natexlab{b}}){Cantat-Gaudin}, {Jordi},
  {Vallenari}, {Bragaglia}, {Balaguer-N{\'u}{\~n}ez}, {Soubiran}, {Bossini},
  {Moitinho}, {Castro-Ginard}, {Krone-Martins}, {Casamiquela}, {Sordo}, \&
  {Carrera}}]{CantatGaudin2018}
{Cantat-Gaudin}, T., {Jordi}, C., {Vallenari}, A., {et~al.} 2018{\natexlab{b}},
  \aap, 618, A93

\bibitem[{{Cassisi} \& {Salaris}(2013)}]{CassisiSalaris2013}
{Cassisi}, S. \& {Salaris}, M. 2013, {Old Stellar Populations: How to Study the
  Fossil Record of Galaxy Formation}

\bibitem[{Clarkson {et~al.}(2011)Clarkson, Sahu, Anderson, Rich, Smith, Brown,
  Bond, Livio, Minniti, Renzini, \& Zoccali}]{Clarkson_2011}
Clarkson, W.~I., Sahu, K.~C., Anderson, J., {et~al.} 2011, ApJ, 735, 37

\bibitem[{{Clement} \& {Shelton}(1999)}]{Clement1999}
{Clement}, C.~M. \& {Shelton}, I. 1999, \apjl, 515, L85

\bibitem[{{Clementini} {et~al.}(2019){Clementini}, {Ripepi}, {Molinaro},
  {Garofalo}, {Muraveva}, {Rimoldini}, {Guy}, {Jevardat de Fombelle},
  {Nienartowicz}, {Marchal}, {Audard}, {Holl}, {Leccia}, {Marconi}, {Musella},
  {Mowlavi}, {Lecoeur-Taibi}, {Eyer}, {De Ridder}, {Regibo}, {Sarro},
  {Szabados}, {Evans}, \& {Riello}}]{Clementini2019}
{Clementini}, G., {Ripepi}, V., {Molinaro}, R., {et~al.} 2019, \aap, 622, A60

\bibitem[{{Cross} {et~al.}(2012){Cross}, {Collins}, {Mann}, {Read}, {Sutorius},
  {Blake}, {Holliman}, {Hambly}, {Emerson}, {Lawrence}, \&
  {Noddle}}]{Cross2012}
{Cross}, N.~J.~G., {Collins}, R.~S., {Mann}, R.~G., {et~al.} 2012, \aap, 548,
  A119

\bibitem[{{Di Matteo} {et~al.}(2015){Di Matteo}, {G{\'o}mez}, {Haywood},
  {Combes}, {Lehnert}, {Ness}, {Snaith}, {Katz}, \& {Semelin}}]{DiMatteo2015}
{Di Matteo}, P., {G{\'o}mez}, A., {Haywood}, M., {et~al.} 2015, \aap, 577, A1

\bibitem[{{Dias} {et~al.}(2021){Dias}, {Palma}, {Minniti},
  {Fern{\'a}ndez-Trincado}, {Alonso-Garc{\'\i}a}, {Barbuy}, {Clari{\'a}},
  {Gomez}, \& {Saito}}]{Dias:submitted-a}
{Dias}, B., {Palma}, T., {Minniti}, D., {et~al.} 2021, arXiv e-prints,
  arXiv:2110.00868

\bibitem[{{Dotter} {et~al.}(2008){Dotter}, {Chaboyer}, {Jevremovi{\'c}},
  {Kostov}, {Baron}, \& {Ferguson}}]{Dotter2008}
{Dotter}, A., {Chaboyer}, B., {Jevremovi{\'c}}, D., {et~al.} 2008, \apjs, 178,
  89

\bibitem[{{Emerson} \& {Sutherland}(2010)}]{2010Msngr.139....2E}
{Emerson}, J. \& {Sutherland}, W. 2010, The Messenger, 139, 2

\bibitem[{{Fern{\'a}ndez-Trincado} {et~al.}(2020){Fern{\'a}ndez-Trincado},
  {Minniti}, {Beers}, {Villanova}, {Geisler}, {Souza}, {Smith}, {Placco},
  {Vieira}, {P{\'e}rez-Villegas}, {Barbuy}, {Alves-Brito}, {Bidin},
  {Alonso-Garc{\'\i}a}, {Tang}, \& {Palma}}]{Fernandez_Trincado2020_UKS1}
{Fern{\'a}ndez-Trincado}, J.~G., {Minniti}, D., {Beers}, T.~C., {et~al.} 2020,
  \aap, 643, A145

\bibitem[{{Fern{\'a}ndez-Trincado}
  {et~al.}(2021{\natexlab{a}}){Fern{\'a}ndez-Trincado}, {Minniti}, {Garro}, \&
  {Villanova}}]{JFT_Pat125:submitted}
{Fern{\'a}ndez-Trincado}, J.~G., {Minniti}, D., {Garro}, E.~R., \& {Villanova},
  S. 2021{\natexlab{a}}, arXiv e-prints, arXiv:2111.04151

\bibitem[{{Fern{\'a}ndez-Trincado}
  {et~al.}(2021{\natexlab{b}}){Fern{\'a}ndez-Trincado}, {Minniti}, {Souza},
  {Beers}, {Geisler}, {Moni Bidin}, {Villanova}, {Majewski}, {Barbuy},
  {P{\'e}rez-Villegas}, {Henao}, {Romero-Colmenares}, {Roman-Lopes}, \&
  {Lane}}]{FernandezTrincado2021}
{Fern{\'a}ndez-Trincado}, J.~G., {Minniti}, D., {Souza}, S.~O., {et~al.}
  2021{\natexlab{b}}, \apjl, 908, L42

\bibitem[{{Fern{\'a}ndez-Trincado} {et~al.}(\noop{3001}2021, in
  press){Fern{\'a}ndez-Trincado}, {Villanova}, {Geisler}, {Barbuya}, {Minniti},
  {Beers}, {Mészáros}, {Tang}, {Cohen}, {Moni Bidin}, {Chaves-Velasquez},
  {Garro}, {Baeza}, \& {Muñoz}}]{JFT_TON1:submitted}
{Fern{\'a}ndez-Trincado}, J.~G., {Villanova}, S., {Geisler}, D., {et~al.}
  \noop{3001}2021, in press, ApJL

\bibitem[{Forbes \& Bridges(2010)}]{Forbes2010}
Forbes, D.~A. \& Bridges, T. 2010, MNRAS, 404, 1203

\bibitem[{{Freeman}(2008)}]{Freeman2008}
{Freeman}, K.~C. 2008, in Formation and Evolution of Galaxy Bulges, ed.
  M.~{Bureau}, E.~{Athanassoula}, \& B.~{Barbuy}, Vol. 245, 3--10

\bibitem[{Froebrich {et~al.}(2008)Froebrich, Meusinger, \&
  Davis}]{Froebrich2008}
Froebrich, D., Meusinger, H., \& Davis, C.~J. 2008, Monthly Notices of the
  Royal Astronomical Society: Letters, 383, L45

\bibitem[{Gadotti(2009)}]{Gadotti2009}
Gadotti, D.~A. 2009, MNRAS, 393, 1531

\bibitem[{{Gaia Collaboration} {et~al.}(2018{\natexlab{a}}){Gaia
  Collaboration}, {Babusiaux}, {van Leeuwen}, {Barstow}, {Jordi}, {Vallenari},
  {Bossini}, {Bressan}, {Cantat-Gaudin}, {van Leeuwen}, {Brown}, {Prusti}, {de
  Bruijne}, {Bailer-Jones}, {Biermann}, {Evans}, {Eyer}, {Jansen}, {Klioner},
  {Lammers}, {Lindegren}, {Luri}, {Mignard}, {Panem}, {Pourbaix}, {Randich},
  {Sartoretti}, {Siddiqui}, {Soubiran}, {Walton}, {Arenou}, {Bastian},
  {Cropper}, {Drimmel}, {Katz}, {Lattanzi}, {Bakker}, {Cacciari},
  {Casta{\~n}eda}, {Chaoul}, {Cheek}, {De Angeli}, {Fabricius}, {Guerra},
  {Holl}, {Masana}, {Messineo}, {Mowlavi}, {Nienartowicz}, {Panuzzo},
  {Portell}, {Riello}, {Seabroke}, {Tanga}, {Th{\'e}venin}, {Gracia-Abril},
  {Comoretto}, {Garcia-Reinaldos}, {Teyssier}, {Altmann}, {Andrae}, {Audard},
  {Bellas-Velidis}, {Benson}, {Berthier}, {Blomme}, {Burgess}, {Busso},
  {Carry}, {Cellino}, {Clementini}, {Clotet}, {Creevey}, {Davidson}, {De
  Ridder}, {Delchambre}, {Dell'Oro}, {Ducourant},
  {Fern{\'a}ndez-Hern{\'a}ndez}, {Fouesneau}, {Fr{\'e}mat}, {Galluccio},
  {Garc{\'\i}a-Torres}, {Gonz{\'a}lez-N{\'u}{\~n}ez}, {Gonz{\'a}lez-Vidal},
  {Gosset}, {Guy}, {Halbwachs}, {Hambly}, {Harrison}, {Hern{\'a}ndez},
  {Hestroffer}, {Hodgkin}, {Hutton}, {Jasniewicz}, {Jean-Antoine-Piccolo},
  {Jordan}, {Korn}, {Krone-Martins}, {Lanzafame}, {Lebzelter}, {L{\"o}ffler},
  {Manteiga}, {Marrese}, {Mart{\'\i}n-Fleitas}, {Moitinho}, {Mora}, {Muinonen},
  {Osinde}, {Pancino}, {Pauwels}, {Petit}, {Recio-Blanco}, {Richards},
  {Rimoldini}, {Robin}, {Sarro}, {Siopis}, {Smith}, {Sozzetti}, {S{\"u}veges},
  {Torra}, {van Reeven}, {Abbas}, {Abreu Aramburu}, {Accart}, {Aerts},
  {Altavilla}, {{\'A}lvarez}, {Alvarez}, {Alves}, {Anderson}, {Andrei},
  {Anglada Varela}, {Antiche}, {Antoja}, {Arcay}, {Astraatmadja}, {Bach},
  {Baker}, {Balaguer-N{\'u}{\~n}ez}, {Balm}, {Barache}, {Barata}, {Barbato},
  {Barblan}, {Barklem}, {Barrado}, {Barros}, {Bartholom{\'e} Mu{\~n}oz},
  {Bassilana}, {Becciani}, {Bellazzini}, {Berihuete}, {Bertone}, {Bianchi},
  {Bienaym{\'e}}, {Blanco-Cuaresma}, {Boch}, {Boeche}, {Bombrun}, {Borrachero},
  {Bouquillon}, {Bourda}, {Bragaglia}, {Bramante}, {Breddels}, {Brouillet},
  {Br{\"u}semeister}, {Brugaletta}, {Bucciarelli}, {Burlacu}, {Busonero},
  {Butkevich}, {Buzzi}, {Caffau}, {Cancelliere}, {Cannizzaro}, {Carballo},
  {Carlucci}, {Carrasco}, {Casamiquela}, {Castellani}, {Castro-Ginard},
  {Charlot}, {Chemin}, {Chiavassa}, {Cocozza}, {Costigan}, {Cowell}, {Crifo},
  {Crosta}, {Crowley}, {Cuypers}, {Dafonte}, {Damerdji}, {Dapergolas}, {David},
  {David}, {de Laverny}, {De Luise}, {De March}, {de Martino}, {de Souza}, {de
  Torres}, {Debosscher}, {del Pozo}, {Delbo}, {Delgado}, {Delgado}, {Diakite},
  {Diener}, {Distefano}, {Dolding}, {Drazinos}, {Dur{\'a}n}, {Edvardsson},
  {Enke}, {Eriksson}, {Esquej}, {Eynard Bontemps}, {Fabre}, {Fabrizio},
  {Faigler}, {Falc{\~a}o}, {Farr{\`a}s Casas}, {Federici}, {Fedorets},
  {Fernique}, {Figueras}, {Filippi}, {Findeisen}, {Fonti}, {Fraile}, {Fraser},
  {Fr{\'e}zouls}, {Gai}, {Galleti}, {Garabato}, {Garc{\'\i}a-Sedano},
  {Garofalo}, {Garralda}, {Gavel}, {Gavras}, {Gerssen}, {Geyer}, {Giacobbe},
  {Gilmore}, {Girona}, {Giuffrida}, {Glass}, {Gomes}, {Granvik}, {Gueguen},
  {Guerrier}, {Guiraud}, {Guti{\'e}}, {Haigron}, {Hatzidimitriou}, {Hauser},
  {Haywood}, {Heiter}, {Helmi}, {Heu}, {Hilger}, {Hobbs}, {Hofmann}, {Holland},
  {Huckle}, {Hypki}, {Icardi}, {Jan{\ss}en}, {Jevardat de Fombelle}, {Jonker},
  {Juh{\'a}sz}, {Julbe}, {Karampelas}, {Kewley}, {Klar}, {Kochoska}, {Kohley},
  {Kolenberg}, {Kontizas}, {Kontizas}, {Koposov}, {Kordopatis},
  {Kostrzewa-Rutkowska}, {Koubsky}, {Lambert}, {Lanza}, {Lasne}, {Lavigne}, {Le
  Fustec}, {Le Poncin-Lafitte}, {Lebreton}, {Leccia}, {Leclerc},
  {Lecoeur-Taibi}, {Lenhardt}, {Leroux}, {Liao}, {Licata}, {Lindstr{\o}m},
  {Lister}, {Livanou}, {Lobel}, {L{\'o}pez}, {Managau}, {Mann}, {Mantelet},
  {Marchal}, {Marchant}, {Marconi}, {Marinoni}, {Marschalk{\'o}}, {Marshall},
  {Martino}, {Marton}, {Mary}, {Massari}, {Matijevi{\v{c}}}, {Mazeh},
  {McMillan}, {Messina}, {Michalik}, {Millar}, {Molina}, {Molinaro},
  {Moln{\'a}r}, {Montegriffo}, {Mor}, {Morbidelli}, {Morel}, {Morris},
  {Mulone}, {Muraveva}, {Musella}, {Nelemans}, {Nicastro}, {Noval},
  {O'Mullane}, {Ord{\'e}novic}, {Ord{\'o}{\~n}ez-Blanco}, {Osborne}, {Pagani},
  {Pagano}, {Pailler}, {Palacin}, {Palaversa}, {Panahi}, {Pawlak},
  {Piersimoni}, {Pineau}, {Plachy}, {Plum}, {Poggio}, {Poujoulet},
  {Pr{\v{s}}a}, {Pulone}, {Racero}, {Ragaini}, {Rambaux}, {Ramos-Lerate},
  {Regibo}, {Reyl{\'e}}, {Riclet}, {Ripepi}, {Riva}, {Rivard}, {Rixon},
  {Roegiers}, {Roelens}, {Romero-G{\'o}mez}, {Rowell}, {Royer}, {Ruiz-Dern},
  {Sadowski}, {Sagrist{\`a} Sell{\'e}s}, {Sahlmann}, {Salgado}, {Salguero},
  {Sanna}, {Santana-Ros}, {Sarasso}, {Savietto}, {Schultheis}, {Sciacca},
  {Segol}, {Segovia}, {S{\'e}gransan}, {Shih}, {Siltala}, {Silva}, {Smart},
  {Smith}, {Solano}, {Solitro}, {Sordo}, {Soria Nieto}, {Souchay}, {Spagna},
  {Spoto}, {Stampa}, {Steele}, {Steidelm{\"u}ller}, {Stephenson}, {Stoev},
  {Suess}, {Surdej}, {Szabados}, {Szegedi-Elek}, {Tapiador}, {Taris}, {Tauran},
  {Taylor}, {Teixeira}, {Terrett}, {Teyssandier}, {Thuillot}, {Titarenko},
  {Torra Clotet}, {Turon}, {Ulla}, {Utrilla}, {Uzzi}, {Vaillant}, {Valentini},
  {Valette}, {van Elteren}, {Van Hemelryck}, {Vaschetto}, {Vecchiato},
  {Veljanoski}, {Viala}, {Vicente}, {Vogt}, {von Essen}, {Voss}, {Votruba},
  {Voutsinas}, {Walmsley}, {Weiler}, {Wertz}, {Wevers}, {Wyrzykowski},
  {Yoldas}, {{\v{Z}}erjal}, {Ziaeepour}, {Zorec}, {Zschocke}, {Zucker},
  {Zurbach}, \& {Zwitter}}]{Babusiaux2018}
{Gaia Collaboration}, {Babusiaux}, C., {van Leeuwen}, F., {et~al.}
  2018{\natexlab{a}}, \aap, 616, A10

\bibitem[{{Gaia Collaboration} {et~al.}(2018{\natexlab{b}}){Gaia
  Collaboration}, {Brown}, {Vallenari}, {Prusti}, {de Bruijne}, {Babusiaux},
  {Bailer-Jones}, {Biermann}, {Evans}, {Eyer}, {Jansen}, {Jordi}, {Klioner},
  {Lammers}, {Lindegren}, {Luri}, {Mignard}, {Panem}, {Pourbaix}, {Randich},
  {Sartoretti}, {Siddiqui}, {Soubiran}, {van Leeuwen}, {Walton}, {Arenou},
  {Bastian}, {Cropper}, {Drimmel}, {Katz}, {Lattanzi}, {Bakker}, {Cacciari},
  {Casta{\~n}eda}, {Chaoul}, {Cheek}, {De Angeli}, {Fabricius}, {Guerra},
  {Holl}, {Masana}, {Messineo}, {Mowlavi}, {Nienartowicz}, {Panuzzo},
  {Portell}, {Riello}, {Seabroke}, {Tanga}, {Th{\'e}venin}, {Gracia-Abril},
  {Comoretto}, {Garcia-Reinaldos}, {Teyssier}, {Altmann}, {Andrae}, {Audard},
  {Bellas-Velidis}, {Benson}, {Berthier}, {Blomme}, {Burgess}, {Busso},
  {Carry}, {Cellino}, {Clementini}, {Clotet}, {Creevey}, {Davidson}, {De
  Ridder}, {Delchambre}, {Dell'Oro}, {Ducourant},
  {Fern{\'a}ndez-Hern{\'a}ndez}, {Fouesneau}, {Fr{\'e}mat}, {Galluccio},
  {Garc{\'\i}a-Torres}, {Gonz{\'a}lez-N{\'u}{\~n}ez}, {Gonz{\'a}lez-Vidal},
  {Gosset}, {Guy}, {Halbwachs}, {Hambly}, {Harrison}, {Hern{\'a}ndez},
  {Hestroffer}, {Hodgkin}, {Hutton}, {Jasniewicz}, {Jean-Antoine-Piccolo},
  {Jordan}, {Korn}, {Krone-Martins}, {Lanzafame}, {Lebzelter}, {L{\"o}ffler},
  {Manteiga}, {Marrese}, {Mart{\'\i}n-Fleitas}, {Moitinho}, {Mora}, {Muinonen},
  {Osinde}, {Pancino}, {Pauwels}, {Petit}, {Recio-Blanco}, {Richards},
  {Rimoldini}, {Robin}, {Sarro}, {Siopis}, {Smith}, {Sozzetti}, {S{\"u}veges},
  {Torra}, {van Reeven}, {Abbas}, {Abreu Aramburu}, {Accart}, {Aerts},
  {Altavilla}, {{\'A}lvarez}, {Alvarez}, {Alves}, {Anderson}, {Andrei},
  {Anglada Varela}, {Antiche}, {Antoja}, {Arcay}, {Astraatmadja}, {Bach},
  {Baker}, {Balaguer-N{\'u}{\~n}ez}, {Balm}, {Barache}, {Barata}, {Barbato},
  {Barblan}, {Barklem}, {Barrado}, {Barros}, {Barstow}, {Bartholom{\'e}
  Mu{\~n}oz}, {Bassilana}, {Becciani}, {Bellazzini}, {Berihuete}, {Bertone},
  {Bianchi}, {Bienaym{\'e}}, {Blanco-Cuaresma}, {Boch}, {Boeche}, {Bombrun},
  {Borrachero}, {Bossini}, {Bouquillon}, {Bourda}, {Bragaglia}, {Bramante},
  {Breddels}, {Bressan}, {Brouillet}, {Br{\"u}semeister}, {Brugaletta},
  {Bucciarelli}, {Burlacu}, {Busonero}, {Butkevich}, {Buzzi}, {Caffau},
  {Cancelliere}, {Cannizzaro}, {Cantat-Gaudin}, {Carballo}, {Carlucci},
  {Carrasco}, {Casamiquela}, {Castellani}, {Castro-Ginard}, {Charlot},
  {Chemin}, {Chiavassa}, {Cocozza}, {Costigan}, {Cowell}, {Crifo}, {Crosta},
  {Crowley}, {Cuypers}, {Dafonte}, {Damerdji}, {Dapergolas}, {David}, {David},
  {de Laverny}, {De Luise}, {De March}, {de Martino}, {de Souza}, {de Torres},
  {Debosscher}, {del Pozo}, {Delbo}, {Delgado}, {Delgado}, {Di Matteo},
  {Diakite}, {Diener}, {Distefano}, {Dolding}, {Drazinos}, {Dur{\'a}n},
  {Edvardsson}, {Enke}, {Eriksson}, {Esquej}, {Eynard Bontemps}, {Fabre},
  {Fabrizio}, {Faigler}, {Falc{\~a}o}, {Farr{\`a}s Casas}, {Federici},
  {Fedorets}, {Fernique}, {Figueras}, {Filippi}, {Findeisen}, {Fonti},
  {Fraile}, {Fraser}, {Fr{\'e}zouls}, {Gai}, {Galleti}, {Garabato},
  {Garc{\'\i}a-Sedano}, {Garofalo}, {Garralda}, {Gavel}, {Gavras}, {Gerssen},
  {Geyer}, {Giacobbe}, {Gilmore}, {Girona}, {Giuffrida}, {Glass}, {Gomes},
  {Granvik}, {Gueguen}, {Guerrier}, {Guiraud}, {Guti{\'e}rrez-S{\'a}nchez},
  {Haigron}, {Hatzidimitriou}, {Hauser}, {Haywood}, {Heiter}, {Helmi}, {Heu},
  {Hilger}, {Hobbs}, {Hofmann}, {Holland}, {Huckle}, {Hypki}, {Icardi},
  {Jan{\ss}en}, {Jevardat de Fombelle}, {Jonker}, {Juh{\'a}sz}, {Julbe},
  {Karampelas}, {Kewley}, {Klar}, {Kochoska}, {Kohley}, {Kolenberg},
  {Kontizas}, {Kontizas}, {Koposov}, {Kordopatis}, {Kostrzewa-Rutkowska},
  {Koubsky}, {Lambert}, {Lanza}, {Lasne}, {Lavigne}, {Le Fustec}, {Le
  Poncin-Lafitte}, {Lebreton}, {Leccia}, {Leclerc}, {Lecoeur-Taibi},
  {Lenhardt}, {Leroux}, {Liao}, {Licata}, {Lindstr{\o}m}, {Lister}, {Livanou},
  {Lobel}, {L{\'o}pez}, {Managau}, {Mann}, {Mantelet}, {Marchal}, {Marchant},
  {Marconi}, {Marinoni}, {Marschalk{\'o}}, {Marshall}, {Martino}, {Marton},
  {Mary}, {Massari}, {Matijevi{\v{c}}}, {Mazeh}, {McMillan}, {Messina},
  {Michalik}, {Millar}, {Molina}, {Molinaro}, {Moln{\'a}r}, {Montegriffo},
  {Mor}, {Morbidelli}, {Morel}, {Morris}, {Mulone}, {Muraveva}, {Musella},
  {Nelemans}, {Nicastro}, {Noval}, {O'Mullane}, {Ord{\'e}novic},
  {Ord{\'o}{\~n}ez-Blanco}, {Osborne}, {Pagani}, {Pagano}, {Pailler},
  {Palacin}, {Palaversa}, {Panahi}, {Pawlak}, {Piersimoni}, {Pineau}, {Plachy},
  {Plum}, {Poggio}, {Poujoulet}, {Pr{\v{s}}a}, {Pulone}, {Racero}, {Ragaini},
  {Rambaux}, {Ramos-Lerate}, {Regibo}, {Reyl{\'e}}, {Riclet}, {Ripepi}, {Riva},
  {Rivard}, {Rixon}, {Roegiers}, {Roelens}, {Romero-G{\'o}mez}, {Rowell},
  {Royer}, {Ruiz-Dern}, {Sadowski}, {Sagrist{\`a} Sell{\'e}s}, {Sahlmann},
  {Salgado}, {Salguero}, {Sanna}, {Santana-Ros}, {Sarasso}, {Savietto},
  {Schultheis}, {Sciacca}, {Segol}, {Segovia}, {S{\'e}gransan}, {Shih},
  {Siltala}, {Silva}, {Smart}, {Smith}, {Solano}, {Solitro}, {Sordo}, {Soria
  Nieto}, {Souchay}, {Spagna}, {Spoto}, {Stampa}, {Steele},
  {Steidelm{\"u}ller}, {Stephenson}, {Stoev}, {Suess}, {Surdej}, {Szabados},
  {Szegedi-Elek}, {Tapiador}, {Taris}, {Tauran}, {Taylor}, {Teixeira},
  {Terrett}, {Teyssand ier}, {Thuillot}, {Titarenko}, {Torra Clotet}, {Turon},
  {Ulla}, {Utrilla}, {Uzzi}, {Vaillant}, {Valentini}, {Valette}, {van Elteren},
  {Van Hemelryck}, {van Leeuwen}, {Vaschetto}, {Vecchiato}, {Veljanoski},
  {Viala}, {Vicente}, {Vogt}, {von Essen}, {Voss}, {Votruba}, {Voutsinas},
  {Walmsley}, {Weiler}, {Wertz}, {Wevers}, {Wyrzykowski}, {Yoldas},
  {{\v{Z}}erjal}, {Ziaeepour}, {Zorec}, {Zschocke}, {Zucker}, {Zurbach}, \&
  {Zwitter}}]{GaiaDR2_2018}
{Gaia Collaboration}, {Brown}, A.~G.~A., {Vallenari}, A., {et~al.}
  2018{\natexlab{b}}, \aap, 616, A1

\bibitem[{{Gaia Collaboration} {et~al.}(2021){Gaia Collaboration}, {Brown},
  {Vallenari}, {Prusti}, {de Bruijne}, {Babusiaux}, {Biermann}, {Creevey},
  {Evans}, {Eyer}, {Hutton}, {Jansen}, {Jordi}, {Klioner}, {Lammers},
  {Lindegren}, {Luri}, {Mignard}, {Panem}, {Pourbaix}, {Randich}, {Sartoretti},
  {Soubiran}, {Walton}, {Arenou}, {Bailer-Jones}, {Bastian}, {Cropper},
  {Drimmel}, {Katz}, {Lattanzi}, {van Leeuwen}, {Bakker}, {Cacciari},
  {Casta{\~n}eda}, {De Angeli}, {Ducourant}, {Fabricius}, {Fouesneau},
  {Fr{\'e}mat}, {Guerra}, {Guerrier}, {Guiraud}, {Jean-Antoine Piccolo},
  {Masana}, {Messineo}, {Mowlavi}, {Nicolas}, {Nienartowicz}, {Pailler},
  {Panuzzo}, {Riclet}, {Roux}, {Seabroke}, {Sordo}, {Tanga}, {Th{\'e}venin},
  {Gracia-Abril}, {Portell}, {Teyssier}, {Altmann}, {Andrae}, {Bellas-Velidis},
  {Benson}, {Berthier}, {Blomme}, {Brugaletta}, {Burgess}, {Busso}, {Carry},
  {Cellino}, {Cheek}, {Clementini}, {Damerdji}, {Davidson}, {Delchambre},
  {Dell'Oro}, {Fern{\'a}ndez-Hern{\'a}ndez}, {Galluccio}, {Garc{\'\i}a-Lario},
  {Garcia-Reinaldos}, {Gonz{\'a}lez-N{\'u}{\~n}ez}, {Gosset}, {Haigron},
  {Halbwachs}, {Hambly}, {Harrison}, {Hatzidimitriou}, {Heiter},
  {Hern{\'a}ndez}, {Hestroffer}, {Hodgkin}, {Holl}, {Jan{\ss}en}, {Jevardat de
  Fombelle}, {Jordan}, {Krone-Martins}, {Lanzafame}, {L{\"o}ffler}, {Lorca},
  {Manteiga}, {Marchal}, {Marrese}, {Moitinho}, {Mora}, {Muinonen}, {Osborne},
  {Pancino}, {Pauwels}, {Petit}, {Recio-Blanco}, {Richards}, {Riello},
  {Rimoldini}, {Robin}, {Roegiers}, {Rybizki}, {Sarro}, {Siopis}, {Smith},
  {Sozzetti}, {Ulla}, {Utrilla}, {van Leeuwen}, {van Reeven}, {Abbas}, {Abreu
  Aramburu}, {Accart}, {Aerts}, {Aguado}, {Ajaj}, {Altavilla}, {{\'A}lvarez},
  {{\'A}lvarez Cid-Fuentes}, {Alves}, {Anderson}, {Anglada Varela}, {Antoja},
  {Audard}, {Baines}, {Baker}, {Balaguer-N{\'u}{\~n}ez}, {Balbinot}, {Balog},
  {Barache}, {Barbato}, {Barros}, {Barstow}, {Bartolom{\'e}}, {Bassilana},
  {Bauchet}, {Baudesson-Stella}, {Becciani}, {Bellazzini}, {Bernet}, {Bertone},
  {Bianchi}, {Blanco-Cuaresma}, {Boch}, {Bombrun}, {Bossini}, {Bouquillon},
  {Bragaglia}, {Bramante}, {Breedt}, {Bressan}, {Brouillet}, {Bucciarelli},
  {Burlacu}, {Busonero}, {Butkevich}, {Buzzi}, {Caffau}, {Cancelliere},
  {C{\'a}novas}, {Cantat-Gaudin}, {Carballo}, {Carlucci}, {Carnerero},
  {Carrasco}, {Casamiquela}, {Castellani}, {Castro-Ginard}, {Castro Sampol},
  {Chaoul}, {Charlot}, {Chemin}, {Chiavassa}, {Cioni}, {Comoretto}, {Cooper},
  {Cornez}, {Cowell}, {Crifo}, {Crosta}, {Crowley}, {Dafonte}, {Dapergolas},
  {David}, {David}, {de Laverny}, {De Luise}, {De March}, {De Ridder}, {de
  Souza}, {de Teodoro}, {de Torres}, {del Peloso}, {del Pozo}, {Delbo},
  {Delgado}, {Delgado}, {Delisle}, {Di Matteo}, {Diakite}, {Diener},
  {Distefano}, {Dolding}, {Eappachen}, {Edvardsson}, {Enke}, {Esquej}, {Fabre},
  {Fabrizio}, {Faigler}, {Fedorets}, {Fernique}, {Fienga}, {Figueras},
  {Fouron}, {Fragkoudi}, {Fraile}, {Franke}, {Gai}, {Garabato},
  {Garcia-Gutierrez}, {Garc{\'\i}a-Torres}, {Garofalo}, {Gavras}, {Gerlach},
  {Geyer}, {Giacobbe}, {Gilmore}, {Girona}, {Giuffrida}, {Gomel}, {Gomez},
  {Gonzalez-Santamaria}, {Gonz{\'a}lez-Vidal}, {Granvik},
  {Guti{\'e}rrez-S{\'a}nchez}, {Guy}, {Hauser}, {Haywood}, {Helmi}, {Hidalgo},
  {Hilger}, {H{\l}adczuk}, {Hobbs}, {Holland}, {Huckle}, {Jasniewicz},
  {Jonker}, {Juaristi Campillo}, {Julbe}, {Karbevska}, {Kervella}, {Khanna},
  {Kochoska}, {Kontizas}, {Kordopatis}, {Korn}, {Kostrzewa-Rutkowska},
  {Kruszy{\'n}ska}, {Lambert}, {Lanza}, {Lasne}, {Le Campion}, {Le Fustec},
  {Lebreton}, {Lebzelter}, {Leccia}, {Leclerc}, {Lecoeur-Taibi}, {Liao},
  {Licata}, {Lindstr{\o}m}, {Lister}, {Livanou}, {Lobel}, {Madrero Pardo},
  {Managau}, {Mann}, {Marchant}, {Marconi}, {Marcos Santos}, {Marinoni},
  {Marocco}, {Marshall}, {Martin Polo}, {Mart{\'\i}n-Fleitas}, {Masip},
  {Massari}, {Mastrobuono-Battisti}, {Mazeh}, {McMillan}, {Messina},
  {Michalik}, {Millar}, {Mints}, {Molina}, {Molinaro}, {Moln{\'a}r},
  {Montegriffo}, {Mor}, {Morbidelli}, {Morel}, {Morris}, {Mulone}, {Munoz},
  {Muraveva}, {Murphy}, {Musella}, {Noval}, {Ord{\'e}novic}, {Orr{\`u}},
  {Osinde}, {Pagani}, {Pagano}, {Palaversa}, {Palicio}, {Panahi}, {Pawlak},
  {Pe{\~n}alosa Esteller}, {Penttil{\"a}}, {Piersimoni}, {Pineau}, {Plachy},
  {Plum}, {Poggio}, {Poretti}, {Poujoulet}, {Pr{\v{s}}a}, {Pulone}, {Racero},
  {Ragaini}, {Rainer}, {Raiteri}, {Rambaux}, {Ramos}, {Ramos-Lerate}, {Re
  Fiorentin}, {Regibo}, {Reyl{\'e}}, {Ripepi}, {Riva}, {Rixon}, {Robichon},
  {Robin}, {Roelens}, {Rohrbasser}, {Romero-G{\'o}mez}, {Rowell}, {Royer},
  {Rybicki}, {Sadowski}, {Sagrist{\`a} Sell{\'e}s}, {Sahlmann}, {Salgado},
  {Salguero}, {Samaras}, {Sanchez Gimenez}, {Sanna}, {Santove{\~n}a},
  {Sarasso}, {Schultheis}, {Sciacca}, {Segol}, {Segovia}, {S{\'e}gransan},
  {Semeux}, {Shahaf}, {Siddiqui}, {Siebert}, {Siltala}, {Slezak}, {Smart},
  {Solano}, {Solitro}, {Souami}, {Souchay}, {Spagna}, {Spoto}, {Steele},
  {Steidelm{\"u}ller}, {Stephenson}, {S{\"u}veges}, {Szabados}, {Szegedi-Elek},
  {Taris}, {Tauran}, {Taylor}, {Teixeira}, {Thuillot}, {Tonello}, {Torra},
  {Torra}, {Turon}, {Unger}, {Vaillant}, {van Dillen}, {Vanel}, {Vecchiato},
  {Viala}, {Vicente}, {Voutsinas}, {Weiler}, {Wevers}, {Wyrzykowski}, {Yoldas},
  {Yvard}, {Zhao}, {Zorec}, {Zucker}, {Zurbach}, \& {Zwitter}}]{GaiaEDR3_2021}
{Gaia Collaboration}, {Brown}, A.~G.~A., {Vallenari}, A., {et~al.} 2021, \aap,
  649, A1

\bibitem[{{Gaia Collaboration} {et~al.}(2017){Gaia Collaboration},
  {Clementini}, {Eyer}, {Ripepi}, {Marconi}, {Muraveva}, {Garofalo}, {Sarro},
  {Palmer}, {Luri}, {Molinaro}, {Rimoldini}, {Szabados}, {Musella}, {Anderson},
  {Prusti}, {de Bruijne}, {Brown}, {Vallenari}, {Babusiaux}, {Bailer-Jones},
  {Bastian}, {Biermann}, {Evans}, {Jansen}, {Jordi}, {Klioner}, {Lammers},
  {Lindegren}, {Mignard}, {Panem}, {Pourbaix}, {Randich}, {Sartoretti},
  {Siddiqui}, {Soubiran}, {Valette}, {van Leeuwen}, {Walton}, {Aerts},
  {Arenou}, {Cropper}, {Drimmel}, {H{\o}g}, {Katz}, {Lattanzi}, {O'Mullane},
  {Grebel}, {Holland}, {Huc}, {Passot}, {Perryman}, {Bramante}, {Cacciari},
  {Casta{\~n}eda}, {Chaoul}, {Cheek}, {De Angeli}, {Fabricius}, {Guerra},
  {Hern{\'a}ndez}, {Jean-Antoine-Piccolo}, {Masana}, {Messineo}, {Mowlavi},
  {Nienartowicz}, {Ord{\'o}{\~n}ez-Blanco}, {Panuzzo}, {Portell}, {Richards},
  {Riello}, {Seabroke}, {Tanga}, {Th{\'e}venin}, {Torra}, {Els},
  {Gracia-Abril}, {Comoretto}, {Garcia-Reinaldos}, {Lock}, {Mercier},
  {Altmann}, {Andrae}, {Astraatmadja}, {Bellas-Velidis}, {Benson}, {Berthier},
  {Blomme}, {Busso}, {Carry}, {Cellino}, {Cowell}, {Creevey}, {Cuypers},
  {Davidson}, {De Ridder}, {de Torres}, {Delchambre}, {Dell'Oro}, {Ducourant},
  {Fr{\'e}mat}, {Garc{\'\i}a-Torres}, {Gosset}, {Halbwachs}, {Hambly},
  {Harrison}, {Hauser}, {Hestroffer}, {Hodgkin}, {Huckle}, {Hutton},
  {Jasniewicz}, {Jordan}, {Kontizas}, {Korn}, {Lanzafame}, {Manteiga},
  {Moitinho}, {Muinonen}, {Osinde}, {Pancino}, {Pauwels}, {Petit},
  {Recio-Blanco}, {Robin}, {Siopis}, {Smith}, {Smith}, {Sozzetti}, {Thuillot},
  {van Reeven}, {Viala}, {Abbas}, {Abreu Aramburu}, {Accart}, {Aguado},
  {Allan}, {Allasia}, {Altavilla}, {{\'A}lvarez}, {Alves}, {Andrei}, {Anglada
  Varela}, {Antiche}, {Antoja}, {Ant{\'o}n}, {Arcay}, {Bach}, {Baker},
  {Balaguer-N{\'u}{\~n}ez}, {Barache}, {Barata}, {Barbier}, {Barblan}, {Barrado
  y Navascu{\'e}s}, {Barros}, {Barstow}, {Becciani}, {Bellazzini}, {Bello
  Garc{\'\i}a}, {Belokurov}, {Bendjoya}, {Berihuete}, {Bianchi},
  {Bienaym{\'e}}, {Billebaud}, {Blagorodnova}, {Blanco-Cuaresma}, {Boch},
  {Bombrun}, {Borrachero}, {Bouquillon}, {Bourda}, {Bragaglia}, {Breddels},
  {Brouillet}, {Br{\"u}semeister}, {Bucciarelli}, {Burgess}, {Burgon},
  {Burlacu}, {Busonero}, {Buzzi}, {Caffau}, {Cambras}, {Campbell},
  {Cancelliere}, {Cantat-Gaudin}, {Carlucci}, {Carrasco}, {Castellani},
  {Charlot}, {Charnas}, {Chiavassa}, {Clotet}, {Cocozza}, {Collins},
  {Costigan}, {Crifo}, {Cross}, {Crosta}, {Crowley}, {Dafonte}, {Damerdji},
  {Dapergolas}, {David}, {David}, {De Cat}, {de Felice}, {de Laverny}, {De
  Luise}, {De March}, {de Souza}, {Debosscher}, {del Pozo}, {Delbo}, {Delgado},
  {Delgado}, {Di Matteo}, {Diakite}, {Distefano}, {Dolding}, {Dos Anjos},
  {Drazinos}, {Dur{\'a}n}, {Dzigan}, {Edvardsson}, {Enke}, {Evans}, {Eynard
  Bontemps}, {Fabre}, {Fabrizio}, {Falc{\~a}o}, {Farr{\`a}s Casas}, {Federici},
  {Fedorets}, {Fern{\'a}ndez-Hern{\'a}ndez}, {Fernique}, {Fienga}, {Figueras},
  {Filippi}, {Findeisen}, {Fonti}, {Fouesneau}, {Fraile}, {Fraser}, {Fuchs},
  {Gai}, {Galleti}, {Galluccio}, {Garabato}, {Garc{\'\i}a-Sedano}, {Garralda},
  {Gavras}, {Gerssen}, {Geyer}, {Gilmore}, {Girona}, {Giuffrida}, {Gomes},
  {Gonz{\'a}lez-Marcos}, {Gonz{\'a}lez-N{\'u}{\~n}ez}, {Gonz{\'a}lez-Vidal},
  {Granvik}, {Guerrier}, {Guillout}, {Guiraud}, {G{\'u}rpide},
  {Guti{\'e}rrez-S{\'a}nchez}, {Guy}, {Haigron}, {Hatzidimitriou}, {Haywood},
  {Heiter}, {Helmi}, {Hobbs}, {Hofmann}, {Holl}, {Holland}, {Hunt}, {Hypki},
  {Icardi}, {Irwin}, {Jevardat de Fombelle}, {Jofr{\'e}}, {Jonker}, {Jorissen},
  {Julbe}, {Karampelas}, {Kochoska}, {Kohley}, {Kolenberg}, {Kontizas},
  {Koposov}, {Kordopatis}, {Koubsky}, {Krone-Martins}, {Kudryashova},
  {Bachchan}, {Lacoste-Seris}, {Lanza}, {Lavigne}, {Le Poncin-Lafitte},
  {Lebreton}, {Lebzelter}, {Leccia}, {Leclerc}, {Lecoeur-Taibi}, {Lemaitre},
  {Lenhardt}, {Leroux}, {Liao}, {Licata}, {Lindstr{\o}m}, {Lister}, {Livanou},
  {Lobel}, {L{\"o}ffler}, {L{\'o}pez}, {Lorenz}, {MacDonald}, {Magalh{\~a}es
  Fernandes}, {Managau}, {Mann}, {Mantelet}, {Marchal}, {Marchant}, {Marinoni},
  {Marrese}, {Marschalk{\'o}}, {Marshall}, {Mart{\'\i}n-Fleitas}, {Martino},
  {Mary}, {Matijevi{\v{c}}}, {McMillan}, {Messina}, {Michalik}, {Millar},
  {Miranda}, {Molina}, {Molinaro}, {Moln{\'a}r}, {Moniez}, {Montegriffo},
  {Mor}, {Mora}, {Morbidelli}, {Morel}, {Morgenthaler}, {Morris}, {Mulone},
  {Narbonne}, {Nelemans}, {Nicastro}, {Noval}, {Ord{\'e}novic},
  {Ordieres-Mer{\'e}}, {Osborne}, {Pagani}, {Pagano}, {Pailler}, {Palacin},
  {Palaversa}, {Parsons}, {Pecoraro}, {Pedrosa}, {Pentik{\"a}inen}, {Pichon},
  {Piersimoni}, {Pineau}, {Plachy}, {Plum}, {Poujoulet}, {Pr{\v{s}}a},
  {Pulone}, {Ragaini}, {Rago}, {Rambaux}, {Ramos-Lerate}, {Ranalli}, {Rauw},
  {Read}, {Regibo}, {Reyl{\'e}}, {Ribeiro}, {Riva}, {Rixon}, {Roelens},
  {Romero-G{\'o}mez}, {Rowell}, {Royer}, {Ruiz-Dern}, {Sadowski}, {Sagrist{\`a}
  Sell{\'e}s}, {Sahlmann}, {Salgado}, {Salguero}, {Sarasso}, {Savietto},
  {Schultheis}, {Sciacca}, {Segol}, {Segovia}, {Segransan}, {Shih},
  {Smareglia}, {Smart}, {Solano}, {Solitro}, {Sordo}, {Soria Nieto}, {Souchay},
  {Spagna}, {Spoto}, {Stampa}, {Steele}, {Steidelm{\"u}ller}, {Stephenson},
  {Stoev}, {Suess}, {S{\"u}veges}, {Surdej}, {Szegedi-Elek}, {Tapiador},
  {Taris}, {Tauran}, {Taylor}, {Teixeira}, {Terrett}, {Tingley}, {Trager},
  {Turon}, {Ulla}, {Utrilla}, {Valentini}, {van Elteren}, {Van Hemelryck}, {van
  Leeuwen}, {Varadi}, {Vecchiato}, {Veljanoski}, {Via}, {Vicente}, {Vogt},
  {Voss}, {Votruba}, {Voutsinas}, {Walmsley}, {Weiler}, {Weingrill}, {Wevers},
  {Wyrzykowski}, {Yoldas}, {{\v{Z}}erjal}, {Zucker}, {Zurbach}, {Zwitter},
  {Alecu}, {Allen}, {Allende Prieto}, {Amorim}, {Anglada-Escud{\'e}},
  {Arsenijevic}, {Azaz}, {Balm}, {Beck}, {Bernstein}, {Bigot}, {Bijaoui},
  {Blasco}, {Bonfigli}, {Bono}, {Boudreault}, {Bressan}, {Brown}, {Brunet},
  {Bunclark}, {Buonanno}, {Butkevich}, {Carret}, {Carrion}, {Chemin},
  {Ch{\'e}reau}, {Corcione}, {Darmigny}, {de Boer}, {de Teodoro}, {de Zeeuw},
  {Delle Luche}, {Domingues}, {Dubath}, {Fodor}, {Fr{\'e}zouls}, {Fries},
  {Fustes}, {Fyfe}, {Gallardo}, {Gallegos}, {Gardiol}, {Gebran}, {Gomboc},
  {G{\'o}mez}, {Grux}, {Gueguen}, {Heyrovsky}, {Hoar}, {Iannicola}, {Isasi
  Parache}, {Janotto}, {Joliet}, {Jonckheere}, {Keil}, {Kim}, {Klagyivik},
  {Klar}, {Knude}, {Kochukhov}, {Kolka}, {Kos}, {Kutka}, {Lainey}, {LeBouquin},
  {Liu}, {Loreggia}, {Makarov}, {Marseille}, {Martayan}, {Martinez-Rubi},
  {Massart}, {Meynadier}, {Mignot}, {Munari}, {Nguyen}, {Nordlander},
  {O'Flaherty}, {Ocvirk}, {Olias Sanz}, {Ortiz}, {Osorio}, {Oszkiewicz},
  {Ouzounis}, {Park}, {Pasquato}, {Peltzer}, {Peralta}, {P{\'e}turaud},
  {Pieniluoma}, {Pigozzi}, {Poels}, {Prat}, {Prod'homme}, {Raison}, {Rebordao},
  {Risquez}, {Rocca-Volmerange}, {Rosen}, {Ruiz-Fuertes}, {Russo}, {Serraller
  Vizcaino}, {Short}, {Siebert}, {Silva}, {Sinachopoulos}, {Slezak}, {Soffel},
  {Sosnowska}, {Strai{\v{z}}ys}, {ter Linden}, {Terrell}, {Theil}, {Tiede},
  {Troisi}, {Tsalmantza}, {Tur}, {Vaccari}, {Vachier}, {Valles}, {Van Hamme},
  {Veltz}, {Virtanen}, {Wallut}, {Wichmann}, {Wilkinson}, {Ziaeepour}, \&
  {Zschocke}}]{2017A&A...605A..79G}
{Gaia Collaboration}, {Clementini}, G., {Eyer}, L., {et~al.} 2017, \aap, 605,
  A79

\bibitem[{{Garro} {et~al.}(2021{\natexlab{a}}){Garro}, {Minniti}, {G{\'o}mez},
  \& {Alonso-Garc{\'\i}a}}]{Garro2021_SGR}
{Garro}, E.~R., {Minniti}, D., {G{\'o}mez}, M., \& {Alonso-Garc{\'\i}a}, J.
  2021{\natexlab{a}}, \aap, 654, A23

\bibitem[{{Garro} {et~al.}(2020){Garro}, {Minniti}, {G{\'o}mez},
  {Alonso-Garc{\'\i}a}, {Barb{\'a}}, {Barbuy}, {Clari{\'a}}, {Chen{\'e}},
  {Dias}, {Hempel}, {Ivanov}, {Lucas}, {Majaess}, {Mauro}, {Moni Bidin},
  {Palma}, {Pullen}, {Saito}, {Smith}, {Surot}, {Ram{\'\i}rez Alegr{\'\i}a},
  {Rejkuba}, {Ripepi}, \& {Fern{\'a}ndez Trincado}}]{Garro2020}
{Garro}, E.~R., {Minniti}, D., {G{\'o}mez}, M., {et~al.} 2020, \aap, 642, L19

\bibitem[{{Garro} {et~al.}(2021{\natexlab{b}}){Garro}, {Minniti}, {G{\'o}mez},
  {Alonso-Garc{\'\i}a}, {Palma}, {Smith}, \& {Ripepi}}]{Garro2021}
{Garro}, E.~R., {Minniti}, D., {G{\'o}mez}, M., {et~al.} 2021{\natexlab{b}},
  \aap, 649, A86

\bibitem[{{Garro} {et~al.}(2021{\natexlab{c}}){Garro}, {Minniti}, {G{\'o}mez},
  {Alonso-Garc{\'\i}a}, {Ripepi}, {Fern{\'a}ndez-Trincado}, \& {Vivanco
  C{\'a}diz}}]{Garro2021b:submitted-a}
{Garro}, E.~R., {Minniti}, D., {G{\'o}mez}, M., {et~al.} 2021{\natexlab{c}},
  arXiv e-prints, arXiv:2111.08317

\bibitem[{{Gonzalez} {et~al.}(2011){Gonzalez}, {Rejkuba}, {Minniti}, {Zoccali},
  {Valenti}, \& {Saito}}]{Gonzalez2011}
{Gonzalez}, O.~A., {Rejkuba}, M., {Minniti}, D., {et~al.} 2011, \aap, 534, L14

\bibitem[{{Gonz{\'a}lez-Fern{\'a}ndez}
  {et~al.}(2018){Gonz{\'a}lez-Fern{\'a}ndez}, {Hodgkin}, {Irwin},
  {Gonz{\'a}lez-Solares}, {Koposov}, {Lewis}, {Emerson}, {Hewett},
  {Yolda{\c{s}}}, \& {Riello}}]{Gonzalez_Fernandez2018}
{Gonz{\'a}lez-Fern{\'a}ndez}, C., {Hodgkin}, S.~T., {Irwin}, M.~J., {et~al.}
  2018, \mnras, 474, 5459

\bibitem[{{Gran} {et~al.}(2021){Gran}, {Zoccali}, {Saviane}, {Valenti},
  {Rojas-Arriagada}, {Ramos}, {Hartke}, {Carballo-Bello}, {Navarrete},
  {Rejkuba}, \& {Carvajal}}]{Gran2021}
{Gran}, F., {Zoccali}, M., {Saviane}, I., {et~al.} 2021, \mnras
  [\eprint[arXiv]{2108.11922}]

\bibitem[{{Gravity Collaboration} {et~al.}(2019){Gravity Collaboration},
  {Abuter}, {Amorim}, {Baub{\"o}ck}, {Berger}, {Bonnet}, {Brandner},
  {Cl{\'e}net}, {Coud{\'e} Du Foresto}, {de Zeeuw}, {Dexter}, {Duvert},
  {Eckart}, {Eisenhauer}, {F{\"o}rster Schreiber}, {Garcia}, {Gao}, {Gendron},
  {Genzel}, {Gerhard}, {Gillessen}, {Habibi}, {Haubois}, {Henning}, {Hippler},
  {Horrobin}, {Jim{\'e}nez-Rosales}, {Jocou}, {Kervella}, {Lacour},
  {Lapeyr{\`e}re}, {Le Bouquin}, {L{\'e}na}, {Ott}, {Paumard}, {Perraut},
  {Perrin}, {Pfuhl}, {Rabien}, {Rodriguez Coira}, {Rousset}, {Scheithauer},
  {Sternberg}, {Straub}, {Straubmeier}, {Sturm}, {Tacconi}, {Vincent}, {von
  Fellenberg}, {Waisberg}, {Widmann}, {Wieprecht}, {Wiezorrek}, {Woillez}, \&
  {Yazici}}]{Gravity2019}
{Gravity Collaboration}, {Abuter}, R., {Amorim}, A., {et~al.} 2019, \aap, 625,
  L10

\bibitem[{{Harris}(1991)}]{Harris_1991}
{Harris}, W.~E. 1991, \araa, 29, 543

\bibitem[{Harris {et~al.}(2013)Harris, Harris, \& Alessi}]{Harris_2013}
Harris, W.~E., Harris, G. L.~H., \& Alessi, M. 2013, ApJ, 772, 82

\bibitem[{{Irwin} {et~al.}(2004){Irwin}, {Lewis}, {Hodgkin}, {Bunclark},
  {Evans}, {McMahon}, {Emerson}, {Stewart}, \& {Beard}}]{Irwin2004}
{Irwin}, M.~J., {Lewis}, J., {Hodgkin}, S., {et~al.} 2004, in (SPIE) Conference
  Series, Vol. 5493, Optimizing Scientific Return for Astronomy through
  Information Technologies, ed. P.~J. {Quinn} \& A.~{Bridger}, 411--422

\bibitem[{{Kharchenko} {et~al.}(2013){Kharchenko}, {Piskunov}, {Roeser},
  {Schilbach}, \& {Scholz}}]{Kharchenko2013}
{Kharchenko}, N.~V., {Piskunov}, A.~E., {Roeser}, S., {Schilbach}, E., \&
  {Scholz}, R.~D. 2013, VizieR Online Data Catalog, J/A+A/558/A53

\bibitem[{{Kharchenko} {et~al.}(2016){Kharchenko}, {Piskunov}, {Schilbach},
  {R{\"o}ser}, \& {Scholz}}]{Kharchenko2016}
{Kharchenko}, N.~V., {Piskunov}, A.~E., {Schilbach}, E., {R{\"o}ser}, S., \&
  {Scholz}, R.~D. 2016, \aap, 585, A101

\bibitem[{{Koposov} {et~al.}(2017){Koposov}, {Belokurov}, \&
  {Torrealba}}]{Koposov2017}
{Koposov}, S.~E., {Belokurov}, V., \& {Torrealba}, G. 2017, \mnras, 470, 2702

\bibitem[{{Kormendy} \& {Kennicutt}(2004)}]{Kormendy2004}
{Kormendy}, J. \& {Kennicutt}, R.~C., J. 2004, \araa, 42, 603

\bibitem[{{Kronberger} {et~al.}(2016){Kronberger}, {Parker}, {Jacoby}, {Acker},
  {Alves}, {Bojicic}, {Eigenthaler}, {Frew}, {Harmer}, {Patchick}, {Reid}, \&
  {Schedler}}]{Kronberger2016}
{Kronberger}, M., {Parker}, Q.~A., {Jacoby}, G.~H., {et~al.} 2016, in Journal
  of Physics Conference Series, Vol. 728, Journal of Physics Conference Series,
  072012

\bibitem[{{Kronberger} {et~al.}(2012){Kronberger}, {Reegen}, {Alessi},
  {Patchick}, \& {Teutsch}}]{Kronberger2012}
{Kronberger}, M., {Reegen}, P., {Alessi}, B.~S., {Patchick}, D., \& {Teutsch},
  P. 2012, Astrophysics and Space Science Proceedings, 29, 105

\bibitem[{{Kronberger} {et~al.}(2006){Kronberger}, {Teutsch}, {Alessi},
  {Steine}, {Ferrero}, {Graczewski}, {Juchert}, {Patchick}, {Riddle},
  {Saloranta}, {Schoenball}, \& {Watson}}]{Kronberger2006}
{Kronberger}, M., {Teutsch}, P., {Alessi}, B., {et~al.} 2006, \aap, 447, 921

\bibitem[{{Lauberts}(1982)}]{1982euse.book.....L}
{Lauberts}, A. 1982, {ESO/Uppsala survey of the ESO(B) atlas}

\bibitem[{{Marigo} {et~al.}(2017){Marigo}, {Girardi}, {Bressan}, {Rosenfield},
  {Aringer}, {Chen}, {Dussin}, {Nanni}, {Pastorelli}, {Rodrigues}, {Trabucchi},
  {Bladh}, {Dalcanton}, {Groenewegen}, {Montalb{\'a}n}, \& {Wood}}]{Marigo2017}
{Marigo}, P., {Girardi}, L., {Bressan}, A., {et~al.} 2017, \apj, 835, 77

\bibitem[{{Massari} {et~al.}(2019){Massari}, {Koppelman}, \&
  {Helmi}}]{Massari2019}
{Massari}, D., {Koppelman}, H.~H., \& {Helmi}, A. 2019, \aap, 630, L4

\bibitem[{{McQuinn} {et~al.}(2019){McQuinn}, {Boyer}, {Skillman}, \&
  {Dolphin}}]{McQuinn2019}
{McQuinn}, K. B.~W., {Boyer}, M., {Skillman}, E.~D., \& {Dolphin}, A.~E. 2019,
  \apj, 880, 63

\bibitem[{{Minniti}(2018)}]{Minniti2018}
{Minniti}, D. 2018, in The Vatican Observatory: 80th Anniversary Celebration,
  ed. G.~{Gionti} \& J.-B. {Kikwaya Eluo}, Vol.~51, 63

\bibitem[{{Minniti} {et~al.}(2021{\natexlab{a}}){Minniti},
  {Fern{\'a}ndez-Trincado}, {G{\'o}mez}, {Smith}, {Lucas}, \& {Contreras
  Ramos}}]{Minniti2021_CL160}
{Minniti}, D., {Fern{\'a}ndez-Trincado}, J.~G., {G{\'o}mez}, M., {et~al.}
  2021{\natexlab{a}}, \aap, 650, L11

\bibitem[{Minniti {et~al.}(2017{\natexlab{a}})Minniti, Geisler,
  Alonso-Garc{\'{\i}}a, Palma, Beam{\'{\i}}n, Borissova, Catelan, Clari{\'{a}},
  Cohen, Ramos, Dias, Fern{\'{a}}ndez-Trincado, G{\'{o}}mez, Hempel, Ivanov,
  Kurtev, Lucas, Moni-Bidin, Pullen, Alegr{\'{\i}}a, Saito, \&
  Valenti}]{Minniti_2017}
Minniti, D., Geisler, D., Alonso-Garc{\'{\i}}a, J., {et~al.}
  2017{\natexlab{a}}, ApJ, 849, L24

\bibitem[{{Minniti} {et~al.}(2011){Minniti}, {Hempel}, {Toledo}, {Ivanov},
  {Alonso-Garc{\'\i}a}, {Saito}, {Catelan}, {Geisler}, {Jord{\'a}n},
  {Borissova}, {Zoccali}, {Kurtev}, {Carraro}, {Barbuy}, {Clari{\'a}},
  {Rejkuba}, {Emerson}, \& {Moni Bidin}}]{Minniti2011}
{Minniti}, D., {Hempel}, M., {Toledo}, I., {et~al.} 2011, \aap, 527, A81

\bibitem[{{Minniti} {et~al.}(2010){Minniti}, {Lucas}, {Emerson}, {Saito},
  {Hempel}, {Pietrukowicz}, {Ahumada}, {Alonso}, {Alonso-Garcia}, {Arias},
  {Bandyopadhyay}, {Barb{\'a}}, {Barbuy}, {Bedin}, {Bica}, {Borissova},
  {Bronfman}, {Carraro}, {Catelan}, {Clari{\'a}}, {Cross}, {de Grijs},
  {D{\'e}k{\'a}ny}, {Drew}, {Fari{\~n}a}, {Feinstein}, {Fern{\'a}ndez
  Laj{\'u}s}, {Gamen}, {Geisler}, {Gieren}, {Goldman}, {Gonzalez}, {Gunthardt},
  {Gurovich}, {Hambly}, {Irwin}, {Ivanov}, {Jord{\'a}n}, {Kerins}, {Kinemuchi},
  {Kurtev}, {L{\'o}pez-Corredoira}, {Maccarone}, {Masetti}, {Merlo},
  {Messineo}, {Mirabel}, {Monaco}, {Morelli}, {Padilla}, {Palma}, {Parisi},
  {Pignata}, {Rejkuba}, {Roman-Lopes}, {Sale}, {Schreiber}, {Schr{\"o}der},
  {Smith}, {}, {Soto}, {Tamura}, {Tappert}, {Thompson}, {Toledo}, {Zoccali}, \&
  {Pietrzynski}}]{Minniti2010}
{Minniti}, D., {Lucas}, P.~W., {Emerson}, J.~P., {et~al.} 2010, \na, 15, 433

\bibitem[{{Minniti} {et~al.}(2021{\natexlab{b}}){Minniti}, {Palma}, {Camargo},
  {Chijani-Saballa}, {Alonso-Garc{\'\i}a}, {Clari{\'a}}, {Dias}, {G{\'o}mez},
  {Pullen}, \& {Saito}}]{Minniti2021_M48}
{Minniti}, D., {Palma}, T., {Camargo}, D., {et~al.} 2021{\natexlab{b}}, \aap,
  652, A129

\bibitem[{Minniti {et~al.}(2017{\natexlab{b}})Minniti, Palma,
  D{\'{e}}k{\'{a}}ny, Hempel, Rejkuba, Pullen, Alonso-Garc{\'{\i}}a,
  Barb{\'{a}}, Barbuy, Bica, Bonatto, Borissova, Catelan, Carballo-Bello,
  Chene, Clari{\'{a}}, Cohen, Ramos, Dias, Emerson, Froebrich, Buckner,
  Geisler, Gonzalez, Gran, Hagdu, Irwin, Ivanov, Kurtev, Lucas, Majaess, Mauro,
  Moni-Bidin, Navarrete, Alegr{\'{\i}}a, Saito, Valenti, \&
  Zoccali}]{Minniti2017_FSR}
Minniti, D., Palma, T., D{\'{e}}k{\'{a}}ny, I., {et~al.} 2017{\natexlab{b}},
  ApJ, 838, L14

\bibitem[{{Moni Bidin} {et~al.}(2011){Moni Bidin}, {Mauro}, {Geisler},
  {Minniti}, {Catelan}, {Hempel}, {Valenti}, {Valcarce}, {Alonso-Garc{\'\i}a},
  {Borissova}, {Carraro}, {Lucas}, {Chen{\'e}}, {Zoccali}, \&
  {Kurtev}}]{MoniBidin2011}
{Moni Bidin}, C., {Mauro}, F., {Geisler}, D., {et~al.} 2011, \aap, 535, A33

\bibitem[{{Muraveva} {et~al.}(2018){Muraveva}, {Delgado}, {Clementini},
  {Sarro}, \& {Garofalo}}]{Muraveva2018}
{Muraveva}, T., {Delgado}, H.~E., {Clementini}, G., {Sarro}, L.~M., \&
  {Garofalo}, A. 2018, \mnras, 481, 1195

\bibitem[{{Muraveva} {et~al.}(2015){Muraveva}, {Palmer}, {Clementini}, {Luri},
  {Cioni}, {Moretti}, {Marconi}, {Ripepi}, \& {Rubele}}]{Muraveva2015}
{Muraveva}, T., {Palmer}, M., {Clementini}, G., {et~al.} 2015, \apj, 807, 127

\bibitem[{{Nataf} {et~al.}(2013){Nataf}, {Gould}, {Fouqu{\'e}}, {Gonzalez},
  {Johnson}, {Skowron}, {Udalski}, {Szyma{\'n}ski}, {Kubiak},
  {Pietrzy{\'n}ski}, {Soszy{\'n}ski}, {Ulaczyk}, {Wyrzykowski}, \&
  {Poleski}}]{Nataf2013}
{Nataf}, D.~M., {Gould}, A., {Fouqu{\'e}}, P., {et~al.} 2013, \apj, 769, 88

\bibitem[{{Navarrete} {et~al.}(2017){Navarrete}, {Catelan}, {Contreras Ramos},
  {Alonso-Garc{\'\i}a}, {Gran}, {D{\'e}k{\'a}ny}, \& {Minniti}}]{Navarrete2017}
{Navarrete}, C., {Catelan}, M., {Contreras Ramos}, R., {et~al.} 2017, \aap,
  604, A120

\bibitem[{{Obasi} {et~al.}(2021){Obasi}, {G{\'o}mez}, {Minniti}, \&
  {Alonso-Garc{\'\i}a}}]{Obasi2021}
{Obasi}, C., {G{\'o}mez}, M., {Minniti}, D., \& {Alonso-Garc{\'\i}a}, J. 2021,
  \aap, 654, A39

\bibitem[{Pallanca {et~al.}(2021)Pallanca, Ferraro, Lanzoni, Crociati,
  Saracino, Dalessandro, Origlia, Rich, Valenti, Geisler, Mauro, Villanova,
  Bidin, \& Beccari}]{Pallanca2021a}
Pallanca, C., Ferraro, F.~R., Lanzoni, B., {et~al.} 2021, 917, 92

\bibitem[{{Pallanca} {et~al.}(2021){Pallanca}, {Lanzoni}, {Ferraro},
  {Casagrande}, {Saracino}, {Purohith Bhaskar Bhat}, {Leanza}, {Dalessandro},
  \& {Vesperini}}]{Pallanca2021b}
{Pallanca}, C., {Lanzoni}, B., {Ferraro}, F.~R., {et~al.} 2021, \apj, 913, 137

\bibitem[{Parzen(1962)}]{Parzen1962}
Parzen, E. 1962, The Annals of Mathematical Statistics, 33, 1065

\bibitem[{{Romero-Colmenares} {et~al.}(2021){Romero-Colmenares},
  {Fern{\'a}ndez-Trincado}, {Geisler}, {Souza}, {Villanova}, {Longa-Pe{\~n}a},
  {Minniti}, {Beers}, {Bidin}, {Perez-Villegas}, {Moreno}, {Garro}, {Baeza},
  {Henao}, {Barbuy}, {Alonso-Garc{\'\i}a}, {Cohen}, {Lane}, \&
  {Mu{\~n}oz}}]{RomeroColmenares2021}
{Romero-Colmenares}, M., {Fern{\'a}ndez-Trincado}, J.~G., {Geisler}, D.,
  {et~al.} 2021, \aap, 652, A158

\bibitem[{Rosenblatt(1956)}]{Rosenblatt1956}
Rosenblatt, M. 1956, The Annals of Mathematical Statistics, 27, 832

\bibitem[{{Ruiz-Dern} {et~al.}(2018){Ruiz-Dern}, {Babusiaux}, {Arenou},
  {Turon}, \& {Lallement}}]{RuizDern2018}
{Ruiz-Dern}, L., {Babusiaux}, C., {Arenou}, F., {Turon}, C., \& {Lallement}, R.
  2018, \aap, 609, A116

\bibitem[{{Ryu} \& {Lee}(2018)}]{Ryu2018}
{Ryu}, J. \& {Lee}, M.~G. 2018, \apjl, 863, L38

\bibitem[{{Saito} {et~al.}(2012){Saito}, {Hempel}, {Minniti}, {Lucas},
  {Rejkuba}, {Toledo}, {Gonzalez}, {Alonso-Garc{\'\i}a}, {Irwin},
  {Gonzalez-Solares}, {Hodgkin}, {Lewis}, {Cross}, {Ivanov}, {Kerins},
  {Emerson}, {Soto}, {Am{\^o}res}, {Gurovich}, {D{\'e}k{\'a}ny}, {Angeloni},
  {Beamin}, {Catelan}, {Padilla}, {Zoccali}, {Pietrukowicz}, {Moni Bidin},
  {Mauro}, {Geisler}, {Folkes}, {Sale}, {Borissova}, {Kurtev}, {Ahumada},
  {Alonso}, {Adamson}, {Arias}, {Bandyopadhyay}, {Barb{\'a}}, {Barbuy},
  {Baume}, {Bedin}, {Bellini}, {Benjamin}, {Bica}, {Bonatto}, {Bronfman},
  {Carraro}, {Chen{\`e}}, {Clari{\'a}}, {Clarke}, {Contreras}, {Corvill{\'o}n},
  {de Grijs}, {Dias}, {Drew}, {Fari{\~n}a}, {Feinstein},
  {Fern{\'a}ndez-Laj{\'u}s}, {Gamen}, {Gieren}, {Goldman},
  {Gonz{\'a}lez-Fern{\'a}ndez}, {Grand}, {Gunthardt}, {Hambly}, {Hanson},
  {He{\l}miniak}, {Hoare}, {Huckvale}, {Jord{\'a}n}, {Kinemuchi}, {Longmore},
  {L{\'o}pez-Corredoira}, {Maccarone}, {Majaess}, {Mart{\'\i}n}, {Masetti},
  {Mennickent}, {Mirabel}, {Monaco}, {Morelli}, {Motta}, {Palma}, {Parisi},
  {Parker}, {Pe{\~n}aloza}, {Pietrzy{\'n}ski}, {Pignata}, {Popescu}, {Read},
  {Rojas}, {Roman-Lopes}, {Ruiz}, {Saviane}, {Schreiber}, {Schr{\"o}der},
  {Sharma}, {Smith}, {Sodr{\'e}}, {Stead}, {Stephens}, {Tamura}, {Tappert},
  {Thompson}, {Valenti}, {Vanzi}, {Walton}, {Weidmann}, \&
  {Zijlstra}}]{Saito2012}
{Saito}, R.~K., {Hempel}, M., {Minniti}, D., {et~al.} 2012, \aap, 537, A107

\bibitem[{{Salaris} \& {Cassisi}(2005)}]{SalarisCassisi2005}
{Salaris}, M. \& {Cassisi}, S. 2005, {Evolution of Stars and Stellar
  Populations}

\bibitem[{Salaris \& Girardi(2002)}]{SalarisGirardi2002}
Salaris, M. \& Girardi, L. 2002, MNRAS, 337, 332

\bibitem[{{Salaris} \& {Weiss}(2002)}]{Salaris2002}
{Salaris}, M. \& {Weiss}, A. 2002, \aap, 388, 492

\bibitem[{{Salaris} {et~al.}(2004){Salaris}, {Weiss}, \&
  {Percival}}]{Salaris2004}
{Salaris}, M., {Weiss}, A., \& {Percival}, S.~M. 2004, \aap, 414, 163

\bibitem[{{Schlafly} \& {Finkbeiner}(2011)}]{Schlafly2011}
{Schlafly}, E.~F. \& {Finkbeiner}, D.~P. 2011, \apj, 737, 103

\bibitem[{{Schlafly} {et~al.}(2018){Schlafly}, {Green}, {Lang}, {Daylan},
  {Finkbeiner}, {Lee}, {Meisner}, {Schlegel}, \& {Valdes}}]{Schlafly2018}
{Schlafly}, E.~F., {Green}, G.~M., {Lang}, D., {et~al.} 2018, \apjs, 234, 39

\bibitem[{{Shao} \& {Li}(2019)}]{Shao2019}
{Shao}, Z. \& {Li}, L. 2019, \mnras, 489, 3093

\bibitem[{{Skrutskie} {et~al.}(2006){Skrutskie}, {Cutri}, {Stiening},
  {Weinberg}, {Schneider}, {Carpenter}, {Beichman}, {Capps}, {Chester},
  {Elias}, {Huchra}, {Liebert}, {Lonsdale}, {Monet}, {Price}, {Seitzer},
  {Jarrett}, {Kirkpatrick}, {Gizis}, {Howard}, {Evans}, {Fowler}, {Fullmer},
  {Hurt}, {Light}, {Kopan}, {Marsh}, {McCallon}, {Tam}, {Van Dyk}, \&
  {Wheelock}}]{Skrutskie2006}
{Skrutskie}, M.~F., {Cutri}, R.~M., {Stiening}, R., {et~al.} 2006, \aj, 131,
  1163

\bibitem[{{Sneden} {et~al.}(1978){Sneden}, {Gehrz}, {Hackwell}, {York}, \&
  {Snow}}]{Sneden1978}
{Sneden}, C., {Gehrz}, R.~D., {Hackwell}, J.~A., {York}, D.~G., \& {Snow},
  T.~P. 1978, \apj, 223, 168

\bibitem[{{Soto} {et~al.}(2019){Soto}, {Barb{\'a}}, {Minniti}, {Kunder},
  {Majaess}, {Nilo-Castell{\'o}n}, {Alonso-Garc{\'\i}a}, {Leone}, {Morelli},
  {Haikala}, {Firpo}, {Lucas}, {Emerson}, {Moni Bidin}, {Geisler}, {Saito},
  {Gurovich}, {Contreras Ramos}, {Rejkuba}, {Barbieri}, {Roman-Lopes},
  {Hempel}, {Alonso}, {Baravalle}, {Borissova}, {Kurtev}, \&
  {Milla}}]{Soto2019}
{Soto}, M., {Barb{\'a}}, R., {Minniti}, D., {et~al.} 2019, \mnras, 488, 2650

\bibitem[{{Souza} {et~al.}(2021){Souza}, {Valentini}, {Barbuy},
  {P{\'e}rez-Villegas}, {Chiappini}, {Ortolani}, {Nardiello}, {Dias}, {Anders},
  \& {Bica}}]{Souza2021}
{Souza}, S.~O., {Valentini}, M., {Barbuy}, B., {et~al.} 2021, \aap, 656, A78

\bibitem[{{Surot} {et~al.}(2019){Surot}, {Valenti}, {Hidalgo}, {Zoccali},
  {S{\"o}kmen}, {Rejkuba}, {Minniti}, {Gonzalez}, {Cassisi}, {Renzini}, \&
  {Weiss}}]{Surot2019}
{Surot}, F., {Valenti}, E., {Hidalgo}, S.~L., {et~al.} 2019, \aap, 623, A168

\bibitem[{{Vasiliev} \& {Baumgardt}(2021)}]{Vasiliev2021}
{Vasiliev}, E. \& {Baumgardt}, H. 2021, \mnras, 505, 5978

\bibitem[{York {et~al.}(2000)York, Adelman, John E.~Anderson, Anderson, Annis,
  Bahcall, Bakken, Barkhouser, Bastian, Berman, Boroski, Bracker, Briegel,
  Briggs, Brinkmann, Brunner, Burles, Carey, Carr, Castander, Chen, Colestock,
  Connolly, Crocker, Csabai, Czarapata, Davis, Doi, Dombeck, Eisenstein,
  Ellman, Elms, Evans, Fan, Federwitz, Fiscelli, Friedman, Frieman, Fukugita,
  Gillespie, Gunn, Gurbani, de~Haas, Haldeman, Harris, Hayes, Heckman,
  Hennessy, Hindsley, Holm, Holmgren, hao Huang, Hull, Husby, Ichikawa,
  Ichikawa, Ivezi{\'{c}}, Kent, Kim, Kinney, Klaene, Kleinman, Kleinman, Knapp,
  Korienek, Kron, Kunszt, Lamb, Lee, Leger, Limmongkol, Lindenmeyer, Long,
  Loomis, Loveday, Lucinio, Lupton, MacKinnon, Mannery, Mantsch, Margon,
  McGehee, McKay, Meiksin, Merelli, Monet, Munn, Narayanan, Nash, Neilsen,
  Neswold, Newberg, Nichol, Nicinski, Nonino, Okada, Okamura, Ostriker, Owen,
  Pauls, Peoples, Peterson, Petravick, Pier, Pope, Pordes, Prosapio,
  Rechenmacher, Quinn, Richards, Richmond, Rivetta, Rockosi, Ruthmansdorfer,
  Sandford, Schlegel, Schneider, Sekiguchi, Sergey, Shimasaku, Siegmund, Smee,
  Smith, Snedden, Stone, Stoughton, Strauss, Stubbs, SubbaRao, Szalay, Szapudi,
  Szokoly, Thakar, Tremonti, Tucker, Uomoto, Berk, Vogeley, Waddell, i~Wang,
  Watanabe, Weinberg, Yanny, \& Yasuda}]{York_2000}
York, D.~G., Adelman, J., John E.~Anderson, J., {et~al.} 2000, AJ, 120, 1579

\end{thebibliography}

\newpage

\begin{appendix}
\section{Optical images for the star clusters}
\begin{figure}[!htb]
\centering
\onecolumn
\includegraphics[width=5cm, height=5cm]{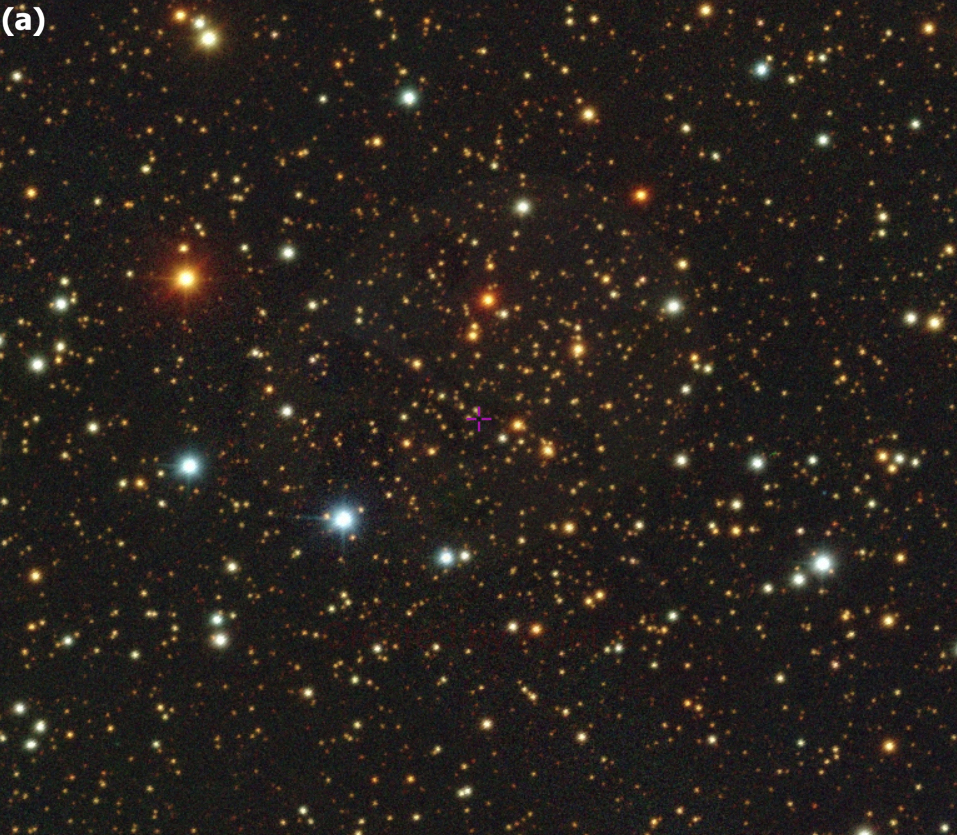} 
\includegraphics[width=5cm, height=5cm]{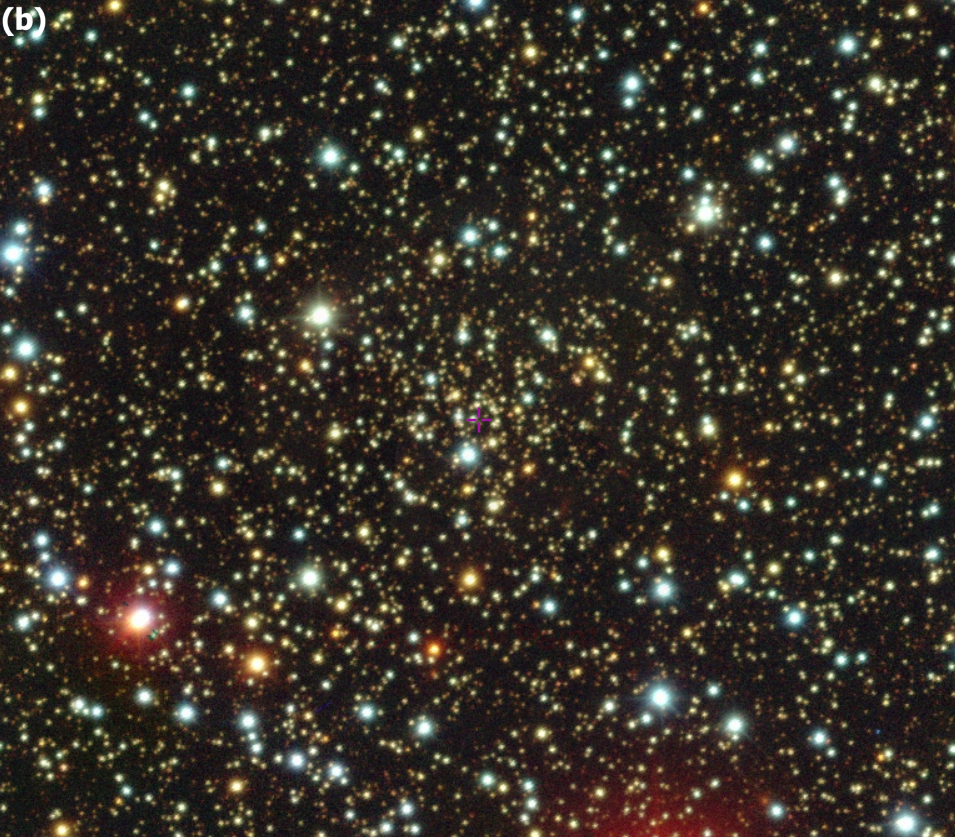} 
\includegraphics[width=5cm, height=5cm]{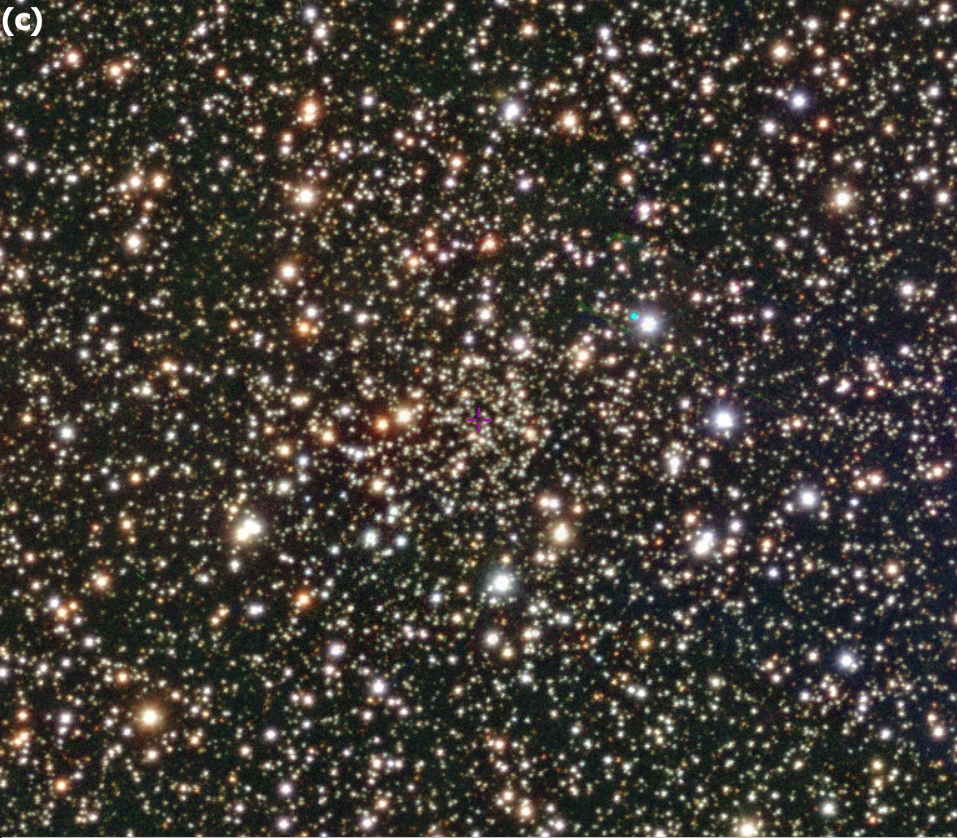} 
\includegraphics[width=5cm, height=5cm]{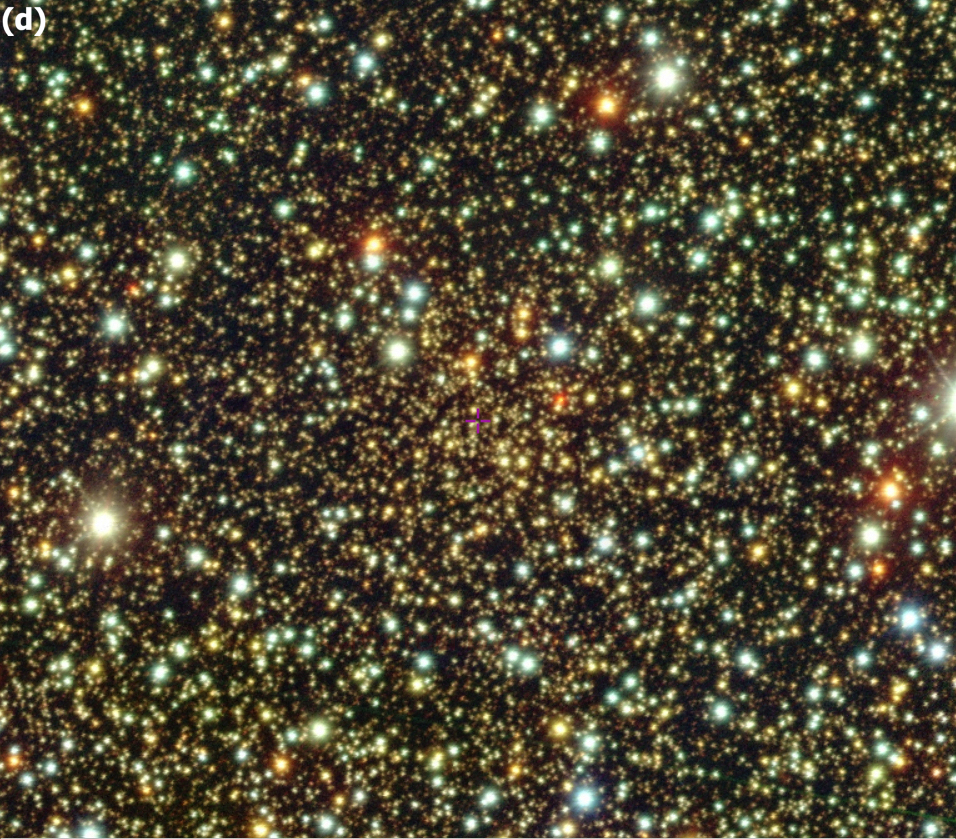} 
\includegraphics[width=5cm, height=5cm]{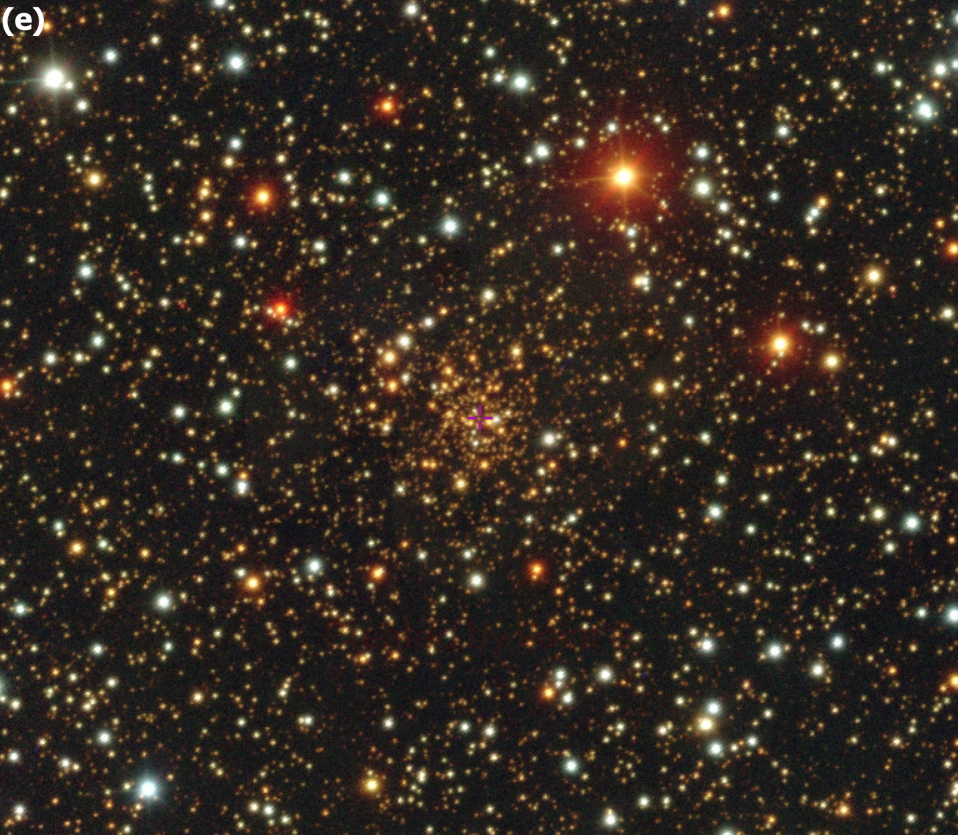} 
\includegraphics[width=5cm, height=5cm]{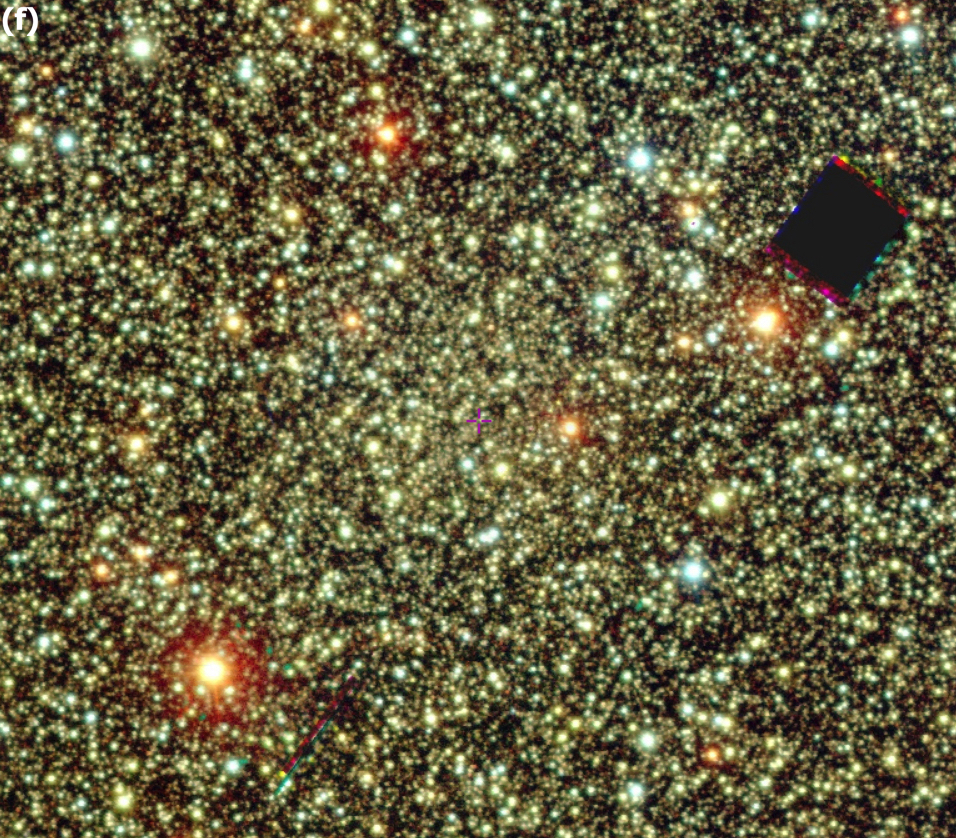} 
\includegraphics[width=5cm, height=5cm]{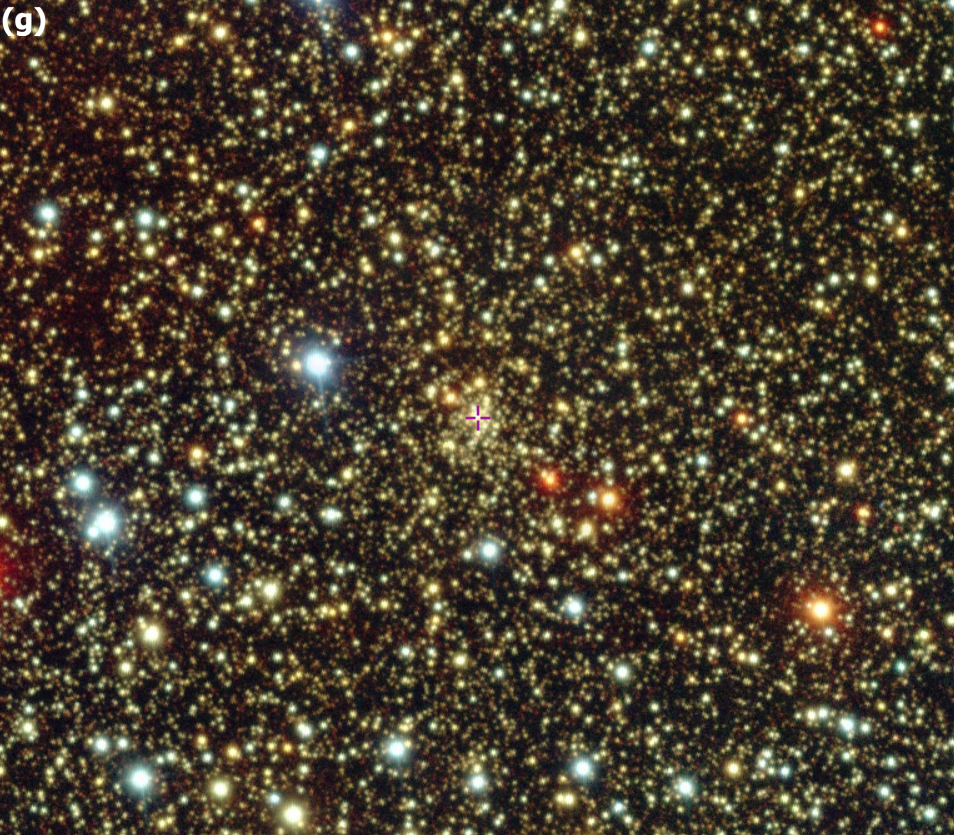} 
\includegraphics[width=5cm, height=5cm]{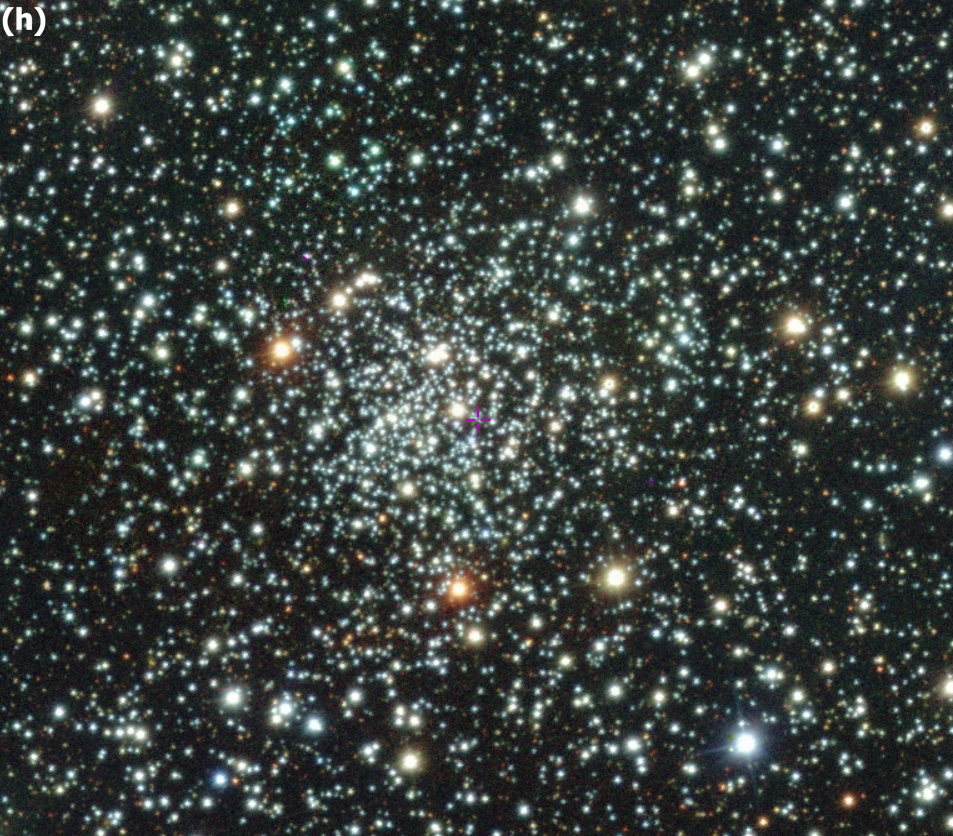} 
\includegraphics[width=5cm, height=5cm]{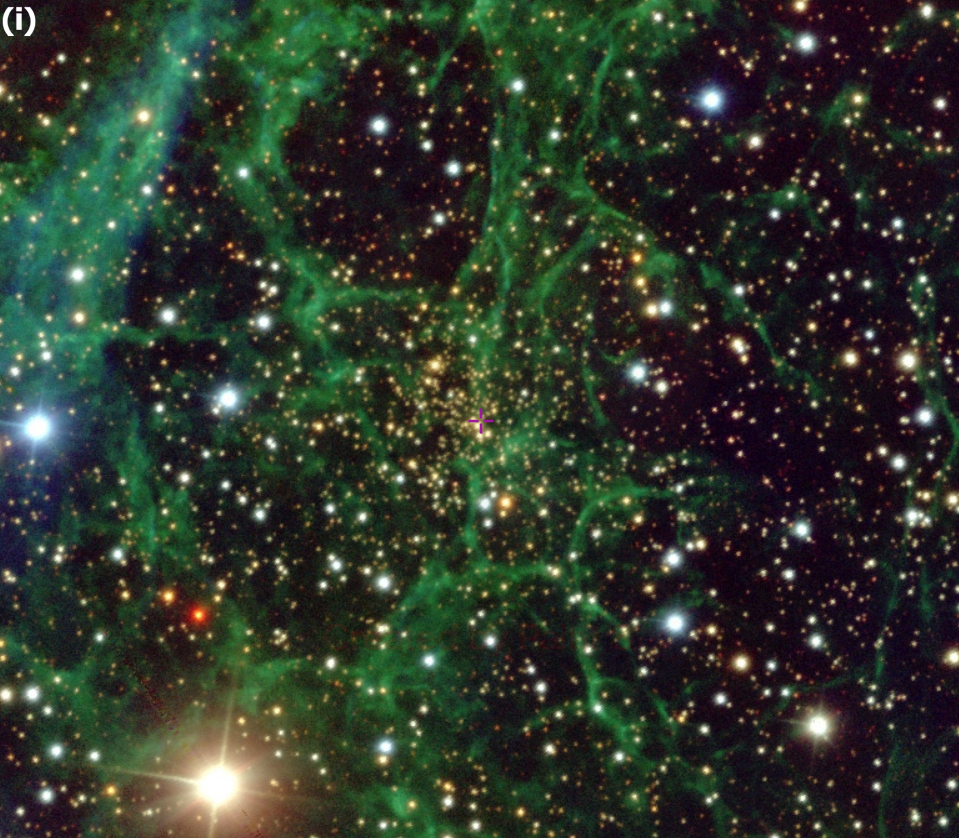} 
\includegraphics[width=5cm, height=5cm]{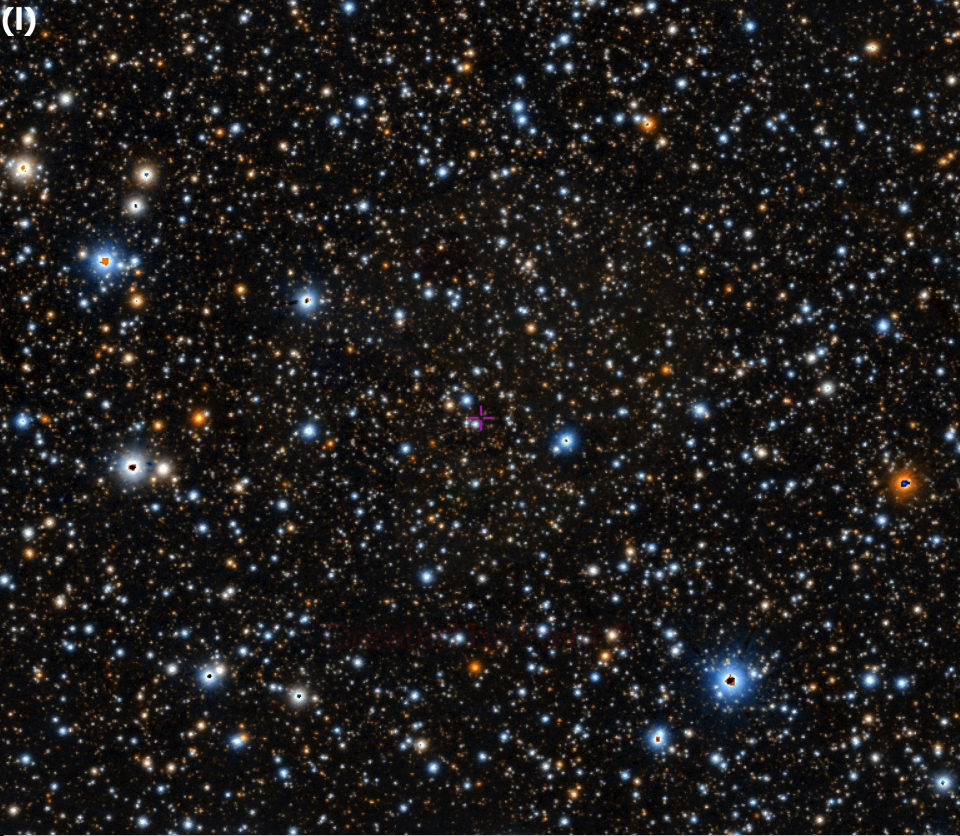} 
\includegraphics[width=5cm, height=5cm]{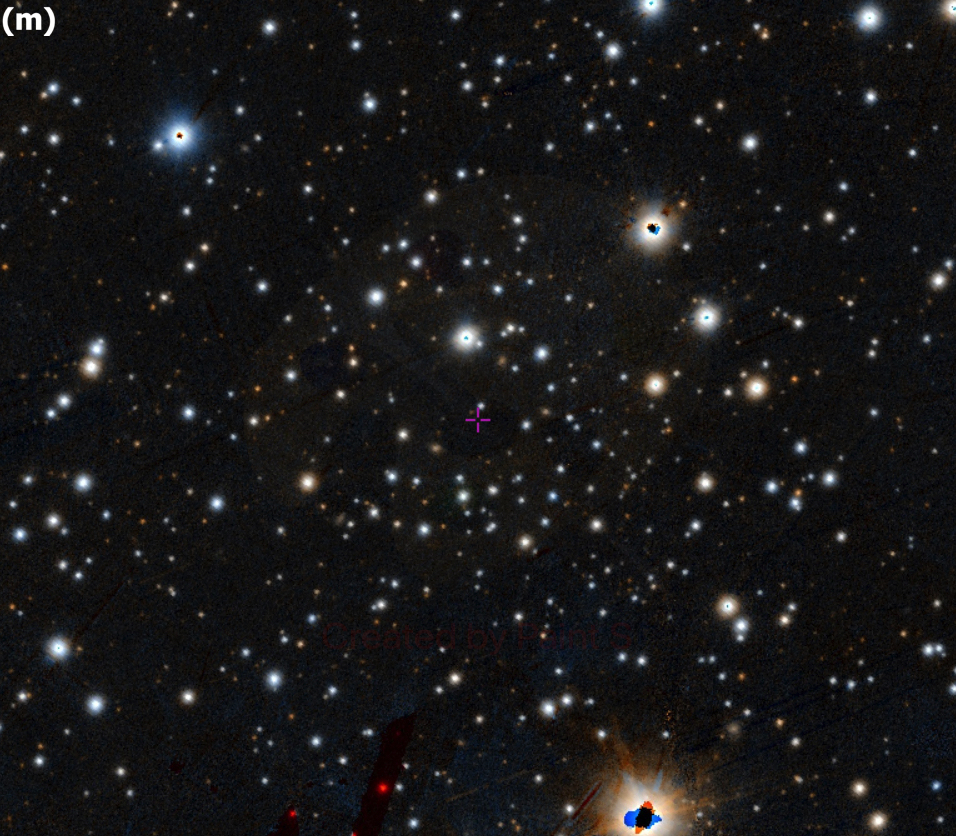} 
\includegraphics[width=5cm, height=5cm]{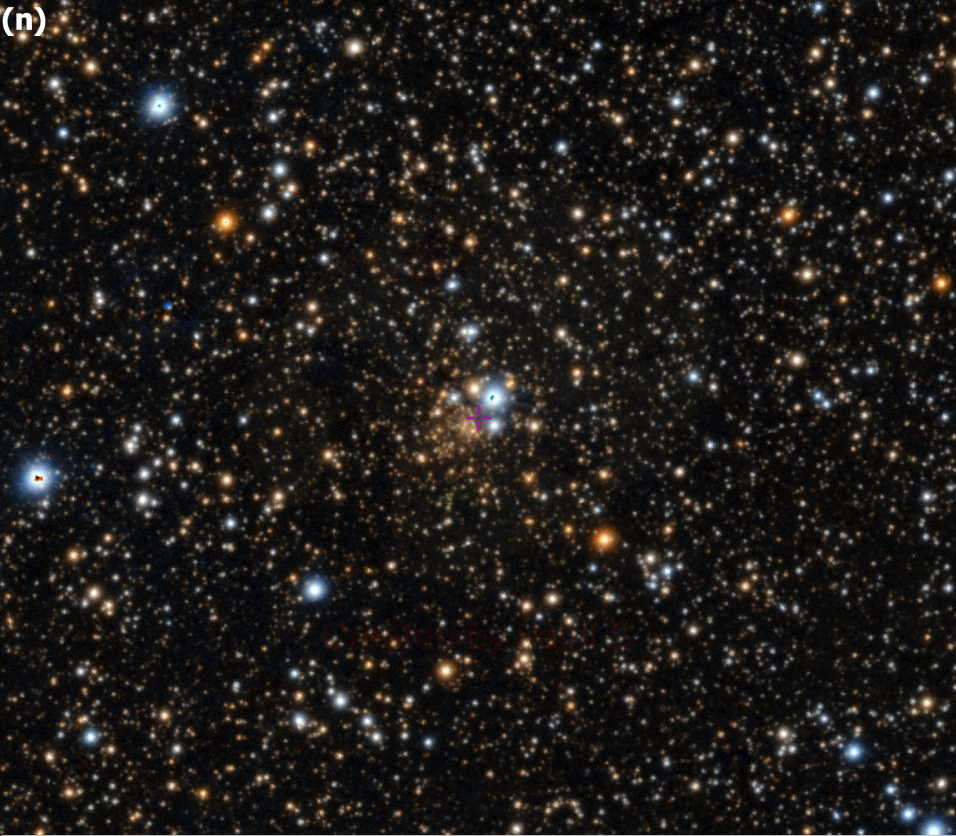}  
\caption{Optical images of the sky regions centred on the star clusters analysed in the present work: (a) Kronberger~99, (b) Kronberger~100, (c) Kronberger~119, (d) Kronberger~143, (e) Patchick~122, (f) Patchick~125, (g) Patchick~126, (h) ESO 92-18, (i) Ferrero~54, (l) FSR~190, (m) Gaia~2, and (n) Riddle~15.
We used DECaPS colours for most of the targets, except for Riddle~15 and FSR~190 for which we used PanSTARRS colours.  We use a coverage of $5.50' \times 6.25'$ for the most of  the clusters, except for FSR~190 ($11.5'\times 13'$). }
\label{images}
\end{figure}


\section{VPMs and CMDs for the cluster sample}

\begin{figure}[!htb]
\centering
\onecolumn
\includegraphics[width=5.5cm, height=5cm]{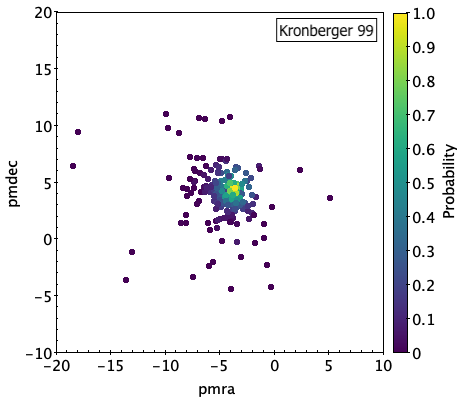} 
\includegraphics[width=5.5cm, height=5cm]{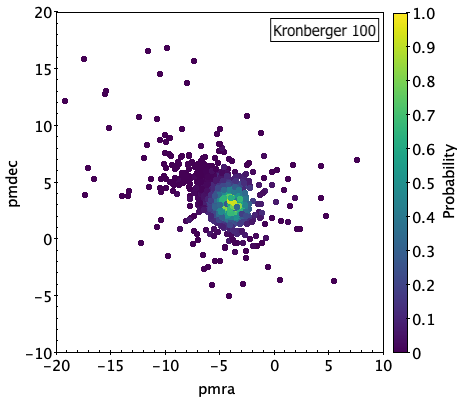} 
\includegraphics[width=5.5cm, height=5cm]{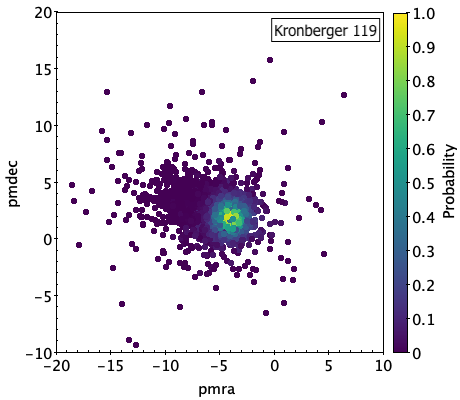} 
\includegraphics[width=5.5cm, height=5cm]{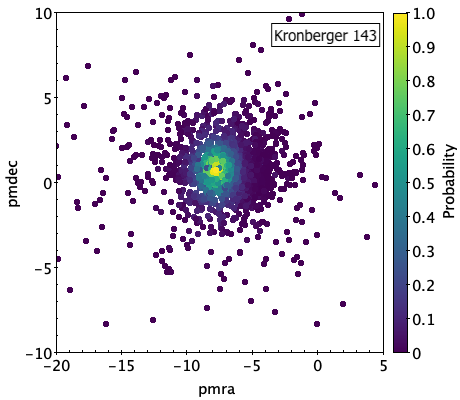} 
\includegraphics[width=5.5cm, height=5cm]{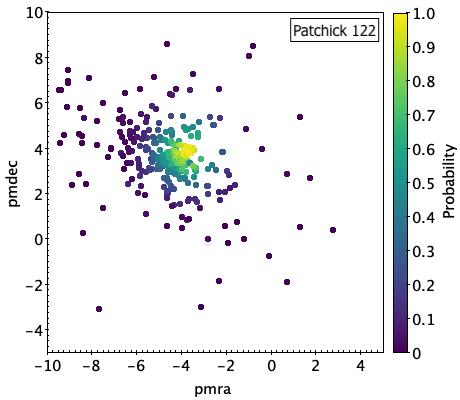} 
\includegraphics[width=5.5cm, height=5cm]{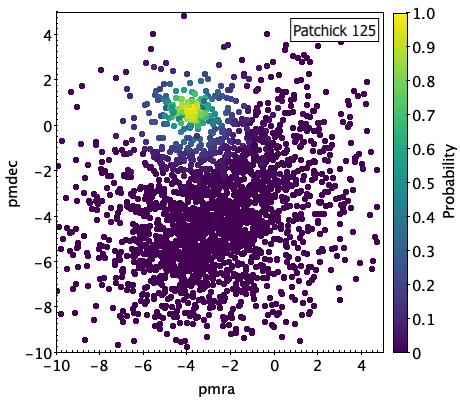} 
\includegraphics[width=5.5cm, height=5cm]{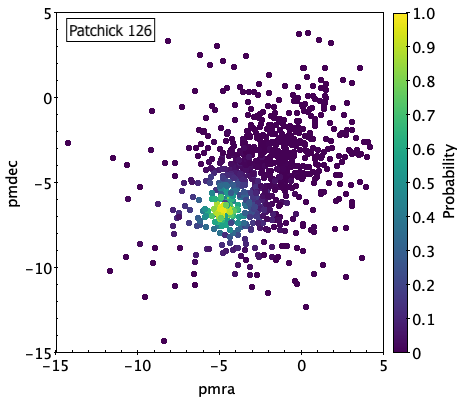} 
\includegraphics[width=5.5cm, height=5cm]{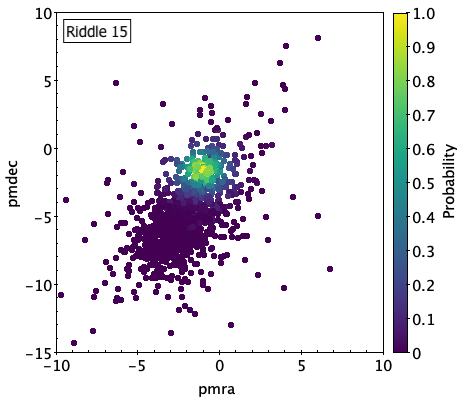} 
\includegraphics[width=5.5cm, height=5cm]{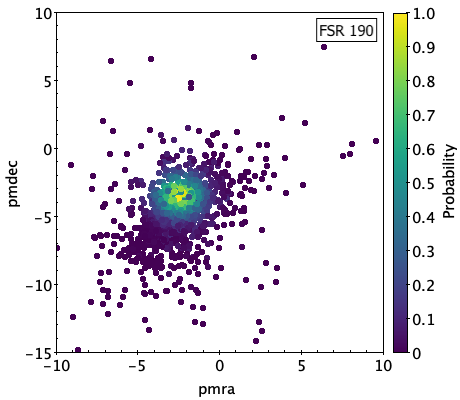} 
\includegraphics[width=5.5cm, height=5cm]{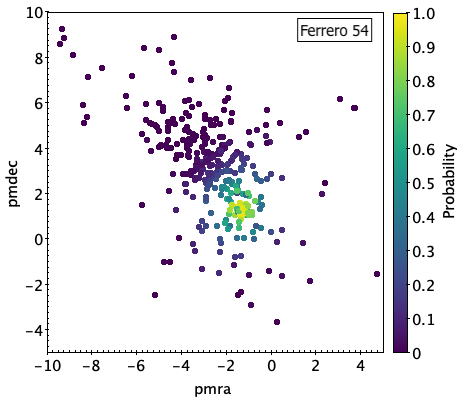} 
\includegraphics[width=5.5cm, height=5cm]{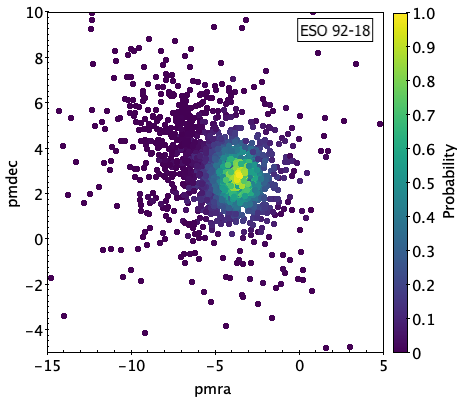} 
\includegraphics[width=5.5cm, height=5cm]{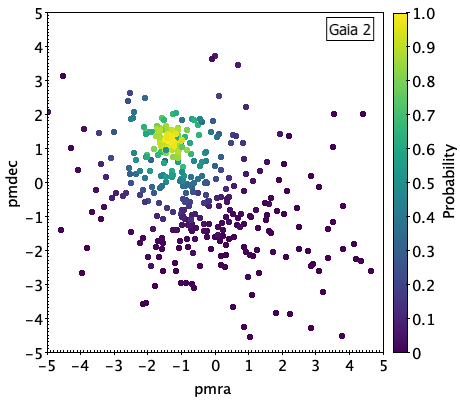} 
\caption{VPM diagrams (Table \ref{DP}).  The brightest cloud represents the highest PM probability (within a radius $<1$ mas yr$^{-1}$), which indicates the presence of the cluster, as described in Section \ref{decoproc}. }
\label{VPM:fig}
\end{figure}

\begin{figure}[!htb]
\centering
\onecolumn
\includegraphics[width=5cm, height=5.2cm]{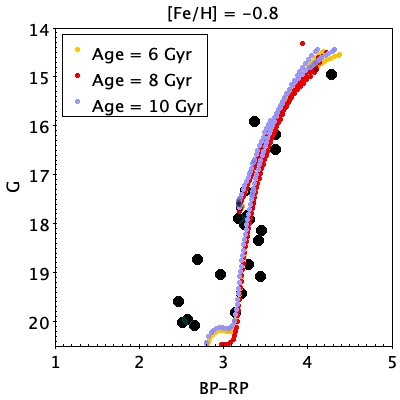} 
\includegraphics[width=5cm, height=5.2cm]{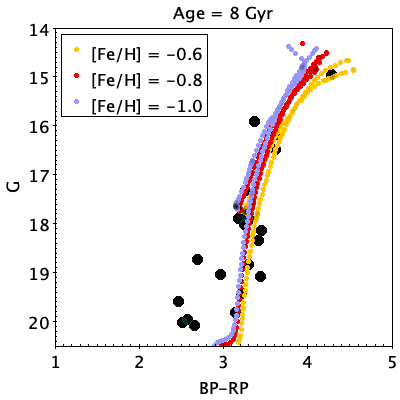} 
\includegraphics[width=5cm, height=5cm]{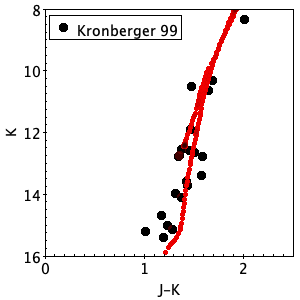} 
\includegraphics[width=5cm, height=5.2cm]{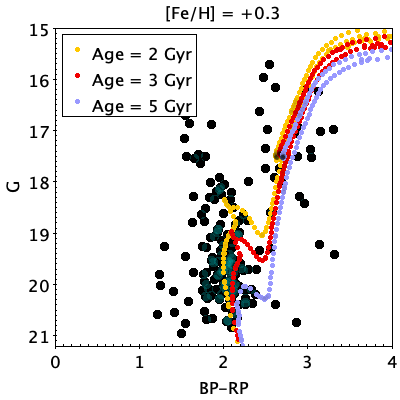} 
\includegraphics[width=5cm, height=5.2cm]{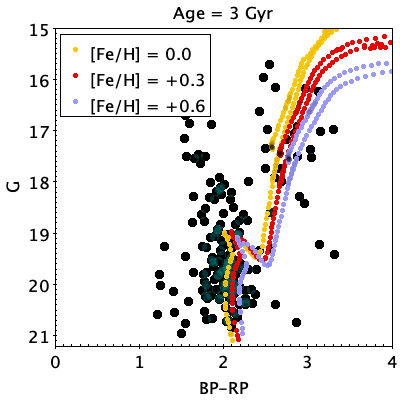} 
\includegraphics[width=5cm, height=5cm]{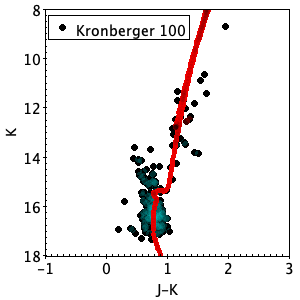} 
\includegraphics[width=5cm, height=5.2cm]{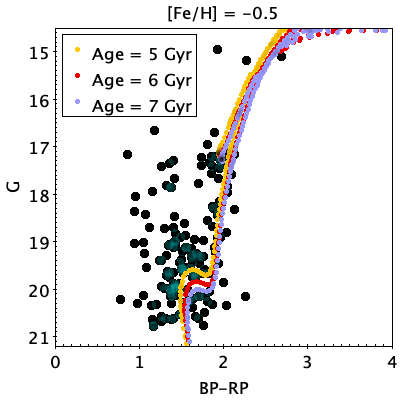} 
\includegraphics[width=5cm, height=5.2cm]{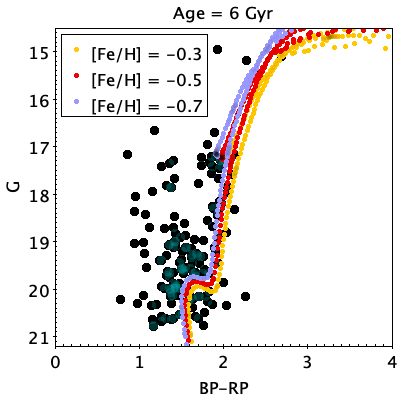} 
\includegraphics[width=5cm, height=5cm]{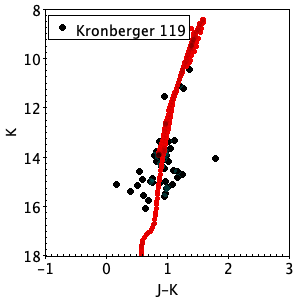} 
\includegraphics[width=5cm, height=5.2cm]{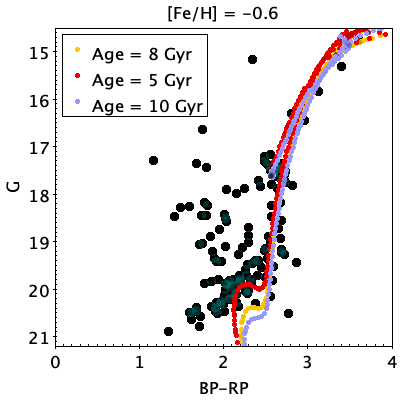} 
\includegraphics[width=5cm, height=5.2cm]{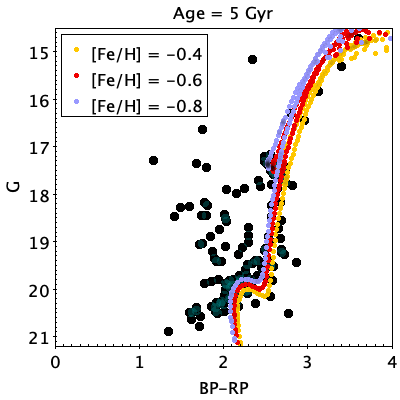} 
\includegraphics[width=5cm, height=5cm]{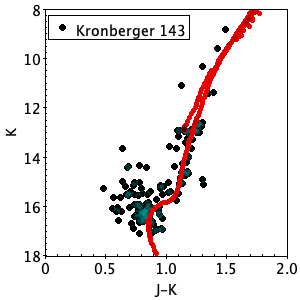}  
\caption{\textit{Left and middle panels:} optical Gaia EDR3 CMDs for the PM-member stars.  We fit a family of PARSEC isochrones in order to derive the uncertainties both in the age and in the metallicity estimates. \textit{Right panel:} NIR (VVVX and/or 2MASS) CMDs for the PM-member stars. The red lines, in all panels, depict the PARSEC isochrones with age and metallicity listed in Table \ref{parameters} for each cluster. In the CMDs, the cyan areas represent over-densities, whereas the black ones are lower-densities. }
\label{CMD}
\end{figure}

\begin{figure}[!htb]
\ContinuedFloat
\captionsetup{list=off, format=continued}
\centering
\onecolumn
\includegraphics[width=5cm, height=5.2cm]{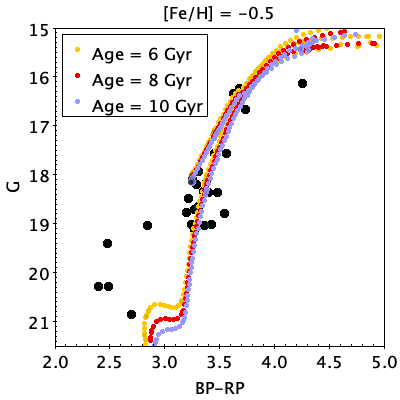} 
\includegraphics[width=5cm, height=5.2cm]{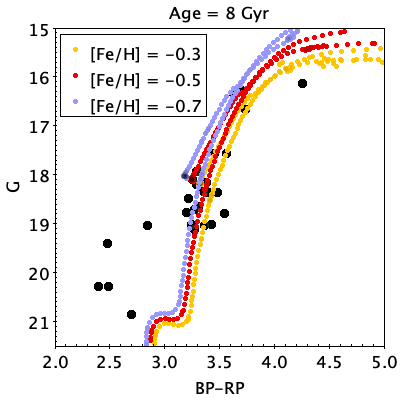} 
\includegraphics[width=5cm, height=5cm]{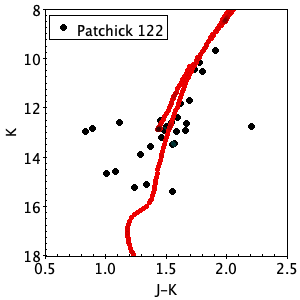}
\includegraphics[width=5cm, height=5.2cm]{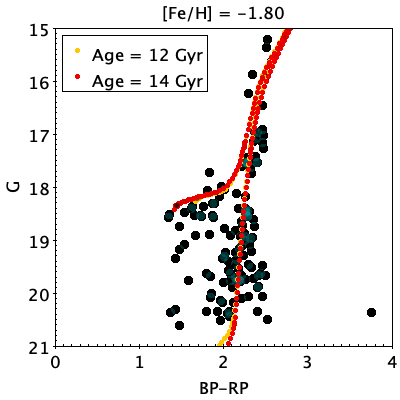} 
\includegraphics[width=5cm, height=5.2cm]{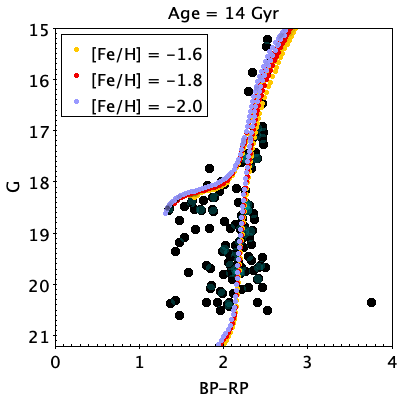} 
\includegraphics[width=5cm, height=5cm]{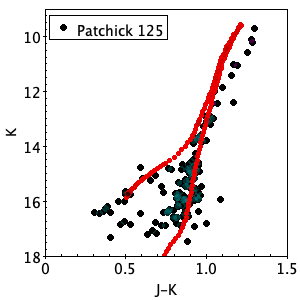} 
\includegraphics[width=5cm, height=5.2cm]{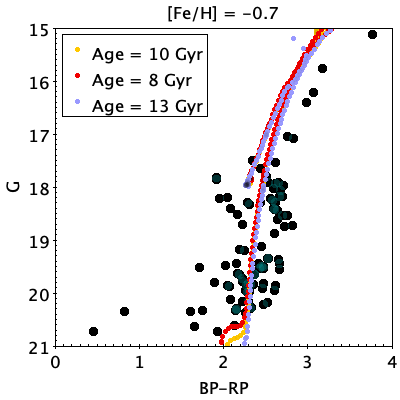} 
\includegraphics[width=5cm, height=5.2cm]{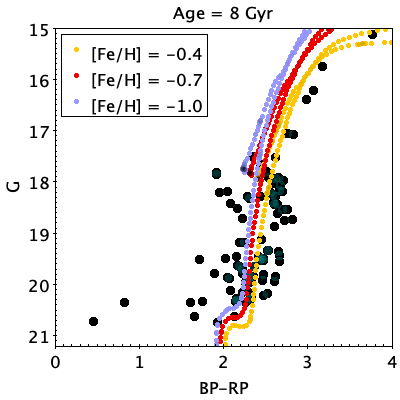} 
\includegraphics[width=5cm, height=5cm]{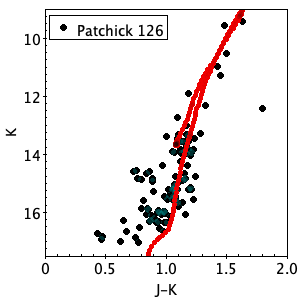} 
\includegraphics[width=5cm, height=5.2cm]{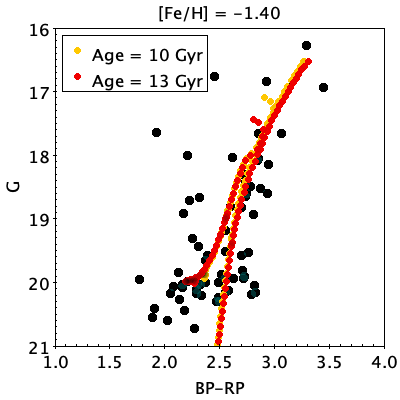} 
\includegraphics[width=5cm, height=5.2cm]{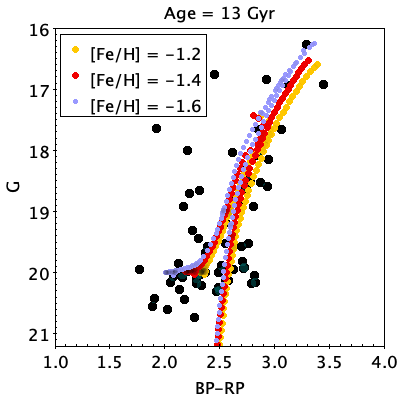} 
\includegraphics[width=5cm, height=5cm]{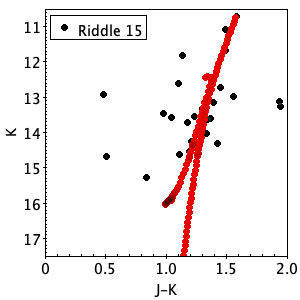} 
\caption{}
\end{figure}

\begin{figure}[!htb]
\ContinuedFloat
\captionsetup{list=off, format=continued}
\centering
\onecolumn
\includegraphics[width=5cm, height=5.2cm]{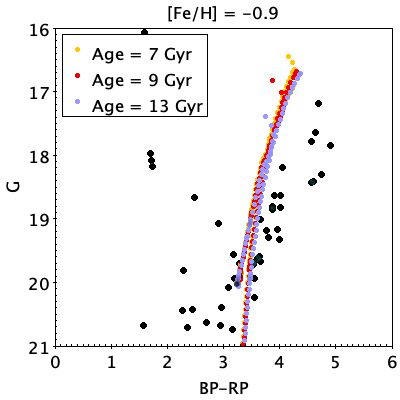} 
\includegraphics[width=5cm, height=5.2cm]{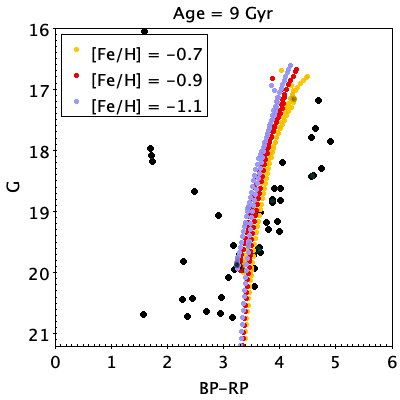} 
\includegraphics[width=5cm, height=5cm]{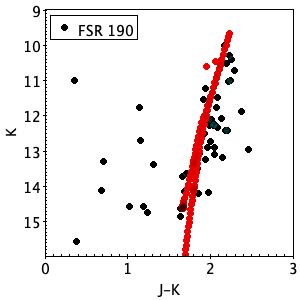} 
\includegraphics[width=5cm, height=5.2cm]{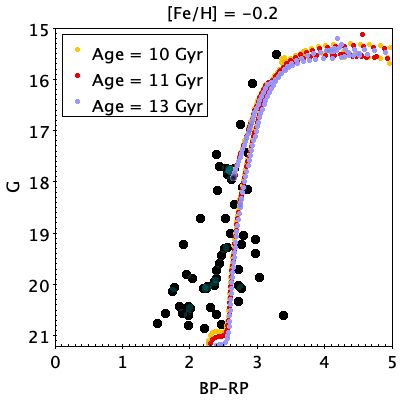} 
\includegraphics[width=5cm, height=5.2cm]{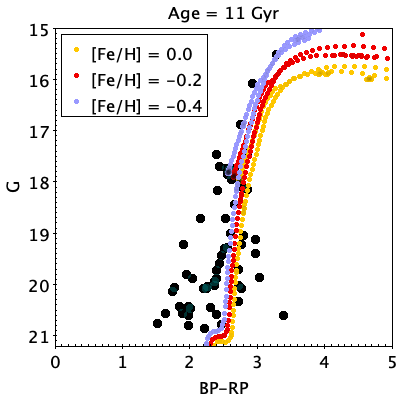} 
\includegraphics[width=5cm, height=5cm]{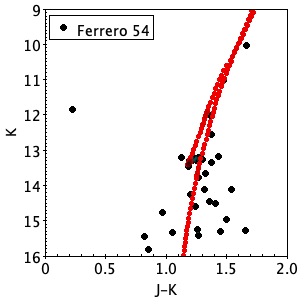} 
\includegraphics[width=5cm, height=5.2cm]{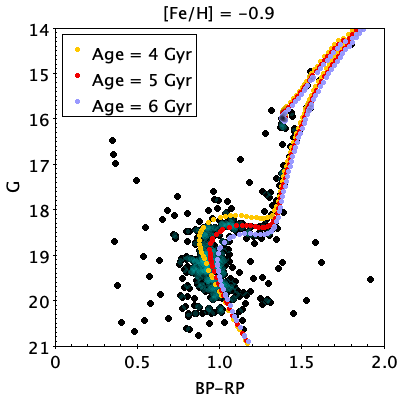} 
\includegraphics[width=5cm, height=5.2cm]{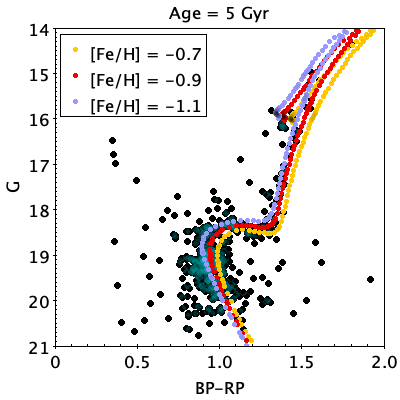} 
\includegraphics[width=5cm, height=5cm]{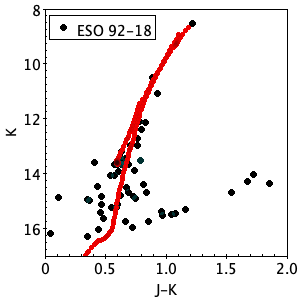} 
\includegraphics[width=5cm, height=5.2cm]{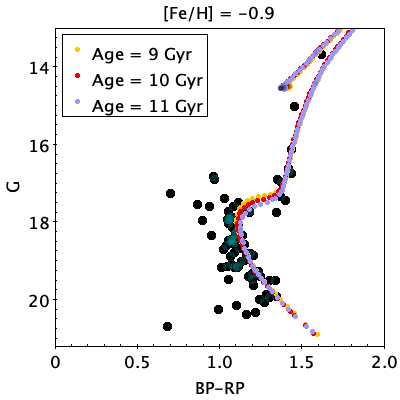} 
\includegraphics[width=5cm, height=5.2cm]{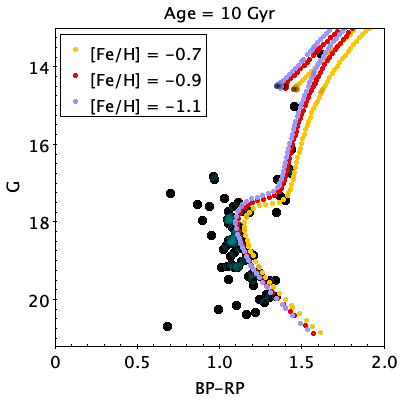} 
\includegraphics[width=5cm, height=5cm]{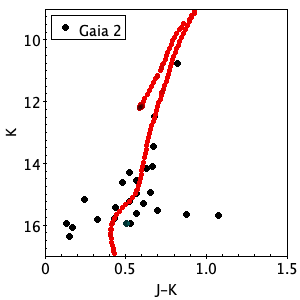} 
\caption{}
\end{figure}



\end{appendix}

\end{document}